\documentclass[longauth,traditabstract]{aa}

\usepackage[nonamebreak]{natbib}
\usepackage[stable]{footmisc}
\usepackage{amsmath}
\usepackage{txfonts}
\usepackage{placeins}
\usepackage{graphicx}
\usepackage{epstopdf}
\usepackage{ifthen}
\usepackage[breaklinks, colorlinks, citecolor=blue]{hyperref}
\usepackage{overpic}
\usepackage{fixltx2e}
\usepackage{rotating}
\usepackage[table,usenames,dvipsnames]{xcolor}
\bibpunct{(}{)}{;}{a}{}{,} %

\usepackage{xspace}
\usepackage{float}
\usepackage{amssymb}% http://ctan.org/pkg/amssymb
\usepackage{pifont}% http://ctan.org/pkg/pifont
\usepackage{booktabs}
\usepackage{color}
\usepackage[labelfont=bf]{caption}
\usepackage{etex}

\newcommand{\hAN}[1]{}
\newcommand{\hDS}[1]{}
\newcommand{\hLK}[1]{}
\newcommand{\hMM}[1]{}
\newcommand{\hMW}[1]{}
\newcommand{\hSG}[1]{}

\def\setsymbol#1#2{\expandafter\def\csname #1\endcsname{#2}}
\def\getsymbol#1{\csname #1\endcsname}

\def\Planck{\textit{Planck}}

\newbox\tablebox    \newdimen\tablewidth
\def\leaderfil{\leaders\hbox to 5pt{\hss.\hss}\hfil}
\def\endPlancktable{\tablewidth=\columnwidth 
    $$\hss\copy\tablebox\hss$$
    \vskip-\lastskip\vskip -2pt}

\def\tablenote#1 #2\par{\begingroup \parindent=0.8em
    \abovedisplayshortskip=0pt\belowdisplayshortskip=0pt
    \noindent
    $$\hss\vbox{\hsize\tablewidth \hangindent=\parindent \hangafter=1 \noindent
    \hbox to \parindent{$^#1$\hss}\strut#2\strut\par}\hss$$
    \endgroup}
\def\doubleline{\vskip 3pt\hrule \vskip 1.5pt \hrule \vskip 5pt}

\def\L2{\ifmmode L_2\else $L_2$\fi}

\def\DeltaT{\ifmmode \Delta T\else $\Delta T$\fi}
\def\deltat{\ifmmode \Delta t\else $\Delta t$\fi}
\def\fknee{\ifmmode f_{\rm knee}\else $f_{\rm knee}$\fi}
\def\Fmax{\ifmmode F_{\rm max}\else $F_{\rm max}$\fi}
\def\solar{\ifmmode{\rm M}_{\mathord\odot}\else${\rm M}_{\mathord\odot}$\fi}
\def\Msolar{\ifmmode{\rm M}_{\mathord\odot}\else${\rm M}_{\mathord\odot}$\fi}
\def\Lsolar{\ifmmode{\rm L}_{\mathord\odot}\else${\rm L}_{\mathord\odot}$\fi}
\def\inv{\ifmmode^{-1}\else$^{-1}$\fi}
\def\mo{\ifmmode^{-1}\else$^{-1}$\fi}
\def\sup#1{\ifmmode ^{\rm #1}\else $^{\rm #1}$\fi}
\def\expo#1{\ifmmode \times 10^{#1}\else $\times 10^{#1}$\fi}
\def\,{\thinspace}
\def\lsim{\mathrel{\raise .4ex\hbox{\rlap{$<$}\lower 1.2ex\hbox{$\sim$}}}}
\def\gsim{\mathrel{\raise .4ex\hbox{\rlap{$>$}\lower 1.2ex\hbox{$\sim$}}}}

\def\simprop{\mathrel{\raise .4ex\hbox{\rlap{$\propto$}\lower 1.2ex\hbox{$\sim$}}}}
\def\deg{\ifmmode^\circ\else$^\circ$\fi}
\def\pdeg{\ifmmode $\setbox0=\hbox{$^{\circ}$}\rlap{\hskip.11\wd0 .}$^{\circ}
          \else \setbox0=\hbox{$^{\circ}$}\rlap{\hskip.11\wd0 .}$^{\circ}$\fi}
\def\arcs{\ifmmode {^{\scriptstyle\prime\prime}}
          \else $^{\scriptstyle\prime\prime}$\fi}
\def\arcm{\ifmmode {^{\scriptstyle\prime}}
          \else $^{\scriptstyle\prime}$\fi}
\newdimen\sa  \newdimen\sb
\def\parcs{\sa=.07em \sb=.03em
     \ifmmode \hbox{\rlap{.}}^{\scriptstyle\prime\kern -\sb\prime}\hbox{\kern -\sa}
     \else \rlap{.}$^{\scriptstyle\prime\kern -\sb\prime}$\kern -\sa\fi}
\def\parcm{\sa=.08em \sb=.03em
     \ifmmode \hbox{\rlap{.}\kern\sa}^{\scriptstyle\prime}\hbox{\kern-\sb}
     \else \rlap{.}\kern\sa$^{\scriptstyle\prime}$\kern-\sb\fi}
\def\ra[#1 #2 #3.#4]{#1\sup{h}#2\sup{m}#3\sup{s}\llap.#4}
\def\dec[#1 #2 #3.#4]{#1\deg#2\arcm#3\arcs\llap.#4}
\def\deco[#1 #2 #3]{#1\deg#2\arcm#3\arcs}
\def\rra[#1 #2]{#1\sup{h}#2\sup{m}}

\def\dots{\relax\ifmmode \ldots\else $\ldots$\fi}
\def\WHzsr{\ifmmode $W\,Hz\mo\,sr\mo$\else W\,Hz\mo\,sr\mo\fi}
\def\mHz{\ifmmode $\,mHz$\else \,mHz\fi}
\def\GHz{\ifmmode $\,GHz$\else \,GHz\fi}
\def\mKs{\ifmmode $\,mK\,s$^{1/2}\else \,mK\,s$^{1/2}$\fi}
\def\muKs{\ifmmode \,\mu$K\,s$^{1/2}\else \,$\mu$K\,s$^{1/2}$\fi}
\def\muKRJs{\ifmmode \,\mu$K$_{\rm RJ}$\,s$^{1/2}\else \,$\mu$K$_{\rm RJ}$\,s$^{1/2}$\fi}
\def\muKHz{\ifmmode \,\mu$K\,Hz$^{-1/2}\else \,$\mu$K\,Hz$^{-1/2}$\fi}
\def\MJysr{\ifmmode \,$MJy\,sr\mo$\else \,MJy\,sr\mo\fi}
\def\MJysrmK{\ifmmode \,$MJy\,sr\mo$\,mK$_{\rm CMB}\mo\else \,MJy\,sr\mo\,mK$_{\rm CMB}\mo$\fi}
\def\microns{\ifmmode \,\mu$m$\else \,$\mu$m\fi}

\def\muK{\ifmmode \,\mu$K$\else \,$\mu$\hbox{K}\fi}
\def\microK{\ifmmode \,\mu$K$\else \,$\mu$\hbox{K}\fi}
\def\muW{\ifmmode \,\mu$W$\else \,$\mu$\hbox{W}\fi}
\def\kms{\ifmmode $\,km\,s$^{-1}\else \,km\,s$^{-1}$\fi}
\def\kmsMpc{\ifmmode $\,\kms\,Mpc\mo$\else \,\kms\,Mpc\mo\fi}
\providecommand{\sorthelp}[1]{}

\newcommand{\mksym}[1]{\ifmmode {\rm #1}\else #1\fi}
\renewcommand{\Planck}{{\it Planck}\xspace}
\newcommand{\planck}{{\it Planck}\xspace}
\newcommand{\omm}{\omega_{\rm m}}
\newcommand{\omb}{\omega_{\rm b}}
\newcommand{\clamp}{A_{\rm s}\,e^{-2\tau}}
\newcommand{\omegal}{\omega_{\Lambda}}
\newcommand{\Mpc}{{\rm Mpc}}
\newcommand{\TT}{\mksym{PlanckTT}}
\newcommand{\planckTTonly}{PlanckTT}

\newcommand{\tauprior}{$\tau$\rm{prior}}

\newcommand{\hunits}{\,{\rm km}\,{\rm s}^{-1}\,{\rm Mpc}^{-1}}

\newcommand{\LCDM}{$\Lambda$CDM\xspace}
\newcommand{\thetaMC}{\theta_{\rm MC}}
\newcommand{\ns}{n_{\rm s}}
\newcommand{\Ombh}{\omega_{\mathrm{b}}}
\newcommand{\Ommh}{\omega_{\mathrm{m}}}

\newcommand{\As}{A_{\rm s}}
\newcommand{\lnAs}{\ln(10^{10} A_{\rm s})}

\newcommand{\be}{\begin{equation}}
\newcommand{\ee}{\end{equation}}
\newcommand{\bea}{\begin{eqnarray}}
\newcommand{\eea}{\end{eqnarray}}

\newcommand\Bell{\ensuremath{\boldsymbol\ell}}

\newcommand{\plikTT}{\texttt{Plik}\rm TT}
\newcommand{\plikEE}{\texttt{Plik}\rm EE}
\newcommand{\plikTE}{\texttt{Plik}\rm TE}

\begin{document}

\title{\textit{Planck} intermediate results. LI.\\
Features in the cosmic microwave background temperature\\
power spectrum and shifts in cosmological parameters}

%This author list corresponds to \title{Author list for PIP\_125\_Galli\_Millea}
%Prepared by M. Lopez-Caniego (Marcos.Lopez.Caniego@sciops.esa.int), ESAC/ESA
%This version is from Thu Jul 28 07:38:44 2016 CET
%\subtitle{There are 152 co-authors in this list}
\author{\small
Planck Collaboration: N.~Aghanim\inst{51}
\and
Y.~Akrami\inst{54, 85}
\and
M.~Ashdown\inst{61, 5}
\and
J.~Aumont\inst{51}
\and
M.~Ballardini\inst{28, 43, 46}
\and
A.~J.~Banday\inst{83, 8}
\and
R.~B.~Barreiro\inst{56}
\and
N.~Bartolo\inst{27, 57}
\and
S.~Basak\inst{72}
\and
K.~Benabed\inst{52, 82}
% ...
}
\institute{\small
APC, AstroParticule et Cosmologie, Universit\'{e} Paris Diderot, CNRS/IN2P3, CEA/lrfu, Observatoire de Paris, Sorbonne Paris Cit\'{e}, 10, rue Alice Domon et L\'{e}onie Duquet, 75205 Paris Cedex 13, France\goodbreak
\and
African Institute for Mathematical Sciences, 6-8 Melrose Road, Muizenberg, Cape Town, South Africa\goodbreak
\and
Agenzia Spaziale Italiana Science Data Center, Via del Politecnico snc, 00133, Roma, Italy\goodbreak
\and
Aix Marseille Universit\'{e}, CNRS, LAM (Laboratoire d'Astrophysique de Marseille) UMR 7326, 13388, Marseille, France\goodbreak
\and
Astrophysics Group, Cavendish Laboratory, University of Cambridge, J J Thomson Avenue, Cambridge CB3 0HE, U.K.\goodbreak
\and
Astrophysics \& Cosmology Research Unit, School of Mathematics, Statistics \& Computer Science, University of KwaZulu-Natal, Westville Campus, Private Bag X54001, Durban 4000, South Africa\goodbreak
\and
CITA, University of Toronto, 60 St. George St., Toronto, ON M5S 3H8, Canada\goodbreak
\and
CNRS, IRAP, 9 Av. colonel Roche, BP 44346, F-31028 Toulouse cedex 4, France\goodbreak
\and
California Institute of Technology, Pasadena, California, U.S.A.\goodbreak
\and
Centre for Theoretical Cosmology, DAMTP, University of Cambridge, Wilberforce Road, Cambridge CB3 0WA, U.K.\goodbreak
\and
Computational Cosmology Center, Lawrence Berkeley National Laboratory, Berkeley, California, U.S.A.\goodbreak
\and
DTU Space, National Space Institute, Technical University of Denmark, Elektrovej 327, DK-2800 Kgs. Lyngby, Denmark\goodbreak
\and
D\'{e}partement de Physique Th\'{e}orique, Universit\'{e} de Gen\`{e}ve, 24, Quai E. Ansermet,1211 Gen\`{e}ve 4, Switzerland\goodbreak
\and
Departamento de Astrof\'{i}sica, Universidad de La Laguna (ULL), E-38206 La Laguna, Tenerife, Spain\goodbreak
\and
Departamento de F\'{\i}sica, Universidad de Oviedo, Avda. Calvo Sotelo s/n, Oviedo, Spain\goodbreak
\and
Department of Astrophysics/IMAPP, Radboud University Nijmegen, P.O. Box 9010, 6500 GL Nijmegen, The Netherlands\goodbreak
\and
Department of Physics \& Astronomy, University of British Columbia, 6224 Agricultural Road, Vancouver, British Columbia, Canada\goodbreak
\and
Department of Physics and Astronomy, Dana and David Dornsife College of Letter, Arts and Sciences, University of Southern California, Los Angeles, CA 90089, U.S.A.\goodbreak
\and
Department of Physics and Astronomy, University College London, London WC1E 6BT, U.K.\goodbreak
\and
Department of Physics and Astronomy, University of Sussex, Brighton BN1 9QH, U.K.\goodbreak
\and
Department of Physics, Gustaf H\"{a}llstr\"{o}min katu 2a, University of Helsinki, Helsinki, Finland\goodbreak
\and
Department of Physics, Princeton University, Princeton, New Jersey, U.S.A.\goodbreak
\and
Department of Physics, University of California, Berkeley, California, U.S.A.\goodbreak
\and
Department of Physics, University of California, One Shields Avenue, Davis, California, U.S.A.\goodbreak
\and
Department of Physics, University of California, Santa Barbara, California, U.S.A.\goodbreak
\and
Department of Physics, University of Illinois at Urbana-Champaign, 1110 West Green Street, Urbana, Illinois, U.S.A.\goodbreak
\and
Dipartimento di Fisica e Astronomia G. Galilei, Universit\`{a} degli Studi di Padova, via Marzolo 8, 35131 Padova, Italy\goodbreak
\and
Dipartimento di Fisica e Astronomia, Alma Mater Studiorum, Universit\`{a} degli Studi di Bologna, Viale Berti Pichat 6/2, I-40127, Bologna, Italy\goodbreak
\and
Dipartimento di Fisica e Scienze della Terra, Universit\`{a} di Ferrara, Via Saragat 1, 44122 Ferrara, Italy\goodbreak
\and
Dipartimento di Fisica, Universit\`{a} La Sapienza, P. le A. Moro 2, Roma, Italy\goodbreak
\and
Dipartimento di Fisica, Universit\`{a} degli Studi di Milano, Via Celoria, 16, Milano, Italy\goodbreak
\and
Dipartimento di Fisica, Universit\`{a} degli Studi di Trieste, via A. Valerio 2, Trieste, Italy\goodbreak
\and
Dipartimento di Fisica, Universit\`{a} di Roma Tor Vergata, Via della Ricerca Scientifica, 1, Roma, Italy\goodbreak
\and
Dipartimento di Matematica, Universit\`{a} di Roma Tor Vergata, Via della Ricerca Scientifica, 1, Roma, Italy\goodbreak
\and
European Space Agency, ESAC, Planck Science Office, Camino bajo del Castillo, s/n, Urbanizaci\'{o}n Villafranca del Castillo, Villanueva de la Ca\~{n}ada, Madrid, Spain\goodbreak
\and
European Space Agency, ESTEC, Keplerlaan 1, 2201 AZ Noordwijk, The Netherlands\goodbreak
\and
Gran Sasso Science Institute, INFN, viale F. Crispi 7, 67100 L'Aquila, Italy\goodbreak
\and
HGSFP and University of Heidelberg, Theoretical Physics Department, Philosophenweg 16, 69120, Heidelberg, Germany\goodbreak
\and
Helsinki Institute of Physics, Gustaf H\"{a}llstr\"{o}min katu 2, University of Helsinki, Helsinki, Finland\goodbreak
\and
INAF - Osservatorio Astronomico di Padova, Vicolo dell'Osservatorio 5, Padova, Italy\goodbreak
\and
INAF - Osservatorio Astronomico di Roma, via di Frascati 33, Monte Porzio Catone, Italy\goodbreak
\and
INAF - Osservatorio Astronomico di Trieste, Via G.B. Tiepolo 11, Trieste, Italy\goodbreak
\and
INAF/IASF Bologna, Via Gobetti 101, Bologna, Italy\goodbreak
\and
INAF/IASF Milano, Via E. Bassini 15, Milano, Italy\goodbreak
\and
INFN - CNAF, viale Berti Pichat 6/2, 40127 Bologna, Italy\goodbreak
\and
INFN, Sezione di Bologna, viale Berti Pichat 6/2, 40127 Bologna, Italy\goodbreak
\and
INFN, Sezione di Ferrara, Via Saragat 1, 44122 Ferrara, Italy\goodbreak
\and
INFN, Sezione di Roma 1, Universit\`{a} di Roma Sapienza, Piazzale Aldo Moro 2, 00185, Roma, Italy\goodbreak
\and
INFN, Sezione di Roma 2, Universit\`{a} di Roma Tor Vergata, Via della Ricerca Scientifica, 1, Roma, Italy\goodbreak
\and
Imperial College London, Astrophysics group, Blackett Laboratory, Prince Consort Road, London, SW7 2AZ, U.K.\goodbreak
\and
Institut d'Astrophysique Spatiale, CNRS, Univ. Paris-Sud, Universit\'{e} Paris-Saclay, B\^{a}t. 121, 91405 Orsay cedex, France\goodbreak
\and
Institut d'Astrophysique de Paris, CNRS (UMR7095), 98 bis Boulevard Arago, F-75014, Paris, France\goodbreak
\and
Institute of Astronomy, University of Cambridge, Madingley Road, Cambridge CB3 0HA, U.K.\goodbreak
\and
Institute of Theoretical Astrophysics, University of Oslo, Blindern, Oslo, Norway\goodbreak
\and
Instituto de Astrof\'{\i}sica de Canarias, C/V\'{\i}a L\'{a}ctea s/n, La Laguna, Tenerife, Spain\goodbreak
\and
Instituto de F\'{\i}sica de Cantabria (CSIC-Universidad de Cantabria), Avda. de los Castros s/n, Santander, Spain\goodbreak
\and
Istituto Nazionale di Fisica Nucleare, Sezione di Padova, via Marzolo 8, I-35131 Padova, Italy\goodbreak
\and
Jet Propulsion Laboratory, California Institute of Technology, 4800 Oak Grove Drive, Pasadena, California, U.S.A.\goodbreak
\and
Jodrell Bank Centre for Astrophysics, Alan Turing Building, School of Physics and Astronomy, The University of Manchester, Oxford Road, Manchester, M13 9PL, U.K.\goodbreak
\and
Kavli Institute for Cosmological Physics, University of Chicago, Chicago, IL 60637, USA\goodbreak
\and
Kavli Institute for Cosmology Cambridge, Madingley Road, Cambridge, CB3 0HA, U.K.\goodbreak
\and
LAL, Universit\'{e} Paris-Sud, CNRS/IN2P3, Orsay, France\goodbreak
\and
LERMA, CNRS, Observatoire de Paris, 61 Avenue de l'Observatoire, Paris, France\goodbreak
\and
Laboratoire Traitement et Communication de l'Information, CNRS (UMR 5141) and T\'{e}l\'{e}com ParisTech, 46 rue Barrault F-75634 Paris Cedex 13, France\goodbreak
\and
Laboratoire de Physique Subatomique et Cosmologie, Universit\'{e} Grenoble-Alpes, CNRS/IN2P3, 53, rue des Martyrs, 38026 Grenoble Cedex, France\goodbreak
\and
Laboratoire de Physique Th\'{e}orique, Universit\'{e} Paris-Sud 11 \& CNRS, B\^{a}timent 210, 91405 Orsay, France\goodbreak
\and
Lawrence Berkeley National Laboratory, Berkeley, California, U.S.A.\goodbreak
\and
Max-Planck-Institut f\"{u}r Astrophysik, Karl-Schwarzschild-Str. 1, 85741 Garching, Germany\goodbreak
\and
Mullard Space Science Laboratory, University College London, Surrey RH5 6NT, U.K.\goodbreak
\and
Nicolaus Copernicus Astronomical Center, Polish Academy of Sciences, Bartycka 18, 00-716 Warsaw, Poland\goodbreak
\and
Nordita (Nordic Institute for Theoretical Physics), Roslagstullsbacken 23, SE-106 91 Stockholm, Sweden\goodbreak
\and
SISSA, Astrophysics Sector, via Bonomea 265, 34136, Trieste, Italy\goodbreak
\and
School of Chemistry and Physics, University of KwaZulu-Natal, Westville Campus, Private Bag X54001, Durban, 4000, South Africa\goodbreak
\and
School of Physics and Astronomy, Sun Yat-Sen University, 135 Xingang Xi Road, Guangzhou, China\goodbreak
\and
School of Physics and Astronomy, University of Nottingham, Nottingham NG7 2RD, U.K.\goodbreak
\and
Simon Fraser University, Department of Physics, 8888 University Drive, Burnaby BC, Canada\goodbreak
\and
Sorbonne Universit\'{e}-UPMC, UMR7095, Institut d'Astrophysique de Paris, 98 bis Boulevard Arago, F-75014, Paris, France\goodbreak
\and
Sorbonne Universit�s, Institut Lagrange de Paris (ILP), 98 bis Boulevard Arago, 75014 Paris, France\goodbreak
\and
Space Sciences Laboratory, University of California, Berkeley, California, U.S.A.\goodbreak
\and
Sub-Department of Astrophysics, University of Oxford, Keble Road, Oxford OX1 3RH, U.K.\goodbreak
\and
The Oskar Klein Centre for Cosmoparticle Physics, Department of Physics,Stockholm University, AlbaNova, SE-106 91 Stockholm, Sweden\goodbreak
\and
UPMC Univ Paris 06, UMR7095, 98 bis Boulevard Arago, F-75014, Paris, France\goodbreak
\and
Universit\'{e} de Toulouse, UPS-OMP, IRAP, F-31028 Toulouse cedex 4, France\goodbreak
\and
University of Granada, Departamento de F\'{\i}sica Te\'{o}rica y del Cosmos, Facultad de Ciencias, Granada, Spain\goodbreak
\and
University of Heidelberg, Institute for Theoretical Physics, Philosophenweg 16, 69120, Heidelberg, Germany\goodbreak
% ...
}

\date{\vglue -1.5mm \today\vglue -5mm}

\abstract{The six parameters of the standard $\Lambda$CDM model have best-fit
values derived from the \Planck\ temperature power spectrum that are shifted
somewhat from the best-fit values derived from WMAP data. These shifts are
driven by features in the \Planck\ temperature power spectrum at angular scales
that had never before been measured to cosmic-variance level precision. We
investigate these shifts to determine whether they are within the range of
expectation and to understand their origin in the data. Taking our parameter set
to be the optical depth of the reionized intergalactic medium $\tau$, the baryon
density $\omega_{\rm b}$, the matter density $\omega_{\rm m}$, the angular size
of the sound horizon $\theta_*$, the spectral index of the primordial power
spectrum, $n_{\rm s}$, and $A_{\rm s}e^{-2\tau}$ (where $A_{\rm s}$ is the
amplitude of the primordial power spectrum), we examine the change in best-fit
values between a WMAP-like large angular-scale data set (with multipole moment
$\ell\,{<}\,800$ in the \Planck\ temperature power spectrum) and an all
angular-scale data set ($\ell\,{<}\,2500$ \Planck\ temperature power spectrum),
each with a prior on $\tau$ of $0.07\pm0.02$. We find that the shifts, in units
of the 1$\,\sigma$ expected dispersion for each parameter,
are $\{\Delta \tau, \Delta
A_{\rm s} e^{-2\tau}, \Delta n_{\rm s}, \Delta \omm, \Delta \omb, \Delta
\theta_*\} = \{-1.7, -2.2, 1.2, -2.0, 1.1, 0.9\}$, with a $\chi^2$ value of 8.0.
We find that this $\chi^2$ value is exceeded in 15\,\% of our simulated data
sets, and that a parameter deviates by more than 2.2$\,\sigma$ in 9\,\% of
simulated data sets, meaning that the shifts are not unusually large.
Comparing $\ell\,{<}\,800$ instead to $\ell\,{>}\,800$, or
splitting at a different multipole, yields similar results. We examine the
$\ell\,{<}\,800$ model residuals in the $\ell\,{>}\,800$ power spectrum data and
find that the features there that drive these shifts are a set of oscillations
across a broad range of angular scales. Although they partly appear like
the effects of enhanced gravitational lensing, the shifts in \LCDM\ parameters
that arise in response to these features correspond to model spectrum changes
that are predominantly due to non-lensing effects; the only exception is $\tau$,
which, at fixed $A_{\rm s}e^{-2\tau}$, affects the $\ell\,{>}\,800$ temperature
power spectrum solely through the associated change in $A_{\rm s}$ and the
impact of that on the lensing potential power spectrum.
We also ask, ``what is it about the power spectrum at $\ell\,{<}\,800$
that leads to somewhat different best-fit parameters than come from the
full $\ell$ range?'' We find that if we discard the data at $\ell\,{<}\,30$,
where there is a roughly $2\,\sigma$ downward fluctuation in power relative
to the model that best fits the full $\ell$ range, the $\ell\,{<}\,800$
best-fit parameters shift significantly toward the $\ell\,{<}\,2500$ best-fit
parameters.  In contrast, including $\ell\,{<}\,30$, this previously noted
``low-$\ell$ deficit'' drives $n_{\rm s}$ up and impacts parameters correlated
with $n_{\rm s}$, such as $\omega_{\rm m}$ and $H_0$. As expected, the
$\ell\,{<}\,30$ data have a much greater impact on the $\ell\,{<}\,800$ best
fit than on the $\ell\,{<}\,2500$ best fit.
So although the shifts are not very significant, we find that they can be
understood through the combined effects of an oscillatory-like set of
high-$\ell$ residuals and the deficit in low-$\ell$ power, excursions
consistent with sample variance that happen to map onto changes in cosmological
parameters.
Finally, we examine agreement between \Planck\ $TT$ data and two other CMB data
sets, namely the \Planck\ lensing reconstruction and the $TT$ power spectrum
measured by the South Pole Telescope, again finding a lack of convincing
evidence of any significant deviations in parameters, suggesting that current
CMB data sets give an internally consistent picture of the \LCDM model.}
 \titlerunning{Parameter shifts}
\authorrunning{Planck Collaboration}

\maketitle

\section{Introduction}
\label{sec:intro}

Probably the most important high-level result from the \Planck\
satellite\footnote{\Planck\
(\url{http://www.esa.int/Planck}) is a project of the European Space Agency
(ESA) with instruments provided by two scientific consortia funded by ESA
member states and led by Principal Investigators from France and Italy,
telescope reflectors provided through a collaboration between ESA and a
scientific consortium led and funded by Denmark, and additional contributions
from NASA (USA).} \citep{planck2014-a01}
is the good agreement of the statistical properties of the cosmic microwave
background anisotropies (CMB) with the predictions of the
6-parameter standard $\Lambda$CDM cosmological model
\citep{planck2013-p08,planck2013-p11,planck2014-a13,planck2014-a15}.
This agreement is quite remarkable, given the very significant increase in
precision of the \Planck\ measurements over those of prior experiments.
The continuing success of the $\Lambda$CDM model has deepened the motivation
for attempts to understand {\em why\/} the Universe is so well-described as
having emerged from Gaussian adiabatic initial conditions with a particular
mix of baryons, cold dark matter (CDM), and a cosmological constant
($\Lambda$).

Since the main message from \Planck, and indeed from the Wilkinson Microwave
Anisotropy Probe \citep[WMAP;][]{bennett2012} before it, has been the continued
success of the 6-parameter $\Lambda$CDM model, attention naturally turns to
precise details of the values of the best-fit parameters of the model. Many
cosmologists have focused on the parameter shifts with respect to the best-fit
values preferred by pre-\Planck\ data.  Compared to the WMAP data, for example,
\Planck\ data prefer a somewhat slower expansion rate, higher dark matter
density, and higher matter power spectrum amplitude, as discussed in several
Planck Collaboration papers
\citep{planck2013-p08,planck2013-p11,planck2014-a13,planck2014-a15}, as well as
in \citet{Addison15}.  These shifts in parameters have increased the degree of
tension between CMB-derived values and those determined from some other
astrophysical data sets, and have thereby motivated discussion of extensions to
the standard cosmological model
\citep[e.g.,][]{Verde13,Marra13,Efstathiou14,Wyman14, MacCrann15,Seehars15,kids2016}.
However, none of these extensions are strongly supported by the \Planck\ data
themselves \citep[e.g., see discussion in][]{planck2014-a15}.

Despite the interest that the shifts in best-fit parameters has generated,
there has not yet been an identification of the particular aspects of the
\Planck\ data, and their differences with WMAP data,
that give rise to the shifts. The main goal of this paper is to identify the
aspects of the data that lead to the shifts, and to understand the physics
that drives $\Lambda$CDM parameters to respond to these differences in the way
they do. We choose to pursue this goal with analysis that is entirely
internal to the \Planck\ data. In carrying out this \Planck-based analysis,
we still shed light on the WMAP-to-\Planck\ parameter shifts,
because when we restrict ourselves to modes that
WMAP measures at high signal-to-noise ratio, the WMAP and \Planck\ temperature
maps agree well \citep[e.g.,][]{Kovacs13,planck2013-p01a}. The qualitatively
new attribute of the \Planck\ data that leads to the parameter shifts is the
high-precision measurement of the temperature power spectrum in the
$600\,{\la}\,\ell\,{\la}\,2000$ range.\footnote{Although the South Pole
Telescope and Atacama Cosmology Telescope had already measured the CMB $TT$
power spectrum over this multipole range \citep[e.g.,][]{Story13,Das14},
\Planck's dramatically increased sky
coverage leads to a much more precise power spectrum determination.}
Restricting our analysis to be internal to \Planck\ has the advantage of
simplicity, without altering the main conclusions.

We also investigate the consistency of the differences in parameters inferred
from different multipole ranges with expectations, given the $\Lambda$CDM model
and our understanding of the sources of error. The consistency of such parameter
shifts has been previously studied in \citet{planck2014-a13}, \citet{Couchot15},
and \citet{Addison15}. In studying the consistency of parameters inferred from
$\ell\,{<}\,1000$ with those inferred from $\ell\,{>}\,1000$ \citet{Addison15}
claim to find significant evidence for internal inconsistencies in the \Planck
data. Our analysis improves upon theirs in several ways, mainly through our use
of simulations to account for covariances between the pair of data sets being
compared, as well as the ``look elsewhere effect,'' and the departure of the
true distribution of the shift statistics away from a $\chi^2$ distribution.

Much has already been demonstrated about the robustness of the \Planck\
parameter results to data processing, data selection, foreground removal, and
instrument modelling choices \cite{planck2014-a13}. We will not revisit all of
that here. However, having identified the power spectrum features that are
causing the shifts in cosmological parameters, we show that these features are
all present in multiple individual frequency channels, as one would expect from
the previous studies. The features in the data therefore appear to be
cosmological in origin.

The \Planck\ polarization maps, and the $TE$ and $EE$ polarization power
spectra determinations they enable, are also new aspects of the
\Planck\ data. These new data are in agreement with the $TT$ results and point
to similar shifts away from the WMAP parameters \citep{planck2014-a15}, although
with less statistical weight. In order to focus on the primary driver of the
parameter shifts, namely
the temperature power spectrum, we ignore polarization data
except for the constraint on the value of the optical depth $\tau$
coming from polarization at the largest angular scales, which in practice we
fold in with a prior on $\tau$.

Our primary analysis is of the shift in best-fit cosmological parameters as
determined from: (1) a prior on the value of $\tau$ (as a proxy for low-$\ell$
polarization data) and \planckTTonly \footnote{In common with other \Planck\
papers, we use \planckTTonly\ to refer to the full \Planck temperature-only
$C_\ell^{TT}$ likelihood. We often omit the ``TT'' when also specifying a
multipole range, e.g. by \Planck $\ell\,{<}\,800$ we mean \planckTTonly\
$\ell\,{<}\,800$.} data restricted to $\ell\,{<}\,800; $\footnote{To avoid
unnecessary detail, we write $\ell_{\rm max}$ of 800, 1000, and 2500, even
though the true $\ell_{\rm max}$ values are 796, 996, and 2509 (since this is
where the nearest data bins happen to fall). For brevity, the implied $\ell_{\rm
min}$ is always 2 unless otherwise stated, e.g. $\ell\,{<}\,800$ means
$2\,{\le}\,\ell\,{<}\,800$.} and (2) the same $\tau$ prior and the full
$\ell$-range ($\ell\,{<}\,2500$) of \planckTTonly\ data.
Taking the former data set as
a proxy for WMAP, these are the parameter shifts that have been of great
interest to the community. There is of course a degree of arbitrariness in the
particular choice of $\ell\,{=}\,800$ for defining the low-$\ell$ data set. One
might argue for a lower $\ell$, based on the fact that the WMAP temperature maps
reach a signal-to-noise ratio of unity by $\ell\,{\simeq}\,600$, and thus above
600 the power spectrum error bars are at least twice as large as the \Planck\
ones. However, we explicitly select $\ell\,{=}\,800$ for our primary analysis
because it splits the weight on $\Lambda$CDM parameters coming from \Planck\ so
that half is from $\ell\,{<}\,800$ and half is from
$\ell\,{>}\,800$,\footnote{More precisely, the product of eigenvalues of the
two Fisher information matrices \citep[see e.g.,][for a
definition]{schervish1996theory}---one for $\ell\,{<}\,800$ and the other for
$\ell\,{>}\,800$---is approximately equal at this multipole split.} Addressing
the parameter shifts from $\ell\,{<}\,800$ versus $\ell\,{>}\,800$ is a related
and interesting issue, and while our main focus is on the comparison of the
full-$\ell$ results to those from $\ell\,{<}\,800$, we compute and show the
low-$\ell$ versus high-$\ell$ results as well. Additionally, in
Appendix~\ref{app:uberstats} we perform an exhaustive search over many different
choices for the multipole at which to split the data.

In addition to the high-$\ell$ \Planck\ temperature data, inferences of the
reionization optical depth obtained from the low-$\ell$ \Planck\ polarization
data also have an important impact on the determination of the other
cosmological parameters. The parameter shifts that have been discussed in the
literature to date have generally assumed
a constraint on $\tau$ coming from \Planck
LFI polarization data \citep{planck2014-a13,planck2014-a15}. During the writing
of this paper, new and tighter constraints on $\tau$ were released using
improved \Planck\ HFI polarization data \citep{planck2014-a10,planck2014-a25}.
These are consistent with the previous ones, shrinking the error by about a
factor of 2 and moving the best fit to slightly lower values of $\tau$. To
make our work more easily comparable to previous discussions, and because the
impact of this updated constraint is not very large, we have chosen to write the
main body of this paper assuming the old $\tau$ prior. This also allows us to
more cleanly isolate and discuss separately the impact of the new prior, which
we do in a later section of this paper.

Our focus here is on the results from \Planck, and so an in-depth study
comparing the \Planck\ results with those from other cosmological data sets is
beyond our scope. Nevertheless, there do exist claims of internal
inconsistencies in CMB data \citep{Addison15,Riess16}, with the parameter
shifts we discuss here playing an important role, since they serve to drive
the \planckTTonly\ best fits away from those of the two other CMB data sets,
namely the \Planck\ measurements of
the $\phi\phi$ lensing potential power spectrum \citep{planck2013-p12,
planck2014-a17} and the South Pole Telescope (SPT) measurement of the $TT$
damping tail \citep{Story13}. Thus, we also briefly examine whether there is
any evidence of discrepancies that are not just internal to the \planckTTonly\
data, but also when comparing with these other two probes.

The features we identify that are driving the changes in parameters are
approximately oscillatory in nature, a part of them with a frequency and
phasing such that they could be caused by a smoothing of the power spectrum,
of the sort that is generated by gravitational lensing. We thus investigate
the role of lensing in the parameter shifts. The impact of lensing in
\planckTTonly\ parameter estimates has previously
been investigated via use of the parameter ``$A_{\rm L}$'' that artificially
scales the lensing power spectrum (as discussed on p.\,28 of
\citealt{planck2013-p11} and p.\,24 of \citealt{planck2014-a15}). Here we
introduce a new method that more directly elucidates the impact of lensing on
cosmological parameter determination.

Given that we regard the $\ell\,{<}\,2500$ \Planck\ data as providing a better
determination of the cosmological parameters than the $\ell\,{<}\,800$ \Planck\
data, it is natural to turn our primary question around and ask: what is it
about the $\ell\,{<}\,800$ data that makes the inferred parameter values differ
from the full $\ell$-range parameters? Addressing this question, we find that
the deficit in low-multipole power at $\ell\,{\la}\,30$, the ``low-$\ell$
deficit,''\footnote{This is the same feature that has sometimes previously
been called the ``low-$\ell$ anomaly.'' We choose to use the name
``low-$\ell$ deficit'' throughout this work to avoid ambiguity with other large
scale ``anomalies'' and because it is more appropriate for a feature of only
moderate significance. See Sect.~\ref{sec:lowellanomaly} for further
discussion.} plays a significant role in driving the $\ell\,{<}\,800$
parameters away from the results coming from the full $\ell$-range.

The paper is organized as follows.  Section~\ref{sec:lowvsfull} introduces the
shifts seen in parameters between using \Planck\ $\ell\,{<}\,800$ data and full-$\ell$
data.  Section~\ref{sec:expectations} describes the extent to which the observed
shifts are consistent with expectations; we make some simplifying assumptions in
our analysis and justify their use here. Section~\ref{sec:physics} represents a
pedagogical summary of the physical effects underlying the various parameter
shifts.  We then turn to a more detailed characterization of the parameter
shifts and their origin. The most elementary, unornamented description of the
shifts is presented in Sect.~\ref{sec:shiftdescription}, followed by a
discussion of the effects of gravitational lensing in Sect.~\ref{sec:lensing}
and the role of the low-$\ell$ deficit in Sect.~\ref{sec:lowellanomaly}. In
Sect.~\ref{sec:systematics} we consider whether there might be systematic
effects significantly impacting the parameter shifts and in
Sect.~\ref{sec:tauprior} we add a discussion of the effect of changing the
$\tau$ prior. Finally, we comment on some differences with respect to other CMB
experiments in Sect.~\ref{sec:compare} and conclude in
Sect.~\ref{sec:conclusions}.

Throughout we work within the context of the 6-parameter, vacuum-dominated, cold
dark matter ($\Lambda$CDM) model. This model is based upon a spatially flat,
expanding Universe whose dynamics are governed by general relativity and
dominated by cold dark matter and a cosmological constant ($\Lambda$). We shall
assume that the primordial fluctuations have Gaussian statistics, with a
power-law power spectrum of adiabatic fluctuations.  Within that framework the
usual set of cosmological parameters used in CMB studies is: $\omega_{\rm
b}\,{\equiv}\,\Omega_{\rm b}h^2$, the physical baryon density; $\omega_{\rm
c}\,{\equiv}\,\Omega_{\rm c}h^2$, the physical density of cold dark matter (or
$\omega_{\rm m}$ for baryons plus cold dark matter plus neutrinos);
$\theta_\ast$, the ratio of sound horizon to angular diameter distance to the
last-scattering surface; $A_{\rm s}$, the amplitude of the (scalar) initial
power spectrum; $n_{\rm s}$, the power-law slope of those initial perturbations;
and $\tau$, the optical depth to Thomson scattering through the reionized
intergalactic medium.  Here the Hubble constant is expressed as
$H_0\,{=}\,100\,h\,{\rm km}\,{\rm s}^{-1}\,{\rm Mpc}^{-1}$. In more detail, we
follow the precise definitions used in \citet{planck2013-p11} and
\citet{planck2014-a15}.

Parameter constraints for our simulations and comparison to data use the
publicly available \texttt{CosmoSlik} package, and the full simulation pipeline
code will be released publicly pending acceptance of this work. Other parameter
constraints are determined using \texttt{cosmomc} \citep{cosmomc}. Theoretical
power spectra are calculated with \texttt{CAMB} \citep{camb}.

\begin{figure}[htbp!]
\begin{center}
\resizebox{\columnwidth}{!}{\includegraphics{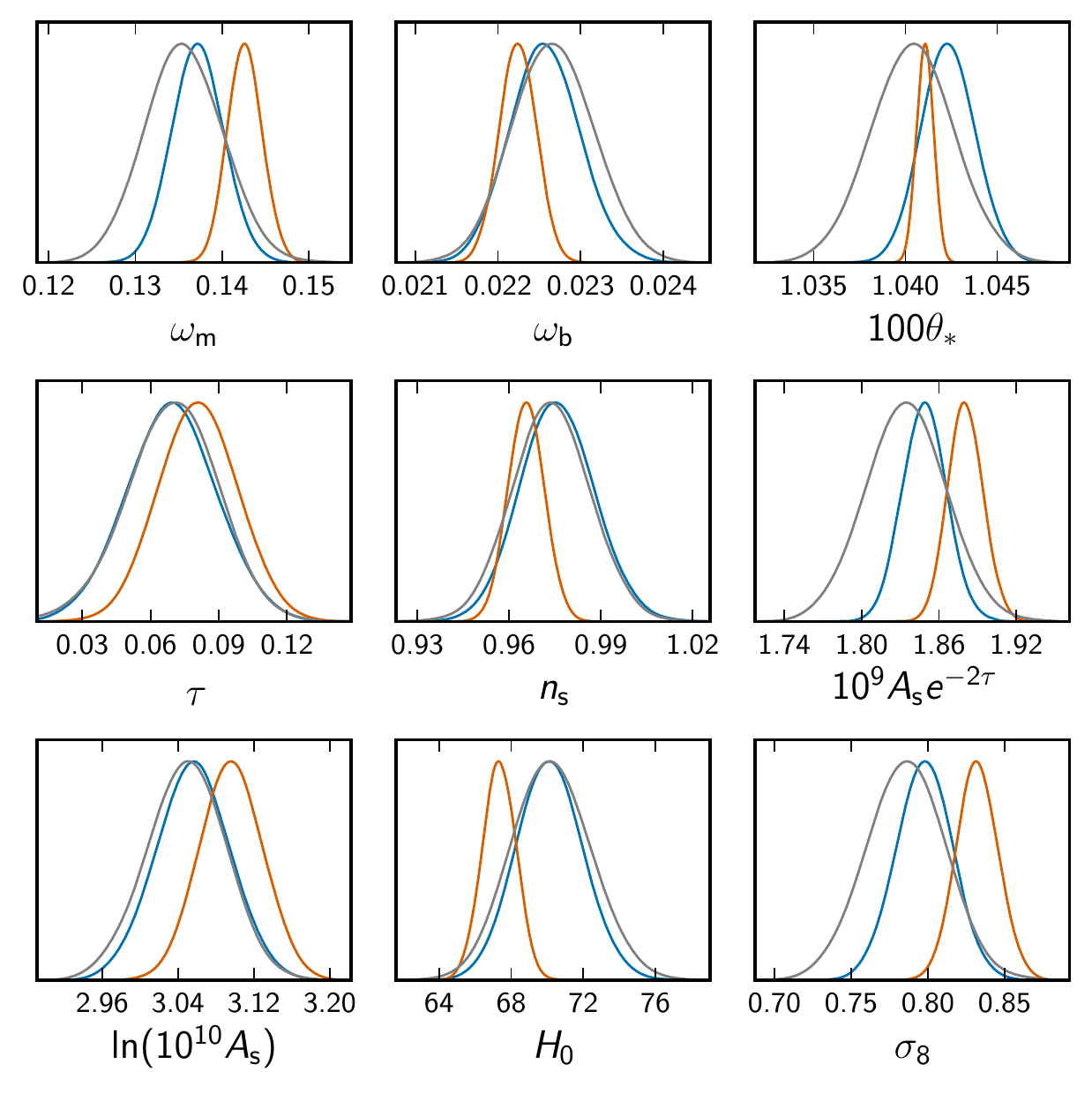}}
\end{center}
\caption {Cosmological parameter constraints from PlanckTT+\tauprior\
for the full multipole range (orange) and for
$\ell\,{<}\,800$ (blue)---see the text for the definitions of the parameters.
Note that the constraints are generally in good agreement, with the full
\Planck\ data providing tighter limits on the parameters; however, the
best-fit values certainly do shift.  It is these shifts that we seek to explain
in this paper. A prior $\tau = 0.07 \pm 0.02$ has been used here as a proxy for
the effect of the low-$\ell$ polarization data (with the impact of a different
prior discussed later). As a comparison, we also show
results for WMAP $TT$ data combined with the same prior on $\tau$ (grey).}
\label{fig:low_vs_full_ell}
\end{figure}

\section{Parameters from low-\Bell\ versus full-\Bell\ \textit{Planck} data}
\label{sec:lowvsfull}

Fig.~\ref{fig:low_vs_full_ell} compares the constraints on six parameters of the
base-$\Lambda$CDM model from the PlanckTT+\tauprior\ data for $\ell\,{<}\,2500$
with those using only the data at $\ell\,{<}\,800$. We have imposed a specific
prior on the optical depth, $\tau=0.07\pm 0.02$, as a proxy for the \Planck\ LFI
low-$\ell$ polarization data, in order to make it easier to compare the
constraints, and to restrict our investigation to the $TT$ power spectrum only.
As mentioned before, we will discuss the impact of the newer HFI polarization
results in Sect.~\ref{sec:tauprior}.

We see that the constraints from the full data set are tighter than those from
using only $\ell\,{<}\,800$, and the best-fit values are slightly shifted. It is
these shifts that we seek to explain in the later sections.
Fig.~\ref{fig:low_vs_full_ell} also shows constraints from the WMAP $TT$
spectrum. As already mentioned, these constraints are qualitatively very similar
to those from \Planck\ $\ell\,{<}\,800$, although not exactly the same, since
WMAP reaches the cosmic variance limit closer to $\ell\,{=}\,600$. Nevertheless,
as was already shown by \cite{Kovacs13}, the CMB maps themselves agree very
well. The small differences in parameter inferences (the largest of which is a
roughly $1\,\sigma$ difference in $\theta_\ast$), are presumably due to small
differences in sky coverage and WMAP instrumental noise. We see that the
dominant source of parameter shifts between \Planck\ and WMAP is the new
information contained in the $\ell\,{>}\,800$ modes, and that by discussing
parameter shifts internal to \Planck\ we are also directly addressing the
differences between WMAP and \Planck.

Fig.~\ref{fig:low_vs_full_ell} shows the shifts for some additional derived
parameters, as well as the basic 6-parameter set.  In particular, one can
choose to use the conventional cosmological parameter $H_0$, rather than the
CMB parameter $\theta_\ast$, as part of a 6-parameter set.  Of course neither
choice is unique, and we could have also focused on other derived quantities
in addition to six that span the space; for the amplitude, we have presented
results for the usual choice $A_{\rm s}$, but added panels for the alternative
choices $A_{\rm s}e^{-2\tau}$ (which will be important later in this paper)
and $\sigma_8$ (the rms density variation in spheres of size $8\,h^{-1}\,{\rm
Mpc}$ in linear theory at $z=0$). The shifts shown in Fig.~\ref{fig:low_vs_full_ell} are fairly
representative of the sorts of shifts that have already been discussed in
previous papers \citep[e.g.,][]{planck2013-p11,planck2014-a13,Addison15},
despite different choices of $\tau$ prior and $\ell$ ranges.

To simplify the analysis as much as possible, throughout most of this paper we
will choose our parametrization of the six degrees of freedom in the \LCDM model
so that we reduce the correlations between parameters, and also so that our
choice maps onto the physically meaningful effects that will be described in
Sect.~\ref{sec:physics}. While a choice of six parameters satisfying both
criteria is not possible, we have settled on $\theta_\ast$, $\omm$, $\omb$,
$\ns$, $A_{\rm s}\,e^{-2\tau}$, and $\tau$.  Most of these choices are standard,
but two are not the same as those focused on in most CMB papers: we have chosen
$\omm$ instead of $\omega_{\rm c}$, because the former governs the size of the
horizon at the epoch of matter-radiation equality, which controls both the
potential-envelope effect and the amplitude of gravitational lensing (see
Sect.~\ref{sec:physics}); and we have chosen to use $A_{\rm s}\,e^{-2\tau}$ in
place of $A_{\rm s}$, because the former is much more precisely determined and
much less correlated with $\tau$. Physically, this arises because at angular
scales smaller than those that subtend the horizon at the epoch of reionization
($\ell\,{\simeq}\,10$) the primary impact of $\tau$ is to suppress power by
$e^{-2\tau}$ (again, see Sect.~\ref{sec:physics}).

As a consequence of this last fact, the temperature power spectrum
places a much tighter constraint on the combination $A_{\rm s}\,e^{-2\tau}$
than it does on $\tau$ or $A_{\rm s}$. Due to the
strong correlation between these two parameters, any extra information on one
will then also translate into a constraint on the other. For this reason, a
change in the prior we use on $\tau$ will be mirrored by a change in $A_{\rm
s}$, given a fixed $A_{\rm s}\,e^{-2\tau}$ combination. Conversely, the extra
information one obtains on $A_{\rm s}$ from the smoothing of the small-scale
power spectrum due to gravitational lensing will be mirrored by a change in the
recovered value of $\tau$ (and this will be important, as we will show later).
As a result, since we will mainly focus on the shifts of $A_{\rm s}\,e^{-2\tau}$
and $\tau$, we will often interpret changes in the value of $\tau$ as a proxy
for changes in $A_{\rm s}$ (at fixed $A_{\rm s}\,e^{-2\tau}$), and thus for the
level of lensing observed in the data (see Sect.~\ref{sec:lensing}).

\section{Comparison of parameter shifts with expectations}
\label{sec:expectations}

\begin{figure*}[htbp!]
\begin{center}
\resizebox{\textwidth}{!}{\includegraphics{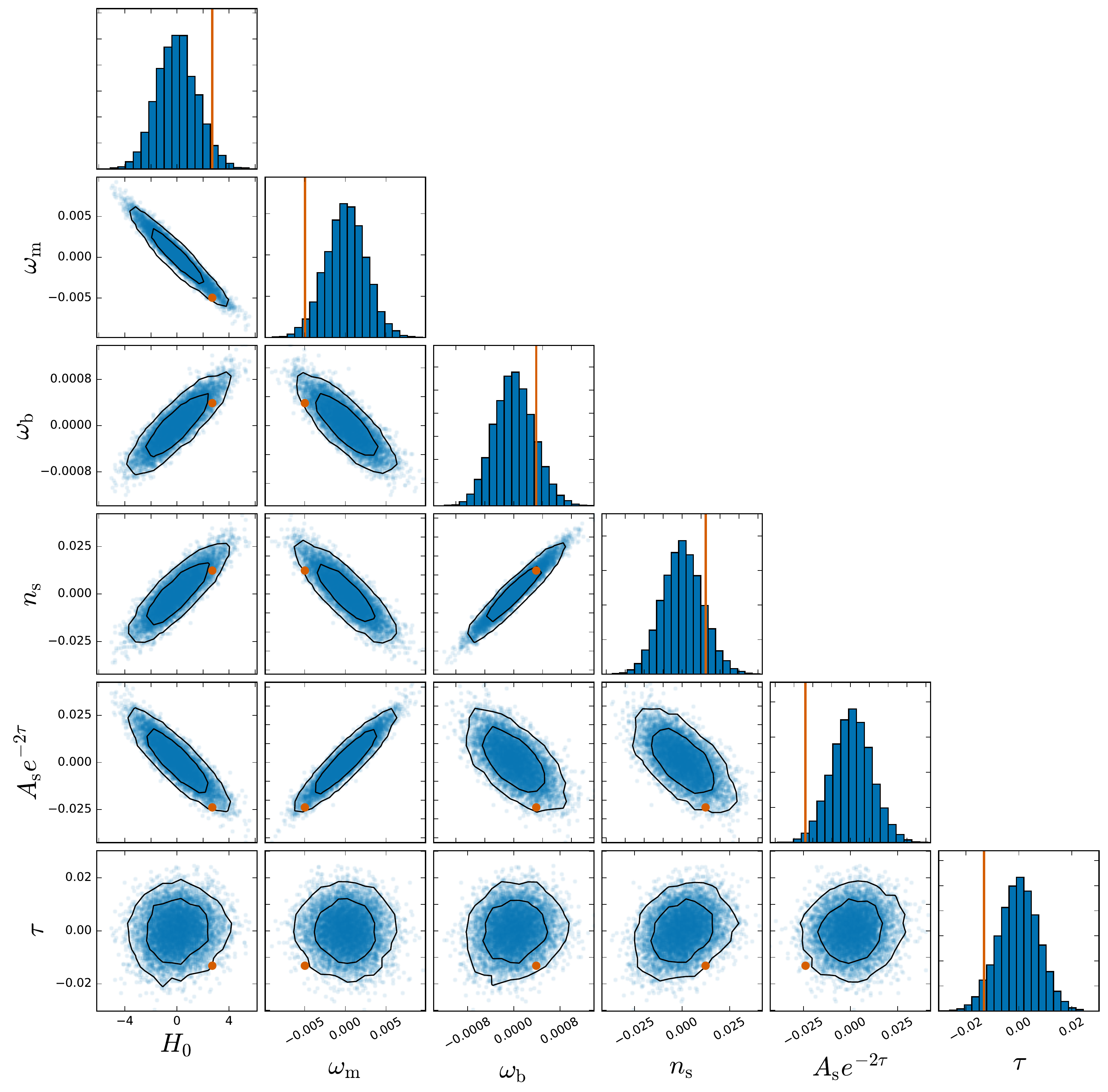}}
\end{center}

\caption{Differences in best-fit parameters between $\ell\,{<}\,800$ and
$\ell\,{<}\,2500$ as compared to expectations from a suite of simulations.  The
cloud of blue points and the histograms are the distribution from simulations
(discussed in Sect.~\ref{sec:expectations}), while the orange points and lines
are the shifts found in the data. Although the shifts may appear to be generally
large for this particular choice of parameter set, it is important to realise
that this is {\it not\/} an orthogonal basis, and that there are strong
correlations among parameters; when this is taken into account, the overall
significance of these shifts is $1.4\,\sigma$, and the significance of the
biggest outlier ($A_{\rm s}e^{-2\tau}$), after accounting for look-elsewhere
effects, is $1.7\,\sigma$. Fig.~\ref{fig:sims_cloud_reparam} shows these same
shifts in a more orthogonal basis that makes judging these significance
levels easier by eye.
Choosing a different multipole at which to split the data, or comparing
low $\ell$s versus high $\ell$s alone, does not change this qualitative level of
agreement. We note that the parameter mode discussed in
Sect.~\ref{sec:results} is {\it not\/} projected out here, since it would
correspond to moving any data point by less than the width of the point itself.}

\label{fig:sims_cloud}
\end{figure*}

\begin{figure}[htbp!]
\begin{center}
\resizebox{\columnwidth}{!}{\includegraphics{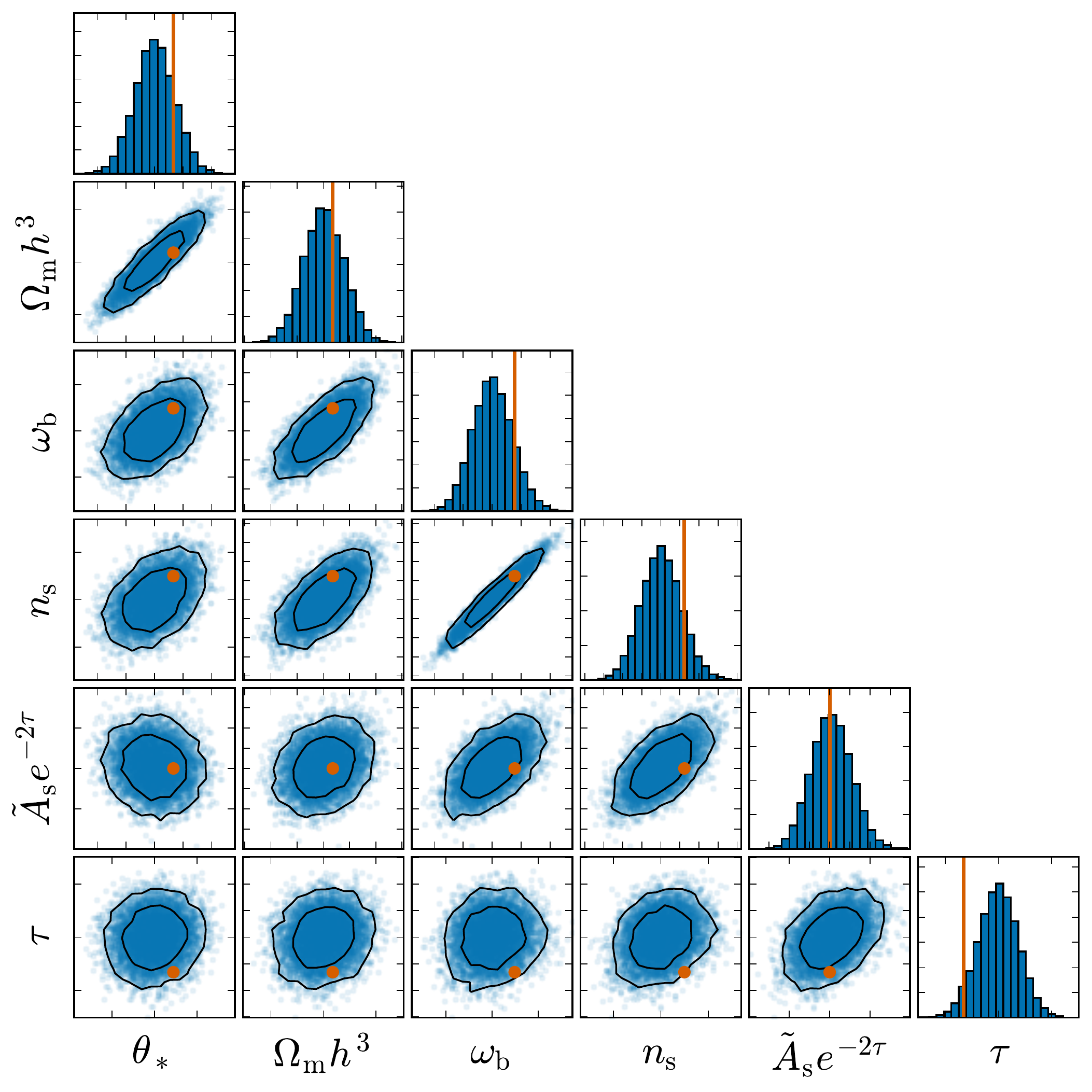}}
\end{center}

\caption{Visually it might seem that the data point in the 6-parameter space
of Fig.~\ref{fig:sims_cloud} is a much worse outlier than only 1.4\,$\sigma$.
One way to see that it really is only 1.4\,$\sigma$ is to transform to another
parameter space, as shown in this figure.
Linear transformations leave the $\chi^2$ unaffected, and while ours here are
not exactly linear, the shifts are small enough that they can be approximated as
linear and the $\chi^2$ is largely unchanged (in fact it is slightly worse,
1.6\,$\sigma$). We have chosen these parameters so the shifts are more
decorrelated while still using physical quantities. The parameter
$\tilde A_{\rm s}$ is the amplitude at a pivot of scale of
$k=0.035\,\Mpc^{-1}$, chosen since there is no shift in
$\tilde A_{\rm s}e^{-2\tau}$. Tick marks are omitted here for clarity.}

\label{fig:sims_cloud_reparam}
\end{figure}

\begin{figure}[htbp!]
\begin{center}
\resizebox{\columnwidth}{!}{\includegraphics{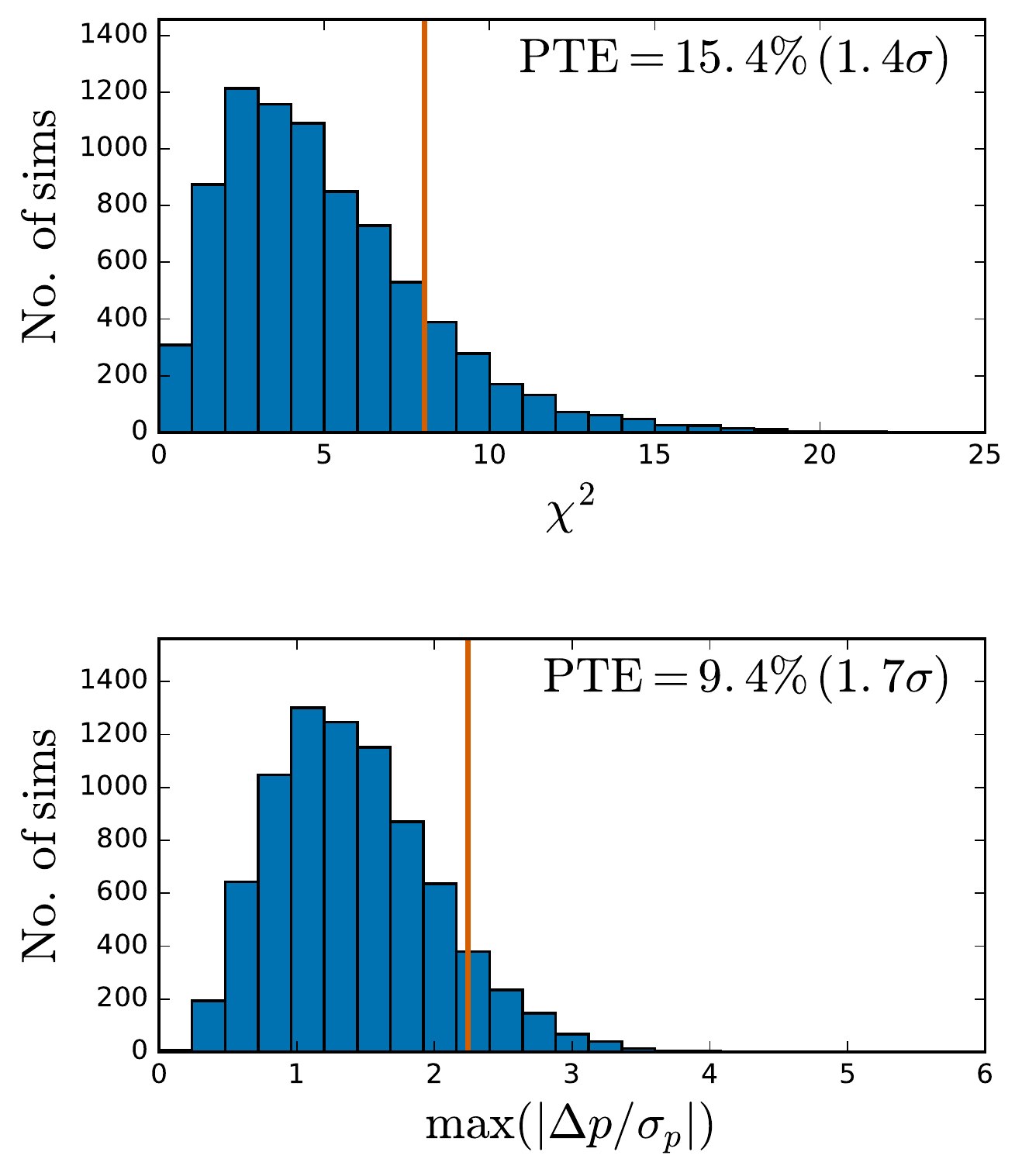}}
\end{center}
\caption{Distribution of two different statistics computed on the simulations
(blue histogram) and on the data (orange line). The first is the $\chi^2$
statistic, where we compute $\chi^2$ for the change in parameters between
$\ell\,{<}\,800$ and $\ell\,{<}\,2500$, with respect to the covariance of the
expected shifts. The second is a ``biggest outlier'' statistic, where we search
for the parameter with the largest change, in units of the standard deviation
of the simulated shifts. We give the probability to exceed (PTE) on each panel.
For both statistics, we find that the observed shifts
are largely consistent with expectations from simulations.}
\label{fig:sims_hist}
\end{figure}

In light of the shifts in parameters described in the previous section, we
would of course like to know whether they are large enough to indicate a
failure of the \LCDM model or the presence of systematic errors in the data,
or if they can
be explained simply as an expected statistical fluctuation arising from
instrumental noise and sample variance. The aim of this section is to give a
precise determination based on simulations, in particular one that
avoids several approximations used by previous analyses.

One of the first attempts to quantify the shifts was performed in appendix~A of
\cite{planck2013-p11}, and was based on a set of Gaussian simulations. More
recent studies using the \Planck\ 2015 data have generally compared posteriors
of disjoint sets of \Planck\ multipole ranges
\citep[e.g.,][]{planck2014-a13,Addison15}. There, the $\chi^2$ is computed,
\be
\label{eq:badchi2}
 \chi^2 = (\vec{\bar{p}}^{(1)} - \vec{\bar{p}}^{(2)}) \tens{\Sigma}^{-1}
 (\vec{\bar{p}}^{(1)} - \vec{\bar{p}}^{(2)}),
\ee
with $\tens{\Sigma} = \tens{C}^{(1)} + \tens{C}^{(2)}$, where $\tens{C}^{(\alpha)}$
are the parameter posterior covariances of the two data sets and
$\vec{\bar{p}}^\alpha$ are the vectors
of parameter means. A probability to exceed $\chi^2$ is then calculated assuming
a $\chi^2$ distribution with degrees of freedom equal to the number of
parameters. This number is usually five, since $\tau$ is ignored because prior
information on $\tau$ went into both sets of estimated parameters.

There are assumptions, both explicit and implicit in previous analyses, which
we avoid with our procedure. We take into account the covariance in the
parameter errors from one data set to the next, and do not assume that
the parameter errors are
normally distributed. Additionally our procedure allows us to include $\tau$ in
the set of compared parameters. As we will see, our more exact procedure shows
that consistency is somewhat better than would have appeared to be the case
otherwise.

\subsection{Description of Simulations}
\label{sec:simulations}

To calculate the expected shifts, we generate a suite of simulated \Planck\
data and, for each data set, compute a likelihood and numerically maximize
it to obtain the best-fit parameters, subject to various multipole range cuts.
The difference in best-fit parameters between different cuts builds up a
distribution of the expected shifts, which can be compared to the shifts seen
in real data. The goal of these
simulations is to be as consistent as possible with the approximations made in
the real analysis (as opposed to, for example, the suite of end-to-end
simulations described in \citealt{planck2014-a13}, which aim to simulate
systematics {\it not\/} directly accounted for by the real likelihood). In this
sense, our simulations are a self-consistency check of \Planck\ data and
likelihood products. We will now describe these simulations in more detail.

For each simulation, we draw a realization of the data independently at
$\ell\,{<}\,30$ and at $\ell\,{>}\,30$.\footnote{We thus ignore $\ell$-to-$\ell$
correlations across this multipole, consistent with what is assumed in the real
likelihood \citep{planck2014-a13}.} At $\ell\,{<}\,30$ we draw realizations
directly at the map level, whereas for $\ell\,{>}\,30$ we use the
\texttt{plik\_lite} CMB covariance \citep[described in][]{planck2014-a13} to
draw power spectrum realizations. For both $\ell\,{<}\,30$ and $\ell\,{>}\,30$,
each realization is drawn assuming a fiducial model. This model is the best-fit
\LCDM model for the \planckTTonly\ data,
with $\tau$ fixed to 0.07, and the \Planck\
calibration parameter, $y_{\rm P}$, fixed to 1. More explicitly, we use
$\{A_{\rm s} e^{-2\tau}, n_{\rm s}, \omm, \omb, \theta_*,\tau, y_{\rm P}\} =
\{1.886, 0.959, 0.1438, 0.02206, 1.04062, 0.07, 1\}$. The reason for fixing
$\tau$ and the calibration in obtaining the fiducial model is that for the
analysis of each simulation, priors on these two parameters are applied, centred
on 0.07 and 1, respectively; if our fiducial model had different values, the
distribution of best-fits across simulations for those and all correlated
parameters would be biased from their fiducial values, and one would need to
recentre the distributions. Our procedure is more straightforward and clearer to
interpret. In any case, our analysis is not very sensitive to the exact fiducial
values and we have checked that for a fiducial model with $\tau\,{=}\,0.055$
the significance levels of the shifts given in the next section change by
$<0.1\,\sigma$.\footnote{In Sect.~\ref{sec:tauprior} we discuss changing the
{\it prior\/} on $\tau$, rather than changing its fiducial value, which
{\it does\/} affect the significance levels somewhat.}

For $\ell\,{>}\,30$, we draw a random Gaussian sample from the
\texttt{plik\_lite} covariance and add it to the fiducial model. This, along
with the covariance itself, forms the simulated likelihood. The
\texttt{plik\_lite} covariance includes in it uncertainties due to foregrounds,
beams, and inter-frequency calibration, hence these are naturally included in
our analysis. We note that the level of uncertainty from these sources is
determined from the \planck $\ell\,{<}\,2500$ data themselves (extracted via a
Gibbs-sampling procedure, assuming only the frequency dependence of the CMB).
Thus, we do not expect exactly the same parameters from \texttt{plik} and
\texttt{plik\_lite} when restricted to an $\ell_{\rm max}$ below 2500
because \texttt{plik\_lite} includes some information on foregrounds from
$\ell_{\rm max}\,{<}\,\ell\,{<}\,2500$.\footnote{Of course, the two likelihoods
are identical when $\ell_{\rm max}=2500$, as demonstrated in
\cite{planck2014-a13}.} For our purposes, this is actually a benefit of using
\texttt{plik\_lite}, since it lets us put well-motivated priors on the
foregrounds for any value of $\ell_{\rm max}$ in a way that does not double
count any data. Regardless of that, the difference between \texttt{plik} and
\texttt{plik\_lite} is not very large. For example, the largest of any parameter
difference at $\ell_{\rm max}\,{=}\,1000$ is $0.15\,\sigma$ (in the $\sigma$ of
that parameter for $\ell_{\rm max}\,{=}\,1000$), growing to $0.35\,\sigma$ at
$\ell_{\rm max}\,{=}\,1500$, and of course back to effectively zero by
$\ell_{\rm max}\,{=}\,2500$. Regardless, since our simulations and analyses of
real data are performed with the same likelihood, our approach is fully
self-consistent.

At $\ell\,{<}\,30$, so as to simulate the correct non-Gaussian shape of the
$C_\ell$ posteriors, we draw a map-level realization of the fiducial CMB power
spectrum. In doing so, we ignore uncertainties due to foregrounds,
inter-frequency calibration, and noise; we will show below that this is a
sufficient approximation. For the likelihood, rather than compute the
\texttt{Commander} \citep{planck2014-a11,planck2014-a12} likelihood for each
simulation (which in practice would be computationally prohibitive),
we instead use the following simple but accurate analytic
approximation. With no masking, the probability distribution of $(2\ell+1)\hat
C_\ell/C_\ell$ is known to be exactly a $\chi^2$ distribution with $2\ell+1$
degrees of freedom (here $\hat C_\ell$ is the observed spectrum and $C_\ell$ is
the theoretical spectrum). Our approximation posits that, for our masked sky,
$f_\ell(2\ell+1)\hat C_\ell/C_\ell$ is drawn from $\chi^2[f_\ell(2\ell+1)]$,
with $f_\ell$ an $\ell$-dependent coefficient determined for our particular mask
via simulations, and with $\hat C_\ell$ being the mask-deconvolved power
spectrum. Approximations very similar to this have been studied previously by
\cite{Benabed09} and \cite{Hamimeche16}. Unlike some of those works, our
approximation here does not aim to be a general purpose low-$\ell$ likelihood,
rather just to work for our specific case of assuming the $\Lambda$CDM model and
when combined with data up to $\ell\,{\simeq}\,800$ or higher. While it
is not a priori obvious that it is sufficient in these cases, we can perform the
following test. We run parameter estimation on the real data, replacing the full
\texttt{Commander} likelihood with our approximate likelihood
using $\hat C_\ell$
and $f_\ell$ as derived from the \texttt{Commander} map and mask.  Note that
this also tests the effect of fixing the foregrounds and inter-frequency
calibrations, since we are using just the best-fit \texttt{Commander} map, and
it also tests the effect of ignoring noise uncertainties, since our likelihood
approximation does not include them. We find that, for both an $\ell\,{<}\,800$
and an $\ell\,{<}\,2500$ run,\footnote{The low $\ell$s have more relative weight
in the $\ell\,{<}\,800$ case, hence that is the more stringent test.} no
parameter deviates from the real results by more than $0.05\,\sigma$, with
several parameters changing much less than that;
hence we find that our approximation is good
enough for our purposes. Additionally, in Appendix~\ref{app:lowl} we describe a
complementary test that scans over many realizations of the CMB sky as well,
also finding the approximation to be sufficient.

The likelihood from each simulation is combined with a prior on $\tau$ of
$0.07\,{\pm}\,0.02$ (with other choices of priors discussed in
Sect.~\ref{sec:tauprior}). It is worth emphasizing that the exact same prior is
imposed on every simulation, and hence implicitly we are not drawing
realizations of different polarization data to go along with the realizations of
temperature data that we have discussed above. This is a valid choice because
the polarization data are close to noise dominated and therefore largely
uncorrelated with the temperature data. We have chosen to do this
because our aim
is to examine parameter shifts between different subsets of temperature data,
rather than between temperature versus polarization, and thus we regard the
polarization data as a fixed external prior. Had we sampled the polarization
data, the significance levels
of shifts would have been slightly smaller because the
expected scatter on $\tau$ and correlated parameters would be slightly larger.
We have explicitly checked this fact by running a subset of the simulations
(ones for $\ell\,{<}\,800$ and $\ell\,{<}\,2500$) with the mean of the $\tau$
prior randomly draw from its prior distribution for each simulation, i.e., we
have implicitly drawn realizations of the polarization data. We find that the
significance levels
of the different statistics discussed in the following section are
reduced by 0.1\,$\sigma$ or less. Note that this same subset of simulations is
described further in Appendix~\ref{app:lowl}, where it is used as an
additional verification of our low-$\ell$ approximation.

\subsection{Results}
\label{sec:results}

With the simulated data and likelihoods in hand, we now numerically maximize the
likelihood for each of the realizations to obtain best-fit parameters. The
maximization procedure uses ``Powell's method'' from the \texttt{SciPy} package
\citep{scipy} and has been tested to be robust to a satisfactory level by running
it on the true data at all $\ell$ splits, beginning from several different
starting points, and ensuring convergence to the same minimum. We find in all
cases that convergence is sufficient to ensure that none of the significance
values given in this section change by more than 0.1\,$\sigma$.

Using the computational power provided by the volunteers at
{\tt Cosmology@Home},\footnote{\url{http://www.cosmologyathome.org}}
whose computers ran a large part of these computations,
we have been able to run simulations not just for $\ell\,{<}\,800$
and $\ell\,{<}\,2500$, but for roughly 100 different subsets of data,
with around 5000 realizations for each. We discuss some of these
results in this section, with a more comprehensive set of tests given in
Appendix~\ref{app:uberstats}.

Figure~\ref{fig:sims_cloud} shows the resulting distribution of parameter
shifts expected between the $\ell\,{<}\,800$ and $\ell\,{<}\,2500$ cases,
compared to
the shift seen in the real data. To quantify the overall consistency, we pick a
statistic, compute its value on the data as well as on the simulations, then
compute the probability to exceed (PTE) the data value based on the
distribution of simulations. We then turn this into the equivalent number of
$\sigma$, such that a 1-dimensional Gaussian has the same 2-tailed PTE.
We use two particular statistics:

\begin{itemize}

\item the $\chi^2$ statistic, computing
$\chi^2=\Delta \vec{p} \, \tens{\Sigma}^{-1} \, \Delta \vec{p}$,
where $\Delta \vec{p}$ is the vector of shifts in parameters between the two
data sets and $\tens{\Sigma}$ is the covariance of these shifts from the set of
simulations;

\item the \texttt{max-param} statistic, where we scan for
${\rm max}(| \Delta \vec{p} / \sigma_p |)$, i.e., the most deviant parameter
from the set $\{\theta_\ast$, $\omega_{\rm m}$, $\omega_{\rm b}$,
$A_{\rm s} e^{-2\tau}$, $n_{\rm s}$, $\tau\}$, in terms of the expected
shifts from the simulations, $\sigma_p$.

\end{itemize}

There are of course an infinite number of statistics one could compute, but
these two are reasonable choices, which test agreement across all
parameters as well on individual outliers.

In the case of the $\chi^2$ statistic, and when one is comparing two nested sets
of data (by ``nested'' we mean that one data set contains the other, i.e.,
$\ell\,{<}\,800$ is part of $\ell\,{<}\,2500$), there is an added caveat. In
cases like this, there is the potential for the existence of one or more
directions in parameter space for which expected shifts are extremely small
compared to the posterior constraint on the same mode. These correspond to
parameter modes where very little new information has been added, and hence one
should see almost no shift. It is thus possible that the $\chi^2$ statistic is
drastically altered by a change to the observed shifts that is in fact
insignificant at our level of interest. Such a mode can be excited by any number
of things, such as systematics, effects of approximations, minimizer errors,
etc., but at a very small level. These modes can be
enumerated by simultaneously diagonalizing the covariance of expected shifts and
the covariance of the posteriors, and ordering them by the ratio of eigenvalues.
For the case of comparing $\ell\,{<}\,800$ and $\ell\,{<}\,2500$, we find that
the worst offending mode corresponds to altering the observed shifts in $\{H_0$,
$\omega_{\rm m}$, $\omega_{\rm b}$, $A_{\rm s} e^{-2\tau}$, $n_{\rm s}$,
$\tau\}$ by $\{0.02$, $-0.01$, $0.02$, $-0.003$, $0.04$, $0.01\}$ in units of
the 1$\,\sigma$ posteriors from $\ell\,{<}\,2500$. This can change the
significance of the $\chi^2$ statistic by an amount that corresponds to
$0.6\,\sigma$, despite no cosmological parameter nor linear combination of them
having changed by more than a few percent of each $\sigma$.
To mitigate this effect
and hence to make the $\chi^2$ statistic more meaningful for our desired goal of
assessing consistency, we quote significance levels after projecting out any
modes whose ratio of eigenvalues is greater than $10$ (which in our case is just
the aforementioned mode). We emphasize that removal of this mode is not meant
to, nor does it, hide any problems; in fact, in some cases the $\chi^2$
becomes worse after removal.
The point is that without removing it we would be sensitive
to shifts in parameters at extremely small levels that we do not care about. In
any case, this mode removal is only necessary for the case of the $\chi^2$
statistic and nested data sets, which is only a small subset of the tests
performed in this paper.

Results for several data splits are summarized in
Table~\ref{tab:significances}, with the comparison of $\ell\,{<}\,800$ to
$\ell\,{<}\,2500$ given in the first row and shown more fully in
Fig.~\ref{fig:sims_hist}. In this case, we find that the parameter shifts are
in fairly good agreement with expectations from simulations, with significance
levels of $1.4\,\sigma$ and $1.7\,\sigma$ from the two statistics,
respectively.  We also note that the qualitative level of agreement is largely
unchanged when considering $\ell\,{<}\,800$ versus
$\ell\,{>}\,800$ or when splitting at $\ell\,{=}\,1000$.

Of the other data splits shown in Table~\ref{tab:significances}, the
$\ell\,{<}\,1000$ versus $\ell\,{>}\,1000$ case may be of particular interest,
since it is discussed extensively in \citet{Addison15}. Although not the
main focus in their paper, those authors find $1.8\,\sigma$ as the
level of the overall agreement by applying the equivalent of our
Eq.~(\ref{eq:badchi2}) to the shifts in five parameters, namely
$\{\theta_\ast, \omega_{\rm c}, \omega_{\rm b}, \log A_{\rm s}, n_{\rm s}\}$.
This is similar to our result, although higher by $0.2\,\sigma$. There are
three main contributors to this difference. Firstly, although Addison et
al.\ drop $\tau$ in the comparison to try to mitigate the effect of the prior
on $\tau$ having induced correlations in the two data sets, they keep
$\log A_{\rm s}$ as a parameter, which is highly correlated with $\tau$. This
means that their comparison fails to remove the correlations, nor does it
take them into account. One could largely remove the correlation by
switching to $A_{\rm s}e^{-2\tau}$ (which is much less correlated with $\tau$);
this has the effect of reducing the significance of the shifts by
0.3\,$\sigma$. Secondly, the Addison et al.\ analysis puts no priors on the
foreground parameters, which is especially important for the $\ell>1000$ part.
For example, fixing the foregrounds to their best-fit levels from
$\ell\,{<}\,2500$ reduces the significance
by an additional 0.2\,$\sigma$. Finally, our result uses six parameters as
opposed to five (since we are able to correctly account for the prior on
$\tau$); this increases the significance back up by around 0.3\,$\sigma$.

There is an additional point that \citet{Addison15} fail to take into account
when quoting significance levels---and the same issue arises in some
other published claims of parameter shifts that focus on a single
parameter.  This is that one should not pick out the most extreme
outlying parameter without assessing how large the largest expected shift is
among the full set of parameters.  In other words,
one should account for what are sometimes called ``look elsewhere'' effects
\citep[see][for a discussion of this issue in a different
context]{planck2014-a18}.
Our simulations allow us to do this easily.  For example, in the
$\ell\,{<}\,1000$ versus $\ell\,{>}\,1000$ case, the biggest change in any
parameter is a $2.3\,\sigma$ shift in $\omega_{\rm m}$; however,
the significance of finding a $2.3\,\sigma$ outlier when searching through
six parameters with our particular correlation structure is only
$1.6\,\sigma$, which is the value we quote in Table~\ref{tab:significances}.

To summarize this section, we do not find strong evidence of inconsistency in
the parameter shifts from $\ell\,{<}\,800$ to those from $\ell\,{<}\,2500$, when
compared with expectations, nor from any of the other data splits shown in
Table~\ref{tab:significances}. We also find that the results of
\cite{Addison15} somewhat exaggerate the significance of tension, for a
number of reasons, as discussed above.

As a final note, we show in Table~\ref{tab:significances_lowtau} the consistency
of various data splits as in Table~\ref{tab:significances}, but using data and
simulations that have a prior of $\tau\,{=}\,0.055\pm0.010$ instead of
$\tau\,{=}\,0.07\pm0.02$. In general the agreement between different splits
changes by between $-0.1$ and $0.3\,\sigma$, thus slightly worse. A detailed
discussion of these results will be presented in Section \ref{sec:tauprior}.

\begin{table}[tb!]
\newdimen\tblskip \tblskip=5pt

\caption{Consistency of various data splits, as determined from two statistics
computed on data and simulations. Fig.~\ref{fig:sims_hist} shows the actual
distribution from simulations for the first row in this table. Entries marked
with a dagger symbol have had a parameter mode projected out, as discussed in
Sect.~\ref{sec:results}.}

\label{tab:significances}
\vskip -3mm
\footnotesize
\setbox\tablebox=\vbox{
 \newdimen\digitwidth
 \setbox0=\hbox{\rm 0}
 \digitwidth=\wd0
 \catcode`*=\active
 \def*{\kern\digitwidth}
  \newdimen\dpwidth
  \setbox0=\hbox{.}
  \dpwidth=\wd0
  \catcode`!=\active
  \def!{\kern\dpwidth}
\halign{\hbox to 2.5cm{#\leaderfil}\hfil\tabskip=0em&
    \hbox to 2.5cm{#\leaderfil}\hfil\tabskip=1em&
    \hfil#\hfil\tabskip=1em&
    \hfil#\hfil\tabskip=0.5em&
    \hfil#\hfil\tabskip=0pt\cr
\noalign{\doubleline}
\omit& \omit& \multispan3\hfil Test\hfil\cr
\noalign{\vskip -4pt}
\omit& \omit&\multispan3\hfil\hrulefill\hfil\cr
\omit Data set 1\hfil& \omit Data set 2\hfil& $\chi^2$& \multispan2\hfil \texttt{max-param}\hfil\cr
\noalign{\vskip 4pt\hrule\vskip 3pt}
$\ell\,{<}\,800$& $\ell\,{<}\,2500$& 1.4\,$\sigma^\dagger$& 1.7\,$\sigma$& ($A_{\rm s}e^{-2\tau}$)\cr
$\ell\,{<}\,800$& $\ell\,{>}\,800$& 1.6\,$\sigma$*& 2.1\,$\sigma$& ($A_{\rm s}e^{-2\tau}$)\cr
$\ell\,{<}\,1000$& $\ell\,{<}\,2500$& 1.8\,$\sigma^\dagger$& 1.5\,$\sigma$& ($A_{\rm s}e^{-2\tau}$)\cr
$\ell\,{<}\,1000$& $\ell\,{>}\,1000$& 1.6\,$\sigma$*& 1.6\,$\sigma$& ($\omega_{\rm m}$)\cr
\noalign{\vskip 4pt\hrule\vskip 3pt}
$30\,{<}\,\ell\,{<}\,800$& $\ell\,{>}\,30$& 1.2\,$\sigma^\dagger$& 1.3\,$\sigma$& ($\tau$)\cr
$30\,{<}\,\ell\,{<}\,800$& $\ell\,{>}\,800$& 1.2\,$\sigma$*& 1.2\,$\sigma$& ($A_{\rm s}e^{-2\tau}$)\cr
$30\,{<}\,\ell\,{<}\,1000$& $\ell\,{>}\,30$& 1.4\,$\sigma^\dagger$& 1.5\,$\sigma$& ($\tau$)\cr
$30\,{<}\,\ell\,{<}\,1000$& $\ell\,{>}\,1000$& 1.2\,$\sigma$*& 0.7\,$\sigma$& ($\omega_{\rm m}$)\cr
\noalign{\vskip 3pt\hrule\vskip 5pt}}}
\endPlancktable
\end{table}

\begin{table}[tb!]
\newdimen\tblskip \tblskip=5pt

\caption{Same as Table.~\ref{tab:significances}, but using data and simulations
that have a prior of $\tau\,{=}\,0.055\pm0.010$ instead of
$\tau\,{=}\,0.07\pm0.02$. See Sect.~\ref{sec:tauprior} for more discussion on
the impact of this updated constraint on $\tau$. Entries marked with a $\dagger$
have had a parameter mode projected out, as discussed in
Sect.~\ref{sec:results}.}

\label{tab:significances_lowtau}
\vskip -3mm
\footnotesize
\setbox\tablebox=\vbox{
 \newdimen\digitwidth
 \setbox0=\hbox{\rm 0}
 \digitwidth=\wd0
 \catcode`*=\active
 \def*{\kern\digitwidth}
  \newdimen\dpwidth
  \setbox0=\hbox{.}
  \dpwidth=\wd0
  \catcode`!=\active
  \def!{\kern\dpwidth}
\halign{\hbox to 2.5cm{#\leaderfil}\hfil\tabskip=0em&
    \hbox to 2.5cm{#\leaderfil}\hfil\tabskip=1em&
    \hfil#\hfil\tabskip=1em&
    \hfil#\hfil\tabskip=0.5em&
    \hfil#\hfil\tabskip=0pt\cr
\noalign{\doubleline}
\omit& \omit& \multispan3\hfil Test\hfil\cr
\noalign{\vskip -4pt}
\omit& \omit&\multispan3\hfil\hrulefill\hfil\cr
\omit Data set 1\hfil& \omit Data set 2\hfil& $\chi^2$& \multispan2\hfil \texttt{max-param}\hfil\cr
\noalign{\vskip 4pt\hrule\vskip 3pt}
$\ell\,{<}\,800$& $\ell\,{<}\,2500$& 1.8\,$\sigma^\dagger$& 2.1\,$\sigma$& ($A_{\rm s} e^{-2\tau}$)\cr
$\ell\,{<}\,800$& $\ell\,{>}\,800$& 1.9\,$\sigma$*& 2.2\,$\sigma$& ($A_{\rm s} e^{-2\tau}$)\cr
$\ell\,{<}\,1000$& $\ell\,{<}\,2500$& 1.9\,$\sigma^\dagger$& 1.9\,$\sigma$& ($A_{\rm s} e^{-2\tau}$)\cr
$\ell\,{<}\,1000$& $\ell\,{>}\,1000$& 1.9\,$\sigma$*& 1.5\,$\sigma$& ($\omega_{\rm m}$)\cr
\noalign{\vskip 3pt\hrule\vskip 5pt}}}
\endPlancktable
\end{table}

\section{Physical explanation of the power spectrum response to changing
$\boldsymbol\Lambda$CDM parameters}
\label{sec:physics}

\begin{figure*}[htbp!]
\begin{center}
\resizebox{\textwidth}{!}{\includegraphics{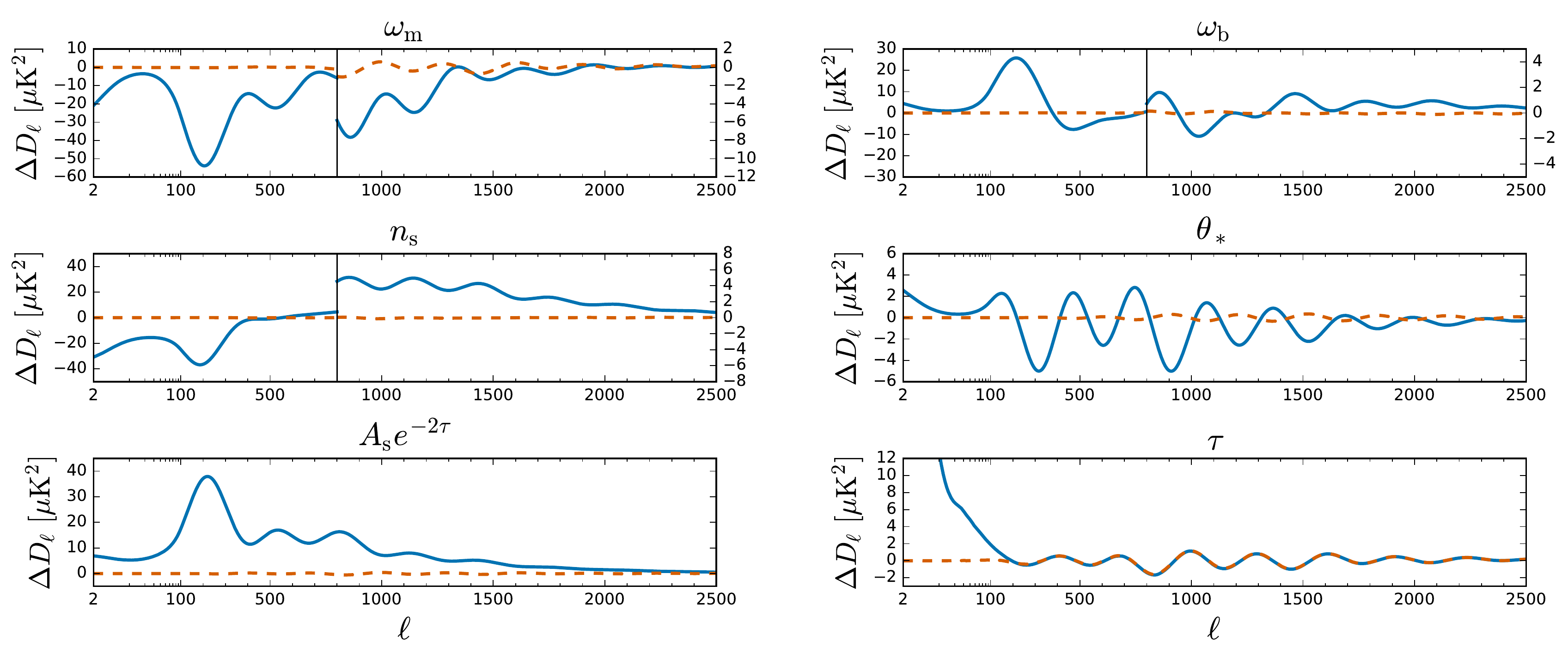}}

\caption{Response of ${\cal D}_l^{TT}$ ($\equiv\ell(\ell+1)C_\ell/2\pi$) to
1$\,\sigma$ increases in each of the parameters \citep[calculated using {\tt
CAMB},][]{Lewis1999}. All changes are made with the other five parameters
pictured here held fixed. The dashed orange line in each panel shows the
contribution from gravitational lensing alone. Note that the $y$-axis scale
changes in some of the panels at $\ell\,{=}\,800$.}

\label{fig:Derivs}
\end{center}
\end{figure*}

Having studied the question of the magnitude of the parameter shifts relative
to expectations, we now turn to an analysis of why the best-fit model
parameters change in the particular way that they do. Understanding this
requires
reviewing exactly how changes to \LCDM parameters affect the CMB power
spectrum, so that these can be matched with the features in the data that
drive the changes.  The material in this section is meant as background for
the narrative that will come later, and readers may want to
skip it on a first reading; nevertheless, the information collected here is
not available in any single source elsewhere, and will be important for
understanding the relationship between parameters and power spectrum
features.  The key information is the response of the angular power spectrum
to changes in parameters, shown in Fig.~\ref{fig:Derivs}.  In
Sect.~\ref{sec:shifts} we will close the loop on how the physics embodied in
the curves of Fig.~\ref{fig:Derivs} interacts with the residual features in the
power spectrum to give the parameter shifts we see in
Fig.~\ref{fig:low_vs_full_ell}.

The structure in the CMB anisotropy spectrum arises from gravity-driven
oscillations in the baryon-photon plasma before recombination
\citep[e.g.,][]{Peebles1970,Zeldovich1972}.
Fortunately our understanding of the CMB spectrum has become highly
developed, so we are able to understand the physical causes
(see Fig.~\ref{fig:Derivs}) of the shifts already discussed
as arising from the interaction of gravitational lensing, the early integrated
Sachs-Wolfe \citep[ISW,][]{SacWol67} effect, the potential envelope,
and diffusion damping.
In this section we review the physics behind the
$\partial C_\ell^{TT}/\partial p_i$ curves and clarify some
interesting interactions by ``turning off'' various effects.
The reader is referred to \citet{Peacock99}, \citet{Liddle00}, and
\citet{Dodelson03} for basic textbook treatments of the physics of CMB
anisotropies.

\subsection{The matter density: $\omm$}
\label{sec:matter}

We begin by considering how changes in the matter density affect the
power spectrum, leading to the rising behaviour seen in the top left
panel of Fig.~\ref{fig:Derivs}.  Note that here we have plotted the
linear response in the quantity
${\cal D}_\ell\equiv\ell(\ell+1)C_\ell/2\pi$ rather than $C_\ell$.

Since much of the relevant action occurs near horizon crossing, a description of
the physics is best accomplished by picking a gauge; we choose the
Newtonian gauge here and focus primarily on the potentials $\Phi$ and $\Psi$ and the density.  Within this picture, the impact of the matter density
comes from the ``early integrated Sachs-Wolfe effect'' (i.e., the evolution of
the potentials immediately after last scattering) and from the ``potential
envelope''.
The effect of main interest to us is the latter---the
enhancement of power above $\ell\,{\simeq}\,100$ arising due to the
near-resonant driving of the acoustic oscillations by decaying potentials as
they cross the horizon near, or earlier than, the epoch of matter-radiation
equality \citep{Hu96b,Hu97,HuSugiyamaSilk}. Overdense modes that enter the horizon during
radiation domination ($\rho_{\rm m}/\rho_{\rm rad}\ll 1$) cannot collapse
rapidly enough
into their potential wells (due to the large pressure of the radiation) to
prevent the potentials from decaying due to the expansion of the Universe. The
time it takes the potential to decay is closely related to the time at which the
photons reach their maximal compression and hence maximal energy density
perturbation.  The near-resonant driving of the oscillator, and the fact that
the photons do not lose (as much) energy climbing out of the potential well (as they gained falling in),
leads to a large increase in observed amplitude of the temperature
perturbation over its initial value. For modes that enter the horizon later,
the matter density perturbations contribute more to the potentials, which are
(partially) stabilized against decay by the contribution of the CDM.  This
reduces the amplitude enhancement. The net result is an $\ell$-dependent
boost to the power spectrum amplitude, transitioning from unity
at low $\ell$ to a factor of over 10 in the high-$\ell$ limit. This boost is
known as the ``potential envelope.'' It is not immediately apparent in the
power spectrum, due to
the effects of damping at high $\ell$, but it imprints a
large dependence on $\omm$ and can be uncovered if the effects of damping and line-of-sight averaging are
removed \citep[e.g., figure~7 of][]{Hu97}.

The characteristic scale of the power boost is set by the angular scale,
$\theta_{\rm eq}$, which is the comoving size of the horizon at the epoch
of matter-radiation equality projected from the last-scattering surface.
Thus the CMB spectra are sensitive to $\theta_{\rm eq}$.
In the $\Lambda$CDM model $\theta_{\rm eq}$ depends almost solely on the
redshift of matter-radiation equality, $z_{\rm eq}$ (with an additional,
very weak, dependence on $\Omega_{\rm m}$).
Higher $\omm$ means higher $z_{\rm eq}$ and thus $\theta_{\rm eq}$ is
smaller; the rise in power from low $\ell$ (modes that entered at
$z<z_{\rm eq}$) to high $\ell$ (modes that entered at $z > z_{\rm eq}$)
gets shifted to higher $\ell$.
This shifting of the transition to higher $\ell$ results in a decrease
in power in the region of the transition and thus the shape of the
change in ${\cal D}_\ell^{TT}$ shown in Fig.~\ref{fig:Derivs}.
As we will see in Sect.~\ref{sec:shiftdescription}, an oscillatory decrease
in lower $\ell$ power (from increasing $\omm$) will be a key part of our
explanation for the parameter shifts.  Indeed, once the impact of the low
multipoles is reduced by the addition of high-$\ell$
data, the increase in power near the first peak from a redder
spectrum must be countered by a higher $\omm$
(and other shifts, see Sect.~\ref{sec:lowellanomaly}).

Additional dependence on $\omm$ comes from the change in the damping scale
and how recombination proceeds.
The damping scale is the geometric mean of the horizon and the mean free
path at recombination, and changing the expansion rate changes this scale
\citep{Silk1968,HuSug95}.
An increase in $\omm$ corresponds to a decrease in the physical damping scale
(which corresponds to a decreased angular scale at fixed distance to last
scattering).  However, within the range of variation in $\omm$ allowed by
\Planck, changes in damping are a sub-dominant effect.

Finally, the anisotropies we observe are modified from their primordial form
due to the effects of lensing by large-scale structure along the line of sight.
One effect of lensing is to ``smear'' the acoustic peaks and troughs, reducing
their contrast \citep{Seljak96}.
The peak smearing by lensing depends on $\omm$ through the decay of
small-scale potentials between horizon crossing and the epoch of equality \citep[see e.g.,][]{Pan14}.
While $\omm$ is an important contributor to the lensing effect, we will see
in Sect.~\ref{sec:lensing} that lensing will primarily drive shifts in
$\tau$ and $A_{\rm s} e^{-2\tau}$.

\subsection{The baryon density: $\omb$}
\label{sec:baryons}

For the nearly scale-invariant, adiabatic perturbations of interest to
us, the presence of baryons causes a modulation in the heights of the
peaks in the power spectrum and a change in the damping scale due to the
change in the mean free path.  Physically a non-zero baryon-photon
momentum density ratio, $R=3\rho_{\rm b}/(4\rho_\gamma)$, alters the zero-point
of the acoustic oscillations away from zero effective temperature
($\Theta_0+\Psi=0$) to $\Theta_0+(1-R)\Psi=0$
\citep[see e.g.,][]{Seljak94,Hu95,HuReview97}.
For non-zero $R\Psi$ this leads to a modulation of even and odd peak heights,
enhancing the odd peaks (corresponding to compression into a potential well)
with $R\Psi<0$ and reducing the even peaks (corresponding to rarefactions in
potential wells).
Given only low-$\ell$ data, such as for WMAP, the relative heights of
the first and second peaks, in particular, are important for determining $R$
and therefore $\omb$.
An increase in $\omb$ boosts the first peak relative to the second,
as is apparent in the $\omb$ panel of Fig.~\ref{fig:Derivs}.
We will see in Sect.~\ref{sec:shiftdescription} that the inclusion of the
high-$\ell$ data will lead to a decrease in $\omb$, which will be required
to better match the ratio of the first and second peaks once the
other parameters have shifted.

A change in $\omb$ also changes the mean free path of photons near
recombination, and the process of recombination itself, thus affecting the
diffusion damping scale.  As with an increase in $\omm$, an increase in $\omb$
decreases the physical damping scale. The angular scale which this corresponds
to depends on the distance to last scattering, which can be altered by changing
$\omb$, depending on what other quantities are held fixed. For the choice shown
in Fig.~\ref{fig:Derivs}, we find that the angular scale decreases as well,
leading to less damping and the excess of power seen at high $\ell$ in the
$\omb$ panel.

\subsection{The optical depth: $\tau$}
\label{sec:tauphysics}

Reionization in the late Universe recouples the CMB photons to the matter
field, but not as tightly as before recombination (since the matter density
has dropped by over six orders of magnitude in the intervening period).
Scattering of photons off electrons in the ionized intergalactic medium
suppresses the power in the primary anisotropies on scales smaller than
the horizon at
reionization ($\ell\,{\ga}\,10$) by $e^{-2\tau}$
\citep{Kaiser84,Efstathiou88,Sugiyama93,Hu96a}.  Because of
this, increasing $\tau$ at fixed $A_{\rm s}\,e^{-2\tau}$
keeps the power spectrum at $\ell\gg10$ nearly constant. The small wiggles
in the $\tau$ panel are entirely from the increased gravitational lensing
power, due to the increase in $A_{\rm s}$ necessary to keep
$A_{\rm s}\,e^{-2\tau}$ constant. At very low $\ell$ this
increase in $A_{\rm s}$ directly boosts anisotropies.

Increasing $A_{\rm s}\,e^{-2\tau}$ at fixed $\tau$ results in changes to
${\cal D}_\ell^{TT}$ that are almost exactly proportional to
${\cal D}_\ell^{TT}$, with small corrections due to the second-order effect of
gravitational lensing.

\subsection{The spectral index, $\ns$, and acoustic scale, $\theta_\ast$}
\label{sec:nstheta}

The final two effects are very easy to understand.
A change in the spectral index of the primordial perturbations yields a
corresponding change to the observed CMB power spectrum
\citep[e.g.,][]{Knox1995}.  Increasing $\ns$ with
the amplitude fixed at the pivot point $k\,{=}\,k_0\,{=}\,0.05\,\Mpc^{-1}$,
increases (decreases) power at $\ell\,{\ga}\, (\,{\la}\,)\,550$, since modes
with $k\,{=}\,k_0$ project into angular scales near $\ell\,{=}\,550$.
We will see in Sect.~\ref{sec:shiftdescription} that a tilt towards redder
spectra (i.e.~a decrease in high-$\ell$ power) will be necessary to best fit
the high-$\ell$ data.  Alternatively, as discussed in
Sect.~\ref{sec:lowellanomaly}, when not tightly constrained by the $\ell>1000$
data, a higher $n_{\rm s}$ allows a better fit to the ``deficit'' of power
at $\ell\,{<}\,30$.

The predominant effect of altering $\theta_\ast$ (which, with the other
parameters held fixed, is performed by modifying $\omegal$) is to stretch the
spectrum in the $\ell$ direction, causing large changes in the
rapidly-varying regions of the spectrum between peaks and troughs.
Note that the high sensitivity of the power spectrum to this scaling
parameter \citep[e.g.,][]{Kosowsky2002} means that small variations in
$\theta_\ast$ can swamp those of other parameters.
In Sect.~\ref{sec:shiftdescription} we will see that one of the differences
between the $\ell\,{<}\,800$ best-fit model and that for $\ell\,{<}\,2500$
is a variation in $\theta_\ast$ that shifts the third peak in the angular power
spectrum slightly to the right, removing some oscillatory residuals.

\subsection{The Hubble constant, $H_0$}
\label{sec:H0physics}

With these effects in hand it is easy to understand how changes in other
parameters, such as $H_0$, impact $\mathcal{D}_\ell^{TT}$. As discussed in
\citet[][section 3.1]{planck2013-p11}, the characteristic angular size of
fluctuations in the CMB ($\theta_\ast$) is exceptionally well and robustly
determined (better than 0.1\,\%). Within the $\Lambda$CDM model this angle is a
ratio of the sound horizon at the time of last scattering and the angular
diameter distance to last scattering.  The sound horizon is determined by the
redshift of recombination, $\omm$, and $\omb$, so the constraint on
$\theta_\ast$ translates into a constraint on the distance to last scattering,
which in turn becomes a constraint on the 3-dimensional subspace
$\omm$--$\omb$--$h$. Marginalizing over $\omb$ gives a strong degeneracy between
$\omm$ and $h$, which can be approximately expressed as $\Omega_{\rm
m}\,h^3={\rm constant}$ (as will be important in Sect.~\ref{sec:lowellanomaly}).
For example, an increase in $\omm$ decreases the sound horizon as
$\omm^{-0.25}$ (softened by the influence of radiation) and hence the distance
to last scattering must decrease, to hold $\theta_\ast$ fixed. This distance is
an integral of $1/H(z)$, with
$H^2(z)\propto\big\{\omm\big[(1+z)^3-1\big]+h^2\big\}$ for the dominant
contribution from $z\ll z_{\rm eq}$.  Thus $h$ must decrease in order for the
distance to last scattering not to decrease too much.

\subsection{Lensing}
\label{sec:lensingphysics}

As mentioned earlier, the anisotropies we observe are modified from their
primordial form by several secondary processes, among them the deflection
of CMB photons by the gravitational lensing associated with large-scale
structure \citep[see e.g.,][for a review]{Lewis06}.
These deflections serve to ``smear'' the last scattering surface, leading
to a smoothing of the peaks and troughs in the angular power spectrum, as
well as generating excess power on small scales, $B$-mode polarization, and
non-Gaussian signatures.  Our focus is on the first effect.

Gradients in the gravitational potential bend the paths of photons by a
few arcminutes, with the bend angles coherent over degree scales,
leading to a pattern of distortion and magnification on the initially
Gaussian CMB sky.
In magnified regions the power is shifted to lower $\ell$, while in
demagnified regions it is shifted to higher $\ell$.  Across the whole sky
this reduces the contrast of the peaks and troughs in the power spectrum
(while conserving the total power), and generates an almost power-law tail to
very high $\ell$.  The amplitude of the peak smearing is set by (transverse
gradients of) the (projected) gravitational
potential and this is sensitive to parameters (such as $A_{\rm s}$ and $\omm$),
which change its amplitude or shape.  The separate topic of CMB lensing
through the 4-point functions (to derive $C_\ell^{\phi\phi}$) is discussed
in Sect.~\ref{sec:Plancklensing}.

\section{Connecting parameter shifts to data to physics}
\label{sec:shifts}

With an understanding of the different ways in which the \LCDM model
parameters can adjust the $TT$ spectrum, we can now begin to try to explain the
parameter shifts of main interest for this paper. We start in
Sect.~\ref{sec:shiftdescription} by showing how the best-fit model has
adjusted from its $\ell\,{<}\,800$ solution to match the new data at
$\ell\,{>}\,800$. This story tracks more or less chronologically how our best
understanding of the \LCDM model has progressed, since the modes at
$\ell\,{\la}\,800$ had mostly been measured first with WMAP. Additionally, it
highlights the features of the \Planck\ data that are important for driving
parameter shifts with respect to the $\ell\,{<}\,800$ best-fit model.

The question answered in Sect.~\ref{sec:shiftdescription} is ``what caused the
parameters to shift from their $\ell\,{<}\,800$ values to their
$\ell\,{<}\,2500$ ones?'' A different, and also useful, question is ``what
causes there to be shifts at all, i.e., where do the {\it differences\/} come
from?'' This puts the $\ell\,{<}\,800$ and $\ell\,{>}\,800$ data on more equal
footing, allowing us to pick aspects of each that generate most of the
difference
between the two. Although the resulting story is not unique, we find that the
particular choice we have made results in a helpful explanation. It leads us to
identify the connection with gravitational lensing, which we discuss in
Sect.~\ref{sec:lensing}, and of the low-$\ell$ deficit, which we discuss in
Sect.~\ref{sec:lowellanomaly}.

\subsection{From $\ell\,{<}\,800$ to $\ell\,{<}\,2500$}
\label{sec:shiftdescription}

\begin{figure*}[htbp!]
\begin{center}
\resizebox{\textwidth}{!}{\includegraphics{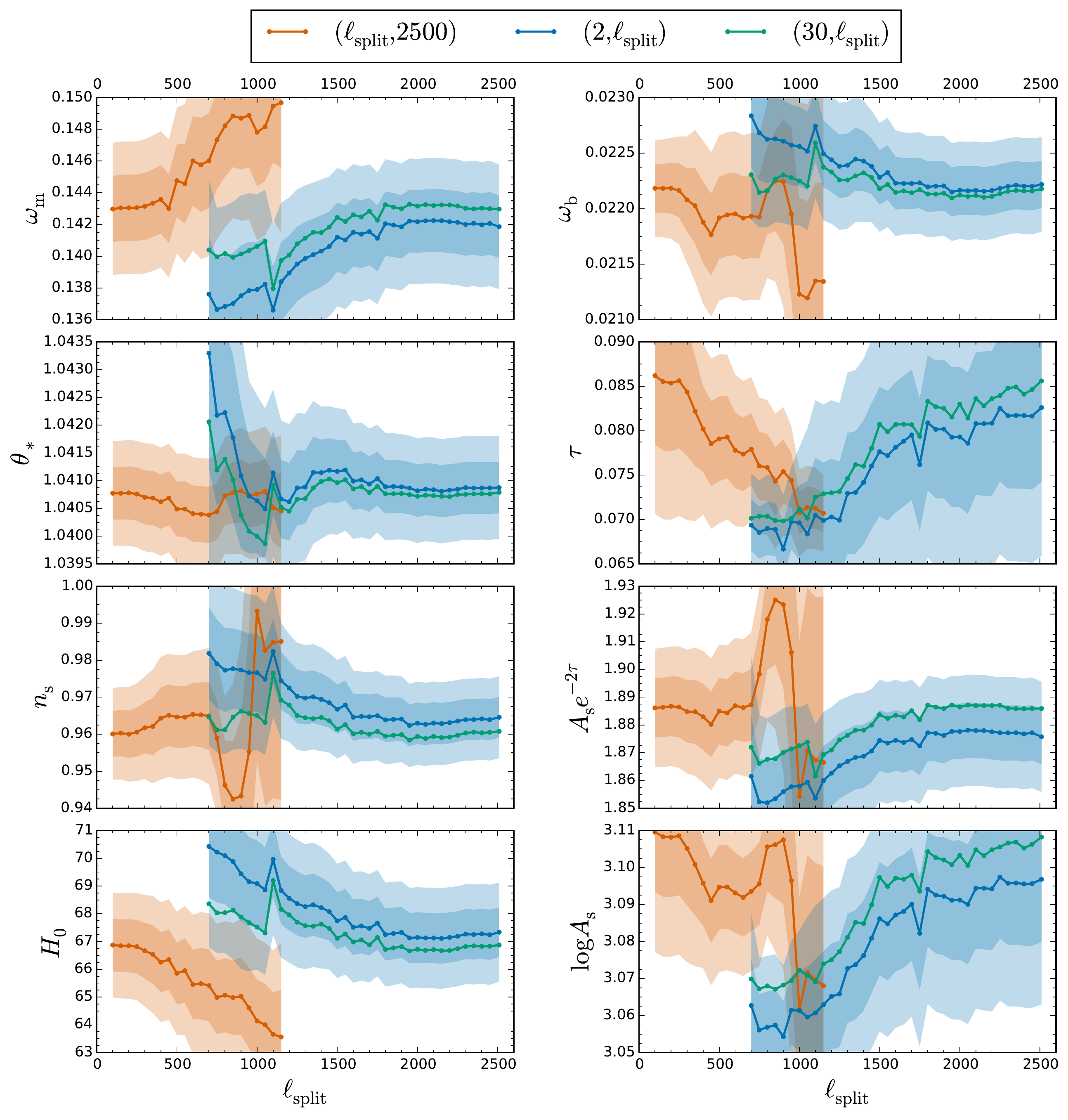}}
\end{center}

\caption{Shifts in the best-fit values of parameters when one considers the
multipole range either below or above different values of $\ell_{\rm split}$.
This uses the PlanckTT+$\tau$prior data combination, with $\ell\,{>}\,30$
computed using \texttt{plik\_lite}. The different lines correspond to
restricting the data to $\ell\,{<}\,\ell_{\rm split}$ (blue),
$30\,{<}\,\ell\,{<}\,\ell_{\rm split}$ (green), and $\ell\,{>}\,\ell_{\rm
split}$ (orange). These shifts are described in
Sect.~\ref{sec:shiftdescription}. One can see here that excising the
$\ell\,{<}\,30$ region moves the low-$\ell$ parameters closer to the high-$\ell$
parameters, as discussed in detail in Sect.~\ref{sec:lowellanomaly}. Error bands
are the $\pm1$ and $\pm2\,\sigma$ scatter in the simulations away from the input
fiducial model. We have chosen to plot this quantity as opposed to posterior
constraints on these parameters (which is different because of our prior on
$\tau$) because it is these bands that are appropriate for comparing the blue
and orange lines against each other. Note that this has the perhaps
counter-intuitive effect of having the error bands in the $\tau$ panel increase
as more data are added. None of the local ``spikes'' are found to be
significant, as can be seen from the bottom panel of
Fig.~\ref{fig:stats_lsplit_scan}.}

\label{fig:lsplit}
\vspace{5cm}
\end{figure*}

\begin{figure*}[htbp!]
\begin{center}
\resizebox{\textwidth}{!}{\includegraphics{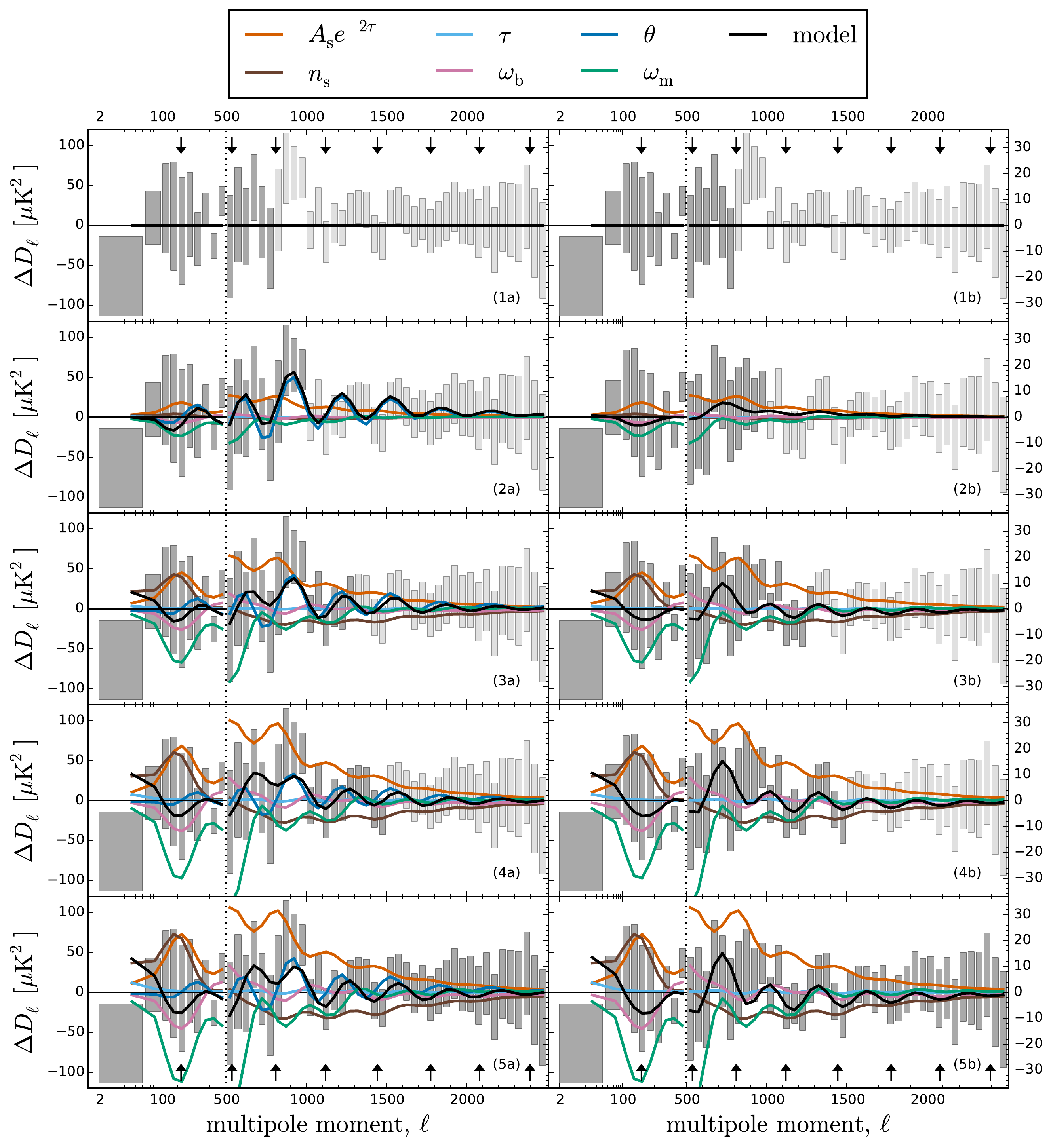}}
\end{center}
\vspace{-0.5cm}

\caption{How the best-fit $\ell\,{<}\,\ell_{\rm max}$ PlanckTT+\tauprior\ \LCDM\
model adjusts as $\ell_{\rm max}$ is increased from 800 to 2500 (going from the
top panels to the bottom panels). \textit{Left column}: all panels show
residuals relative to the $\ell\,{<}\,800$ model. \Planck\ power spectrum binned
estimates and $\pm1\,\sigma$ errors on the CMB spectrum, as extracted with
\texttt{plik\_lite}, are shown as grey boxes. Note the change in $y$-axis
scale at
$\ell\,{=}\,500$, indicated by the vertical dotted line. The solid black line is
the best-fit model for $\ell\,{<}\,\ell_{\rm max}$, where $\ell_{\rm max}$ is
different for each panel, as indicated by which of the boxes are shaded darker.
The various coloured lines indicate the {\it linear\/} response to the shift in
individual parameters between their $\ell\,{<}\,800$ best-fit value and their
$\ell\,{<}\,\ell_{\rm max}$ one. \textit{Right column}: identical to the left
column, except that the contribution from $\theta_\ast$ (i.e., the blue line
from the corresponding left panel) has been subtracted from the sums, as well
as from the actual model and from the data. For reference, the arrows in the
top and bottom panels show the locations of the peaks in the power spectrum.}

\label{fig:residuals_grid}
\vspace{1cm}
\end{figure*}

We begin by examining how parameters shift as we increase $\ell_{\rm max}$ from
800 to 2500.  The best-fit parameters from the range $\ell\,{<}\,\ell_{\rm max}$
are shown by the solid blue curve in Fig.~\ref{fig:lsplit} (where $\ell_{\rm
split}$ is, in this case, $\ell_{\rm max}$). Although eight parameters are
displayed in this figure, for the purpose of explaining shifts it is important
to consider only six parameters at a time (since there are only six degrees of
freedom in the \LCDM model). We will use the set of six discussed in
Sect.~\ref{sec:lowvsfull}, for the reasons described there. As a reminder, they
are $\theta_\ast$, $\omm$, $\omb$, $\ns$, $A_{\rm s}\,e^{-2\tau}$, and $\tau$.
Focusing on these parameters, one can see in Fig.~\ref{fig:lsplit} the following
changes:

\begin{itemize}
\item a sharp drop in $\theta_\ast$ between $\ell_{\rm max}\,{=}\,800$ and
1000;
\item a highly correlated gradual drop in $\omega_{\rm b}$, drop in $n_{\rm s}$,
increase in $\omega_{\rm m}$, and increase in $A_{\rm s}e^{-2\tau}$ across the
whole multipole range;
\item an increase in $\tau$ between $\ell_{\rm max}\,{=}\,1000$ and 1500.
\end{itemize}

Figure~\ref{fig:residuals_grid} illustrates even more explicitly how these
different multipole ranges cause the parameter shifts. This figure compresses
a large amount of information into a combination of 10 panels, the full
understanding of which requires a slow stepwise explanation.
Each of the panels in
the left column shows residuals of the data relative to the best-fit
$\ell\,{<}\,800$ model. The thick black line is the best-fit model for
$\ell\,{<}\,\ell_{\rm max}$, with $\ell_{\rm max}$ increased in each
subsequent panel and represented by the darker data points (varying from
$\ell_{\rm max}\,{=}\,800$ in the top panels to $\ell_{\rm max}\,{=}\,2500$ in
the bottom panels).

In panel (1a) of Fig.~\ref{fig:residuals_grid}
we have $\ell_{\rm max}\,{=}\,800$ and thus we see directly the
residuals in the $\ell\,{>}\,800$ data with respect to the $\ell\,{<}\,800$
model that cause the parameter shifts of main interest for this work. We will
sometimes refer to these features as the ``oscillatory residuals''; for
definiteness, we are referring to the upward trends at $\ell\,{\simeq}\,\{900,
1300, 1600, 1800\}$ and downward ones at $\ell\,{\simeq}\,\{1100, 1400, 1700\}$.
Note that these oscillations are (roughly)
out of phase with the CMB
peaks themselves, a point which will be important for future discussion.

We can assess the significance of the residuals at the power spectrum level by
computing their $\chi^2$. With the same $\Delta\ell\,{=}\,50$ bins as in
Fig.~\ref{fig:residuals_grid}, we find $\chi^2\,{=}\,36.4$ for 34 bins,
equivalent to a $0.6\,\sigma$ Gaussian fluctuation. This lack of significance in
the residuals in the power spectrum itself underscores the fact that we are not
talking about large residuals here, even if they happen to appear more
significant in the cosmological parameter space. Finally, we point out that
these residuals are of course not inherent to the $\ell\,{>}\,800$ data
themselves, rather to the difference with the best-fit model predicted from the
$\ell\,{<}\,800$ data; in Sect.~\ref{sec:lowellanomaly} we will comment on how
the $\ell\,{<}\,30$ data in particular threw off this model from the best
estimate coming from the full $\ell$ range.

Beginning now to increase $\ell_{\rm max}$ up to 1000, in panel (2a) we see the
model adjusting to match the data in the 800 to 1000 region. We would also like
to understand why and how the various parameters have shifted to incorporate
these data, which we can do in the following way. Under the approximation of
linear response, it is possible to break apart the total change in the model
into the contribution from each individual parameter. This is given by the
quantity $\Delta p_i\, dC_\ell/dp_i$, where $p_i$ represents each of the
parameters and $\Delta p_i$ is the shift in each parameter's value between the
two cases being compared. If the linear approximation were perfect, the sum of
the contributions from each parameter would give exactly the total shift; here
we find that the approximation is accurate to 10\,\% of the total shift, which
is sufficient for our discussion here.  We have computed these
derivatives for the best-fit $\ell\,{<}\,800$ model. Because these are
linear responses, the model can only change their amplitudes.

Panel (2a) of Fig.~\ref{fig:residuals_grid} shows
that the only response with significant support on the 800--1000
region is $\theta_\ast$, which indeed shifts to almost perfectly pick up the
difference there.  The effect is essentially that the third peak has shifted
slightly to the right.  With the other parameters held fixed, this change in
$\theta_\ast$ alone is responsible for lowering $H_0$ by
$0.5\,{\rm km}\,{\rm s}^{-1}\,{\rm Mpc}^{-1}$.
An additional decrease in $H_0$, by about the same amount, can be ascribed to
an increase of the matter density, which, in combination with an
increased $A_{\rm s}e^{-2\tau}$, better fits the position of the second trough
at $\ell\,{\simeq}\,650$.

Because no further increase in $\ell_{\rm max}$ changes $\theta_\ast$ by much
(and because \Planck's measurement of $\theta_\ast$ is so sensitive that the
oscillation caused by changing $\theta_\ast$ can be accommodated by only a small
shift in its value), we subtract its effect from the model and data to better
see the effects of the other parameters and we plot the result in the right
column of Fig.~\ref{fig:residuals_grid}. With this shift in $\theta_\ast$
subtracted, panel (2b) shows that qualitatively this makes the oscillatory
features that we have already seen become slightly more pronounced.

The first way in which the parameters adjust to fit the remaining data is via
movement along a parameter direction involving $\omega_{\rm b}$, $\omega_{\rm
m}$, $A_{\rm s}e^{-2\tau}$, and $n_{\rm s}$. Although this is a fairly
complicated combination, the biggest change in the spectrum comes from the
increase in primordial power that results in an {\it oscillatory increase\/} in
the CMB spectrum, and an increase in the matter density that results in an {\it
oscillatory decrease\/} in power.  This leaves an oscillatory pattern
oscillating about zero when we consider $\ell_{\rm max}\,{=}\,1000$. As we
increase $\ell_{\rm max}$ between panels (2b) and (5b), this same parameter mode
grows in amplitude. Furthermore, the effect of the change in the primordial
power spectrum, both the increase in amplitude and tilt towards redder spectra,
is also necessary to match the oscillations. This combination of parameters, and
in particular the decrease in $\ns$, also drives disagreement with the very
lowest bin in this figure, $\ell\,{<}\,30$ (as we discuss in
Sect.~\ref{sec:lowellanomaly}).

Finally, we observe an increase in $\tau$ and a corresponding increase in
$\As$, which, although barely visible in Fig.~\ref{fig:residuals_grid},
does also track the same oscillatory features. We discuss this shift further
in Sect.~\ref{sec:lensing}.

To summarize, the features in the $\ell\,{>}\,800$ data that are primarily
responsible for the shifts in parameters are largely oscillatory, as seen in
e.g., panel (1a) of Fig.~\ref{fig:residuals_grid}. After an initial
shift in $\theta_\ast$ to pick up the excess between $\ell\,{=}\,800$ and
1000, the remaining residuals are tracked by two directions in parameter
space, namely an increase in $\tau$ and a movement along the
$A_{\rm s}e^{-2\tau}$--$n_{\rm s}$--$\omega_{\rm b}$--$\omega_{\rm m}$
degeneracy direction, both of which serve to increase the amplitude of the
oscillations.

\begin{table*}[tb!]
\newdimen\tblskip \tblskip=5pt

\caption{Comparison of the expected dispersion (``Exp.'') and observed
(``Obs.'') parameter shifts between pairs of datasets. We show results for the
case where we use all the lowest multipoles, where we excise the $\ell\,{<}\,30$
multipoles, and where we also fix the lensing potential, as described in
Sect.~\ref{sec:lensing}. The shifts are shown in units of standard deviation of
the respective $\ell\,{<}\,800$ runs for each case. The ratio between observed
shifts and expected dispersions becomes smaller when excising the
$\ell\,{<}\,30$ multipoles, and even more when factoring out the impact of
lensing. Note the final column has expected shifts calculated as in equation~53
of \cite{planck2014-a13} rather than using simulations.}

\label{tab:shifts}
\vskip -3mm
\footnotesize
\setbox\tablebox=\vbox{
 \newdimen\digitwidth
 \setbox0=\hbox{\rm 0}
 \digitwidth=\wd0
 \catcode`*=\active
 \def*{\kern\digitwidth}
  \newdimen\signwidth
  \setbox0=\hbox{+}
  \signwidth=\wd0
  \catcode`!=\active
  \def!{\kern\signwidth}
  \newdimen\dpwidth
  \setbox0=\hbox{.}
  \dpwidth=\wd0
  \catcode`?=\active
  \def?{\kern\dpwidth}
\halign{\hbox to 3.5cm{#\leaderfil}\hfil\tabskip=1em&
    \hfil#\hfil\tabskip=1em&
    \hfil#\hfil\tabskip=0.5em&
    \hfil#\hfil\tabskip=2em&
    \hfil#\hfil\tabskip=1em&
    \hfil#\hfil\tabskip=0.5em&
    \hfil#\hfil\tabskip=2em&
    \hfil#\hfil\tabskip=1em&
    \hfil#\hfil\tabskip=0.5em&
    \hfil#\hfil\tabskip=0pt\cr
\noalign{\doubleline}
\omit& \multispan3\hfil $(2,800)$ vs. $(2,2500)$ \hfil& \multispan3\hfil $(30,800)$ vs. $(30,2500)$\hfil& \multispan3\hfil $(30,800)$ vs. $(30,2500)$, fixlens\hfil\cr
\noalign{\vskip -4pt}
\omit& \multispan3\hfil\hrulefill\hfil& \multispan3\hfil\hrulefill\hfil& \multispan3\hfil\hrulefill\hfil\cr
\omit Parameters\hfil& Exp.& *Obs.& $\lvert$Obs./Exp.$\rvert$& Exp.& *Obs.& $\lvert$Obs./Exp.$\rvert$& Exp.& *Obs.& $\lvert$Obs./Exp.$\rvert$\cr
\omit& [$\sigma$]& *[$\sigma$]& & [$\sigma$]& [$\sigma$]& & *[$\sigma$]& [$\sigma$]&\cr
	\noalign{\vskip 4pt\hrule\vskip 3pt}
   $\Ombh$& 0.8& $!0.9$& 1.1&   0.8& $!0.0$& 0.0&   0.8& $-0.5$& 0.6\cr
   $\Ommh$& 0.8& $-1.6$& 2.0&   0.8& $-0.7$& 0.9&   0.7& $-0.3$& 0.4\cr
$\thetaMC$& 0.9& $!0.9$& 0.9&	0.9& $!0.4$& 0.4&   0.9& $!0.2$& 0.2\cr
    $\tau$& 0.4& $-1.0$& 1.7&	0.4& $-0.7$& 1.9&   0.2& $-0.0$& 0.2\cr
   $\lnAs$& 0.4& $-1.0$& 2.4&	0.4& $-1.0$& 2.2&   0.1& $-0.2$& 1.7\cr
     $\ns$& 0.8& $!1.0$& 1.2&	0.9& $!0.0$& 0.0&   0.9& $-0.5$& 0.5\cr
     $H_0$& 0.8& $!1.4$& 1.8&	0.8& $!0.5$& 0.6&   0.8& $!0.1$& 0.1\cr
  $\clamp$& 0.7& $-1.5$& 2.2&	0.7& $-0.9$& 1.3&   0.6& $-0.7$& 1.1\cr
\noalign{\vskip 3pt\hrule\vskip 5pt}}}
\endPlancktable
\end{table*}

\begin{figure}[htbp!]
\begin{center}
\resizebox{\columnwidth}{!}{\includegraphics{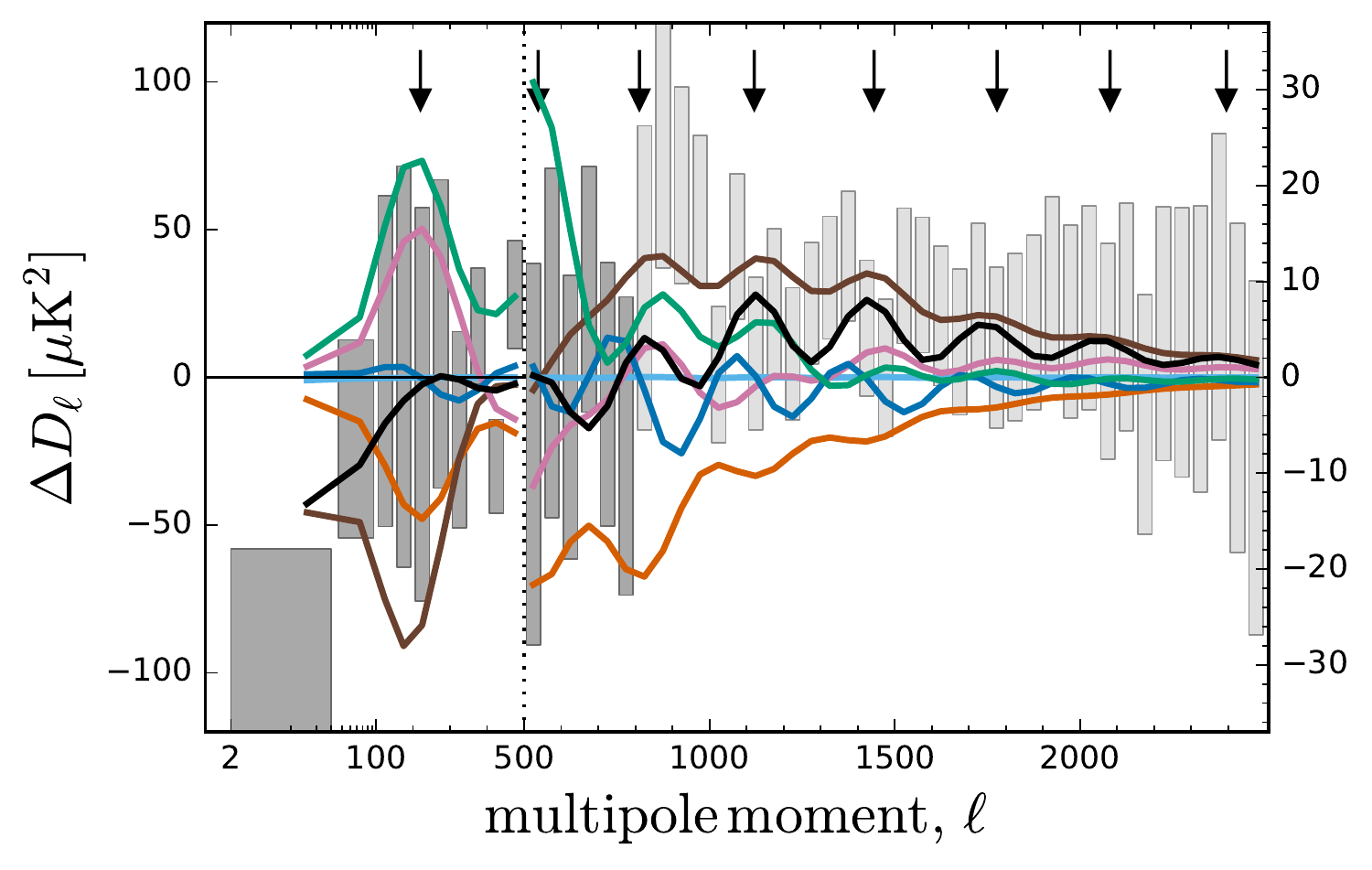}}
\resizebox{\columnwidth}{!}{\includegraphics{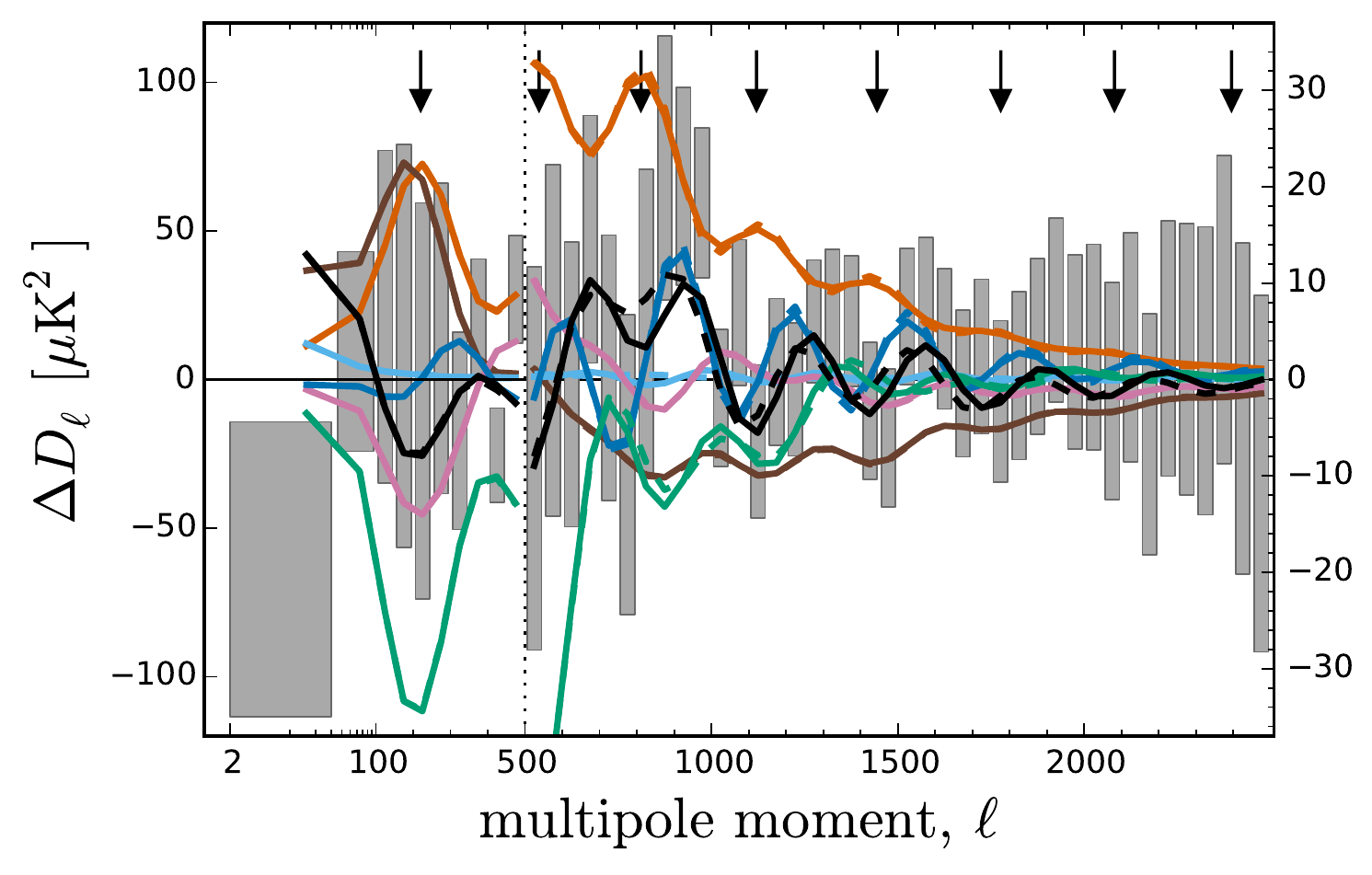}}
\resizebox{\columnwidth}{!}{\includegraphics{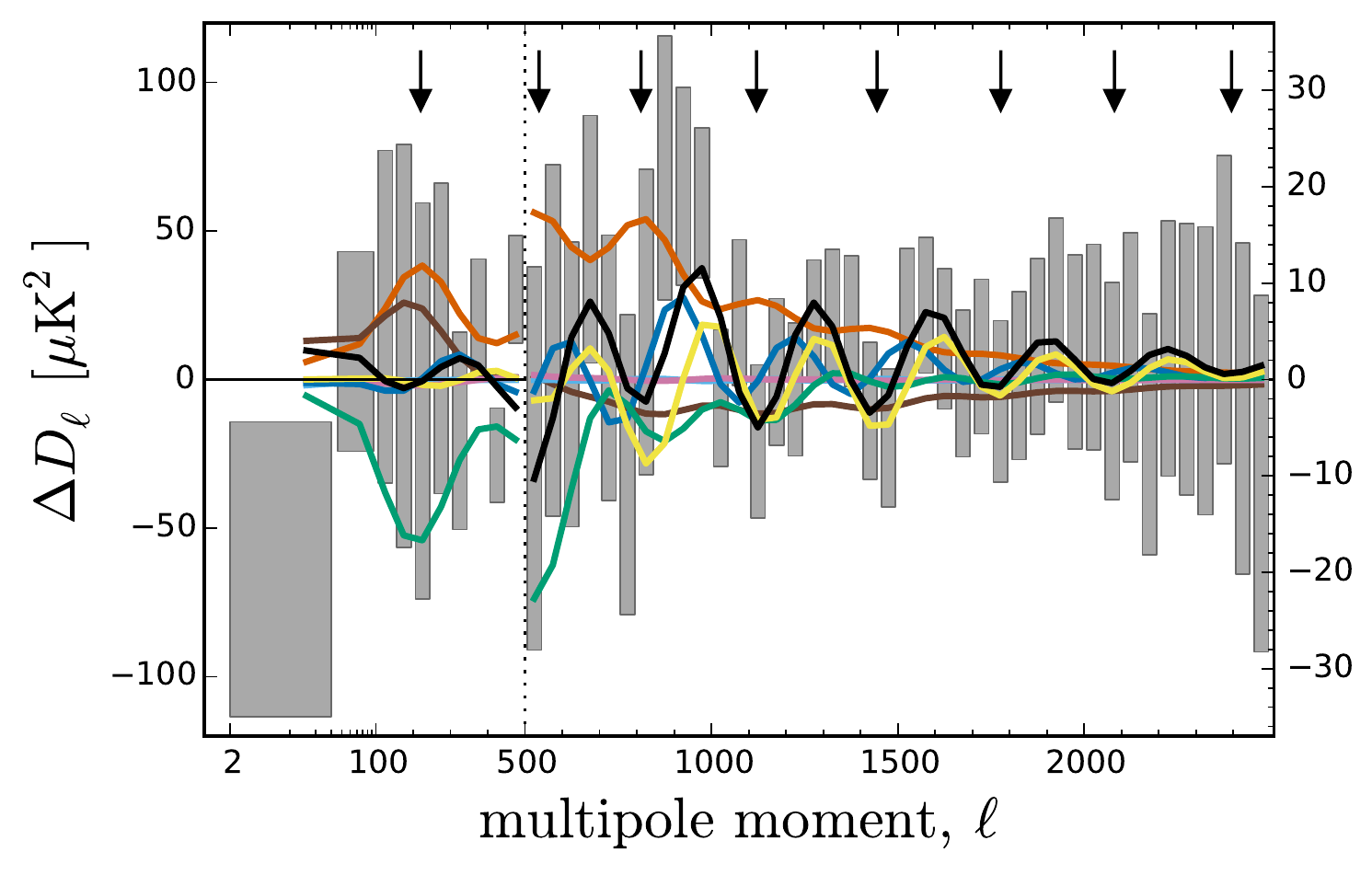}}

\end{center}

\caption {Power spectrum residuals for
a few additional cases, in the same format as
Fig.~\ref{fig:residuals_grid}. Note that for the top panel, the fiducial model
is the best-fit from $30\,{<}\,\ell\,{<}\,800$, as opposed to from
$\ell\,{<}\,800$, as is the case in Fig.~\ref{fig:residuals_grid} and in the
bottom two panels of this figure. In all cases the black line is the best-fit
\LCDM model in the range indicated by the shaded data boxes. The coloured lines
are the linear responses to the shifts in parameters between these two
best-fit solutions. {\it Top}: The way in which the best-fit model from
$30\,{<}\,\ell\,{<}\,800$ is ``thrown off'' by inclusion of $\ell\,{<}\,30$
data. Note that although visually the $\ell\,{>}\,800$ data appears to be a
{\it better\/} fit with $\ell\,{<}\,30$, the $\chi^2$ is worse by $\Delta
\chi^2\,{=}\,3.2$. {\it Middle}: Same as panel (5a) of
Fig.~\ref{fig:residuals_grid}, but with dashed lines showing the responses with
the gravitational potential fixed. {\it Bottom}: Same as panel (5a) of
Fig.~\ref{fig:residuals_grid}, but with an additional free parameter, $A_{\rm
L}$, shown in yellow. This added degree of freedom tracks reasonably well the
oscillatory residuals, leaving smaller shifts for the other parameters and a
reduced low-$\ell$ deficit.} \label{fig:residuals_one_panel}

\end{figure}

\begin{figure*}[htbp!]
\begin{center}
\resizebox{\textwidth}{!}{\includegraphics{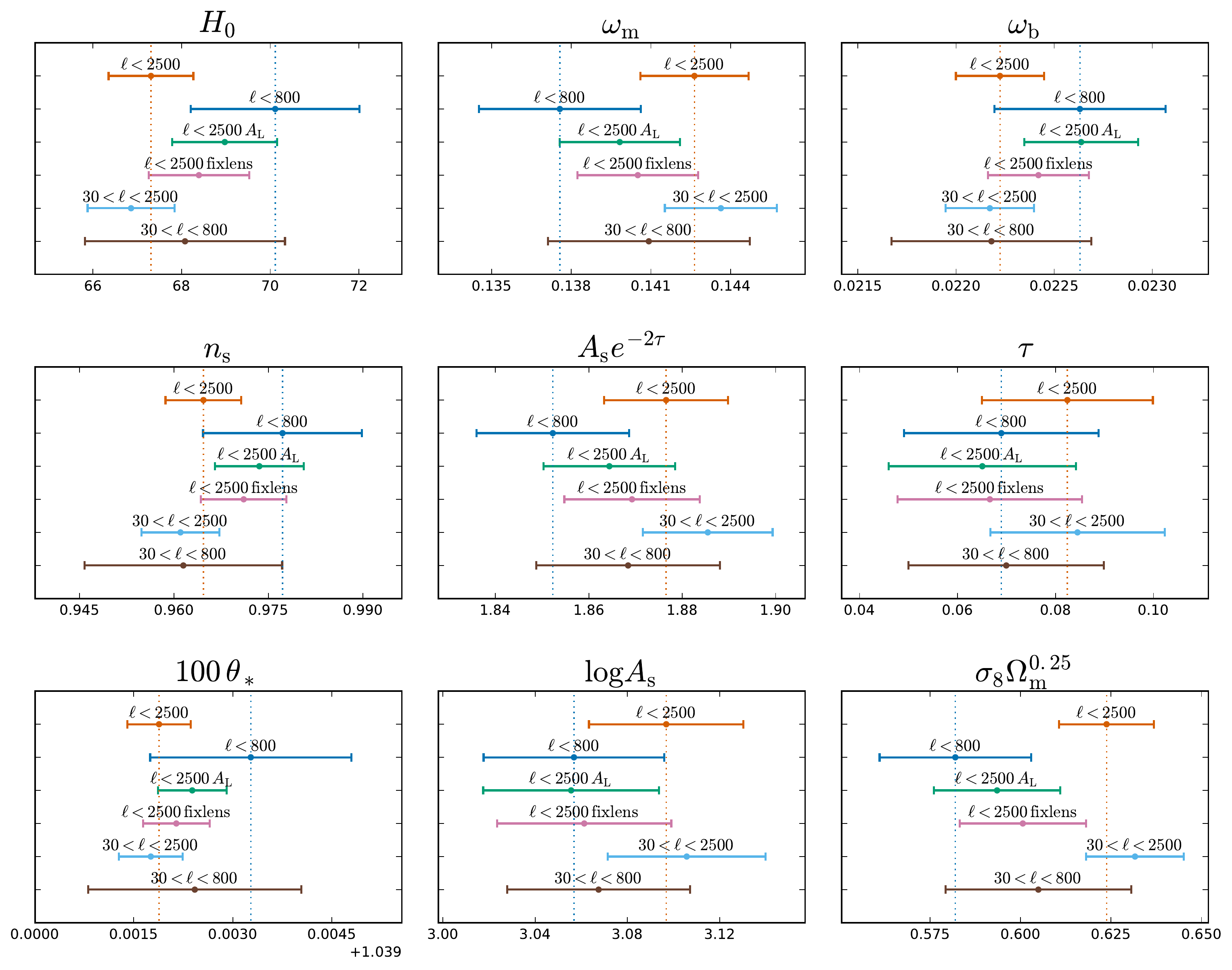}}
\end{center}

\caption {Marginal mean and 68\,\% error bars on cosmological parameters
estimated with different data choices, assuming the $\Lambda$CDM model (unless
otherwise labelled). We use the \TT\ likelihood in combination with a prior
$\tau=0.07\pm0.02$. Excising the low multipoles, i.e., $\ell\,{<}\,30$,
substantially improves the agreement between the parameters from
$\ell\,{<}\,800$ and the $\ell\,{<}\,2500$ range. Further agreement is then
achieved when removing the effect of gravitational lensing.}

\label{fig:whisker}
\end{figure*}

\subsection{Gravitational lensing}
\label{sec:lensing}

Having described the shifts fairly pragmatically, we now turn to trying to
understand what, physically, is driving them. It is clear that the oscillatory
residuals are important, and qualitatively we can see that they look like extra
smoothing of the peaks and hence resemble the effects of gravitational lensing.
Indeed, along with the parameter shifts themselves, much attention has been
given in the literature to the fact that the \Planck\ high-$\ell$ data appear to
favour an overly enhanced gravitational lensing potential with respect to that
expected from \LCDM \citep{planck2014-a15,Couchot15,Addison15}. Given this, and
noting that the parameters shift to increase $A_{\rm s}$ and $\omega_{\rm m}$
(both of which
increase the gravitational lensing potential) it may be tempting to think that
the parameter shifts are dominantly driven by a desire to increase lensing and
hence increase peak smoothing at high $\ell$. We will see,
however, that this only
explains about a third of the total shifts and instead most of the change in the
best-fit model spectrum is related to non-lensing effects such as changing the
matter envelope (Sect.~\ref{sec:matter}) and the primordial tilt
(Sect.~\ref{sec:nstheta}).

The effect of lensing of the $TT$ spectrum has traditionally been studied by
introducing an additional phenomenological parameter, $A_{\rm L}$, which
artificially scales the lensing potential power spectrum used to calculate the
lensed CMB spectra. By definition $A_{\rm L}\,{=}\,1$ corresponds to \LCDM. The
\Planck\ $\ell\,{<}\,2500$ data prefer a value higher than unity, $A_{\rm
L}\,{=}\,1.22\pm0.10$ \cite{planck2014-a15}.
The bottom panel of Fig.~\ref{fig:residuals_one_panel}
shows the same power spectrum residual and linear responses of
Fig.~\ref{fig:residuals_grid}, now with $A_{\rm L}$ as an additional free
parameter. As we see, the response from increasing $A_{\rm L}$ on its own does a
somewhat good job of fitting the data, particularly at $\ell\,{>}\,1000$,
leaving
smaller shifts in the other parameters. We do note, however, that although some
of the other cosmological parameters shift closer to the values preferred by the
$\ell\,{<}\,800$ case,\footnote{The best-fit cosmology of the $\ell\,{<}\,800$
case is not significantly influenced by the impact of lensing.} differences
remain. For example, as shown in Fig.~\ref{fig:whisker}, about half of the
shifts (in e.g., $\omega_m$ and $H_0$) remain even in the $\Lambda$CDM+$A_{\rm
L}$ case. Thus, the shift in parameters between $\ell\,{<}\,800$ and
$\ell\,{<}\,2500$ cannot be entirely explained through an extra peak-smoothing
effect at high $\ell$; other aspects of the data are also independently pointing
to similar shifts.

In terms of understanding physically how the features in the $\ell\,{>}\,800$
data are fit by the \LCDM model, the $A_{\rm L}$ test is, however, not entirely
useful. The \LCDM model, unlike $\Lambda$CDM+$A_{\rm L}$,
is of course {\it not\/}
free to arbitrarily increase the lensing potential; it must do so through other
parameters that also have non-lensing related effects. Thus the particular way
in which \LCDM chooses to optimally fit the features will be a balance between
lensing and non-lensing effects. It is now useful to define more exactly the
question we are seeking to answer. Ascertaining what aspects of the data ``are
lensing'' is an ill-defined question; conversely, ascertaining which parts of
the change between two model power spectra come from
lensing is perfectly well defined
because we can theoretically calculate the two spectra with and without lensing
included. This is what is shown in the middle panel of
Fig.~\ref{fig:residuals_one_panel}. Here we plot the same power spectrum linear
responses as in Fig.~\ref{fig:residuals_grid}, but additionally (as the dashed
lines) we remove the contribution from changing the lensing potential; more
precisely, the dashed lines are $dC_\ell/dp$, with $C_\ell$ being the
{\it unlensed\/}
power spectrum. Thus, even without affecting the lensing potential, the shifts
in parameters we have been discussing cause the spectrum to largely match the
oscillatory features we see in the data.

In terms of cosmological parameters, we can verify that
most of the shifts are still
there even in the absence of changes to the gravitational lensing potential with
the following test. We again look at shifts between $\ell\,{<}\,800$ and
$\ell\,{<}\,2500$, but for the $\ell\,{<}\,2500$ case we fix the lensing
potential to its own best-fit from $\ell\,{<}\,2500$. In doing so, the
cosmological parameters no longer impact the amplitude of the lensing potential,
which is already at the value favoured by the full $\ell$-range fit. Any
remaining shifts must reflect features in the data that are not accounted for by
the change to the lensing potential alone, and are instead fit by non-lensing
effects of changing the cosmological parameters. We find, as shown in
Fig.~\ref{fig:whisker}, that the majority of the shifts are still present. For
example, $H_0$ still moves from ($70.0 \pm 1.9$) with $\ell\,{<}\,800$ to
$(68.4\pm 1.1) \,{\rm km}\,{\rm s}^{-1}\,{\rm Mpc}^{-1}$ with $\ell\,{<}\,2500$
and fixed lensing.  Roughly speaking, about two thirds of the shift in the
Hubble constant and other parameters comes from non-lensing effects.

The only exception to lensing being a sub-dominant part of the shifts is $\tau$
and the corresponding change in $\As$, whose entire shift is explained by
lensing. This confirms what we might expect, since at $\ell\,{>}\,100$ the {\it
only\/} effect of changing $\tau$ (at fixed $A_{\rm s}e^{-2\tau}$) is via
lensing effects, and if the non-lensing effect of $\tau$ at $\ell\,{<}\,100$
would have been driving its shift, it is clear from
Fig.~\ref{fig:residuals_grid} that it would have shifted in the other direction.
We have gone further and also investigated whether the part of shifts in $\As$
and $\Ommh$ that {\it are\/} related to lensing are due to the fact that both of
these parameters directly impact the lensing amplitude, or whether this is
rather through the correlation between the two due to non-lensing effects in the
power spectrum. We checked this by fixing the lensing potential to the
$30\,{<}\,\ell\,{<}\,800$ best-fit case, and letting only $\As$ change its
amplitude. We find that in this case, $\As$ and $\tau$ are forced to values even
higher than in the standard $30\,{<}\,\ell\,{<}\,2500$ case, while the posterior
of $\Ommh$ remains very close to the best-fit of the $30\,{<}\,\ell\,{<}\,800$
case. We thus conclude that it is indeed the direct impact of both $\Ommh$ and
$\As$ on the lensing amplitude that is important.

One reason the sub-dominant impact of lensing discussed in this section is
subtle is because of a coincidental parameter degeneracy. As discussed in the
previous section, fitting the oscillatory features increases $\omega_{\rm m}$
and $A_{\rm s}e^{-2\tau}$. By coincidence, these shifts both increase the
lensing potential {\it and\/} increase the amplitude of the peak smoothing via
non-lensing effects, but it is the latter that is more important.

\subsection{The low-$\ell$ deficit}
\label{sec:lowellanomaly}

With part of the shifts explained by a preference, albeit sub-dominant, for an
increased lensing potential, we now seek to explain the rest of the differences.
If we are free to attribute the variations to specific
multipoles in either of the two data sets we are comparing, there is not a
unique way to tell this story. For example, one could look further at the
$\ell\,{>}\,800$ data and isolate what, aside from the lensing piece we have
just described, is causing the shifts. We choose here a different path,
which we believe is more elucidating and attributes
the remaining difference to the $\ell\,{<}\,800$ data instead. It also has the
advantage that it likely explains, chronologically, why the parameters have
shifted (since, again, these modes were measured first with WMAP). The specific
explanation is that a large remaining part of the differences is due to
multipoles at $\ell\,{<}\,30$ having ``thrown off'' the $\ell\,{<}\,800$ result.

In the previous section, it was noted that as the model adjusted to fit the data
in the $1000\,{<}\,\ell\,{<}\,1500$ region, the fit at $\ell\,{<}\,30$ became
much worse. This is evidence that the $\ell\,{<}\,30$ region might play a major
role in driving disagreement between the low and high multipoles. Indeed,
``anomalies'' related to the low-$\ell$'s have been discussed extensively in the
literature, for example the low quadrupole or the localized ``dip'' near
$\ell\,{\simeq}\,20$ \citep{Bennett96, Hinshaw03, Spergel03, Peiris03,
Mortonson09, Cai15}. Here we are interested mainly in the overall deficit in
power across the entire $\ell\,{\la}\,30$ region (which does of course gain
{\it some\/} contributions from the low quadrupole and the
$\ell\,{\simeq}\,20$ dip, but
also from other multipoles); we refer to this as the ``low-$\ell$ deficit.''
This is exactly the same deficit in power discussed previously in
\citet{planck2013-p08}, \citet{planck2013-p11}, \citet{planck2014-a15}, and
others papers, where it is sometimes called the ``low-$\ell$ anomaly.''
We explicitly call it a ``power deficit'' here to avoid
confusion with any other ``anomalies'' at low-$\ell$, and because it is a
more appropriate name for a feature of only moderate significance.  Indeed, if
one models the deficit simply as an overall power rescaling at $\ell\,{<}\,30$
with respect to the \LCDM model, its significance is 1.1\,$\sigma$ when
considering the $\ell\,{<}\,800$ data, growing to 1.6\,$\sigma$ for the
full-$\ell$ range (since the \LCDM model prediction is moved
higher).\footnote{See sections~8 and 9 of \citet{planck2014-a24} for
alternative investigations of the significance of the power deficit using
$P(k)$ reconstruction and parameterized model fits.  Inflationary models with
features are not found to give sufficiently improved fits (compared to a
featureless power spectrum) to justify adding the additional parameters.}
Assuming
\LCDM, the low-$\ell$ deficit is thus most likely a sample-variance fluctuation
in $C_\ell$ that happens to be concentrated at the lowest multipoles. Despite
interpretation of the deficit from different perspectives
\citep[e.g.][]{Contaldi03, Iqbal15, Chen16}, up until now, its effect on the
parameter shifts has not been thoroughly explored.

Indeed, when excising the range $\ell\,{<}\,30$, we observe a relatively large,
correlated shift in parameters, as shown in Fig.~\ref{fig:whisker}.  For
example, $H_0$ shifts from $(70.0\,{\pm}\,1.9)\hunits$ when using
$\ell\,{<}\,800$ to $(68.0\,{\pm}\,2.2)\hunits$ when using
$30\,{<}\,\ell\,{<}\,800$, much closer to the value preferred by the full
multipole \Planck\ cosmology, which is $(67.3\,{\pm}\,1.0)\hunits$. This shift
is 1.8 times larger than the expected shift from simulations for the two data
sets, in line with its somewhat anomalous nature. Although the deviations
induced by these low multipoles are not statistically very significant, they are
one of the main sources of difference between the $\ell\,{<}\,800$ and
$\ell\,{<}\,2500$ parameters, as also shown in Table~\ref{tab:shifts}.
Furthermore, if one considers this ``deficit'' as a mere statistical fluctuation
in the power spectrum, the fact that it happens to occur at the lowest
multipoles gives it greater weight in shifting parameter like $n_{\rm s}$ than
if it had occurred elsewhere.  In detail we find that the shifts between the two
ranges $\{\Delta A_{\rm s} e^{-2\tau}, \Delta n_{\rm s}, \Delta \omm, \Delta
\omb, \Delta H_0, \Delta \tau \}$ in units of the $1\,\sigma$ expected shifts
are $\{-2.2, 1.2, -2.0, 1.1, 1.8, -1.7\}$; without $\ell\,{<}\,30$ in either
data set, they become $\{-1.3, 0.0, -0.9, -0.0, 0.6, -1.9\}$.

We now turn to understanding in more detail the way that the low-$\ell$ deficit
sources these parameter differences. This discussion follows closely the top
panel of Fig.~\ref{fig:residuals_one_panel}, which shows how one goes from the
$30\,{<}\,\ell\,{<}\,800$ best-fit (the fiducial model against which the points
in the figure are differenced) to the $\ell\,{<}\,800$ best-fit (the black
line). Here we see how the low amplitude of the first 30 multipoles can be fit
by a correlated change in $\ns$, $\Ombh$, $\Ommh$, and $\clamp$. In particular,
with the $30\,{<}\,\ell\,{<}\,800$ best-fit as a starting point, the model needs
to decrease power at $\ell\,{<}\,30$ to fit the low-$\ell$ deficit; this can be
achieved with an increase in $\ns$, which tilts the spectrum and decreases power
at the lowest multipoles. However, this has three additional effects
that trigger the
response of the other cosmological parameters. Firstly, since the increase in
$n_{\rm s}$ reduces power not just at $\ell\,{<}\,30$ but over the entire
$\ell\,{\la}\,550$ part of the power spectrum (because our pivot scale
corresponds to $\ell\,{\simeq}\,550$), $\Ommh$ decreases to compensate by
shifting the matter envelope and increasing the early ISW effect (see
Sect.~\ref{sec:matter}). The change in $\Ommh$ in turn raises the value of $H_0$
due to the angular diameter distance degeneracy discussed in
Sect.~\ref{sec:H0physics}. Secondly, the increase in $\ns$ increases the
amplitude of the power spectrum at $\ell\,{\ga}\,550$; this can be compensated
by a lower value of $\clamp$. Thirdly, this shift in $\clamp$ also reduces power
around the first peak, and so yields an increase in $\Ombh$, which increases the
amplitude to partially compensate (through the modulation effect described in
Sect.~\ref{sec:baryons}). Finally, some further adjustments are achieved by
selecting a larger value of $\theta_\ast$, which shifts the position of the
peaks to the left. Comparatively speaking, excising $\ell\,{<}\,30$ from
$\ell\,{<}\,2500$ leads to shifts that are similar to those just described but
of smaller amplitude, since the excised region is a smaller fraction of the
data. Hence, the parameter shifts are smaller without $\ell\,{<}\,30$, as can be
seen in Fig.~\ref{fig:whisker}.

\begin{figure}[htbp!]
\begin{center}
\resizebox{\columnwidth}{!}{\includegraphics{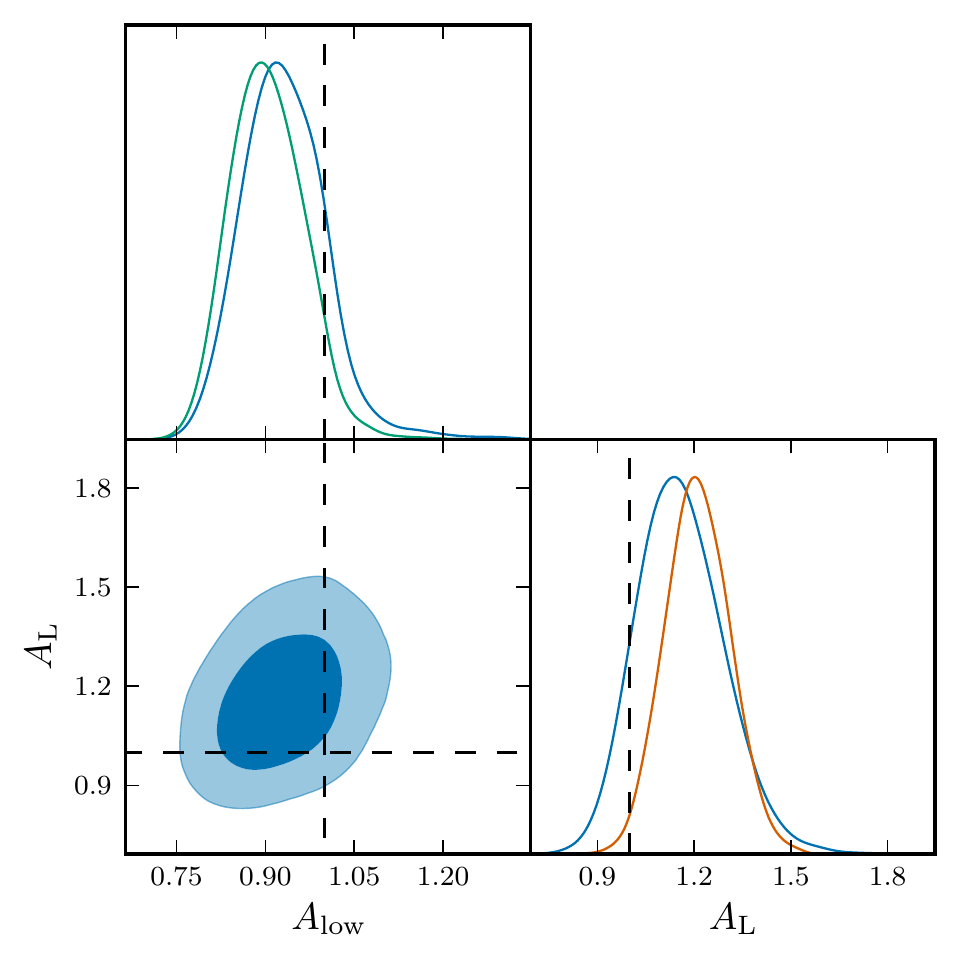}}
\end{center}
\caption {Posterior distributions for $A_{\rm low}$ (which phenomenologically
parametrizes the low-$\ell$ deficit by multiplying the amplitude of the power
spectrum at multipoles smaller than $\ell\,{<}\,30$) and for $A_{\rm L}$
(which parametrizes the peak smoothing effect). We show the results for a
$\Lambda$CDM$+A_{\rm low}+A_{\rm L}$ model (black solid line), for
$\Lambda$CDM$+A_{\rm low}$ (blue) and for $\Lambda$CDM$+A_{\rm L}$ (red).
Although a degeneracy is present between the two parameters, small deviations
with respect to the $\Lambda$CDM expectations remain even when varying both
parameters at the same time. }
\label{fig:triangle}
\end{figure}

As a final check, we have tested the degeneracy between the
low-$\ell$ deficit and the peak smoothing effect. The purpose of this test
is to verify that these are two {\it different\/} effects, and that one
cannot be explained with the other through
degeneracies among cosmological parameters. In order to perform this test,
we use an additional parameter $A_{\rm low}$ that multiplies the amplitude of
the power spectrum at $\ell\,{<}\,30$.
This parametrization does not fully capture the feature at
low-$\ell$, but should be enough for our purpose here, since we verified that
the results we obtain in the \LCDM${+}A_{\rm low}$ case overlap those from
excising completely the $\ell\,{<}\,30$ region.
We then estimate parameters for a \LCDM+$A_{\rm L}$+$A_{\rm low}$ case.
Fig.~\ref{fig:triangle} shows the results of
this exercise. As expected, we find a moderate degeneracy between $A_{\rm L}$
and $A_{\rm low}$, at the level of 30\,\%, which reduces the deviations of both
these parameters. Therefore, when looking at parameter shifts due to one of
these two effects, one has to keep in mind that they are somewhat correlated.
At the same time, since in Fig.~\ref{fig:triangle} both parameters remain
deviant at more than about the $1\,\sigma$ level, this test suggests that
{\it both\/} effects are present and cannot mutually explain each other.

\subsection{Robustness tests}
\label{sec:systematics}

A large number of tests were performed in \cite{planck2014-a13} in order to
validate the robustness of the \Planck\ likelihood against possible systematics
(for more details, see section~5 in that paper). We recall here briefly the
tests performed on the high-$\ell$ $TT$ likelihood, and describe an additional
one that has been added specifically for this work.

The \Planck\ likelihood was tested against methodological (e.g.,
incorrect likelihood
approximations), instrumental (e.g., incorrect instrument characterization) and
astrophysical (e.g., incorrect foreground modelling) systematics,
through specific
tests and the use of simulations. These three sources were shown, to the best of
our knowledge, to introduce a possible bias on cosmological parameters smaller
than about $0.2\,\sigma$.

More specifically, a number of tests were performed to assess the impact of
the use of: ``detset'' cross-spectra in place of ``half-mission'' ones (the
former are less affected by systematics that are uncorrelated between
detectors, the latter by systematics with timescales shorter than half of the
mission); smaller Galactic masks (less contaminated by foregrounds);
Galactic dust template and amplitude priors; beam uncertainties; and frequency
cross-spectra. All of these showed consistent results.

The latter test is particularly interesting. The baseline {\tt Plik}
likelihood at $\ell\,{>}\,30$ uses half-mission cross-spectra from the 100,
143, and 217-GHz frequency
channels.  Consistent results are obtained if one takes out one frequency at a
time. For example, using two frequencies at a time with $\ell\,{>}\,30$,
a prior on $\tau=0.07\pm0.02$, and leaving foregrounds free to vary,
for the Hubble parameter we obtain: $(67.0\pm1.1)\hunits$ for
100 and 143\,GHz; $(67.1\pm 1.1)\hunits$ for
100 and 217\,GHz; and $(66.9\pm1.0)\hunits$ for 143 and 217\,GHz.  These are in
excellent agreement with the final result using all three frequencies,
$(66.9\pm0.95)\hunits$. This indicates that
if the \Planck\ results are affected by systematic effects, then all the main
CMB channels must be affected in a similar way.

Another consistency check comes from the comparison of the results from the
$TT$ spectrum with those obtained from the high-$\ell$ polarization power
spectra. Although known to be affected by small levels of residual
systematics, both $TE$ and $EE$ provide cosmological parameters that are
consistent with those from $TT$. We discuss this point further in Sect.~\ref{sec:polarization}.

\begin{figure*}[htbp!]
\begin{center}
\resizebox{\textwidth}{!}{\includegraphics{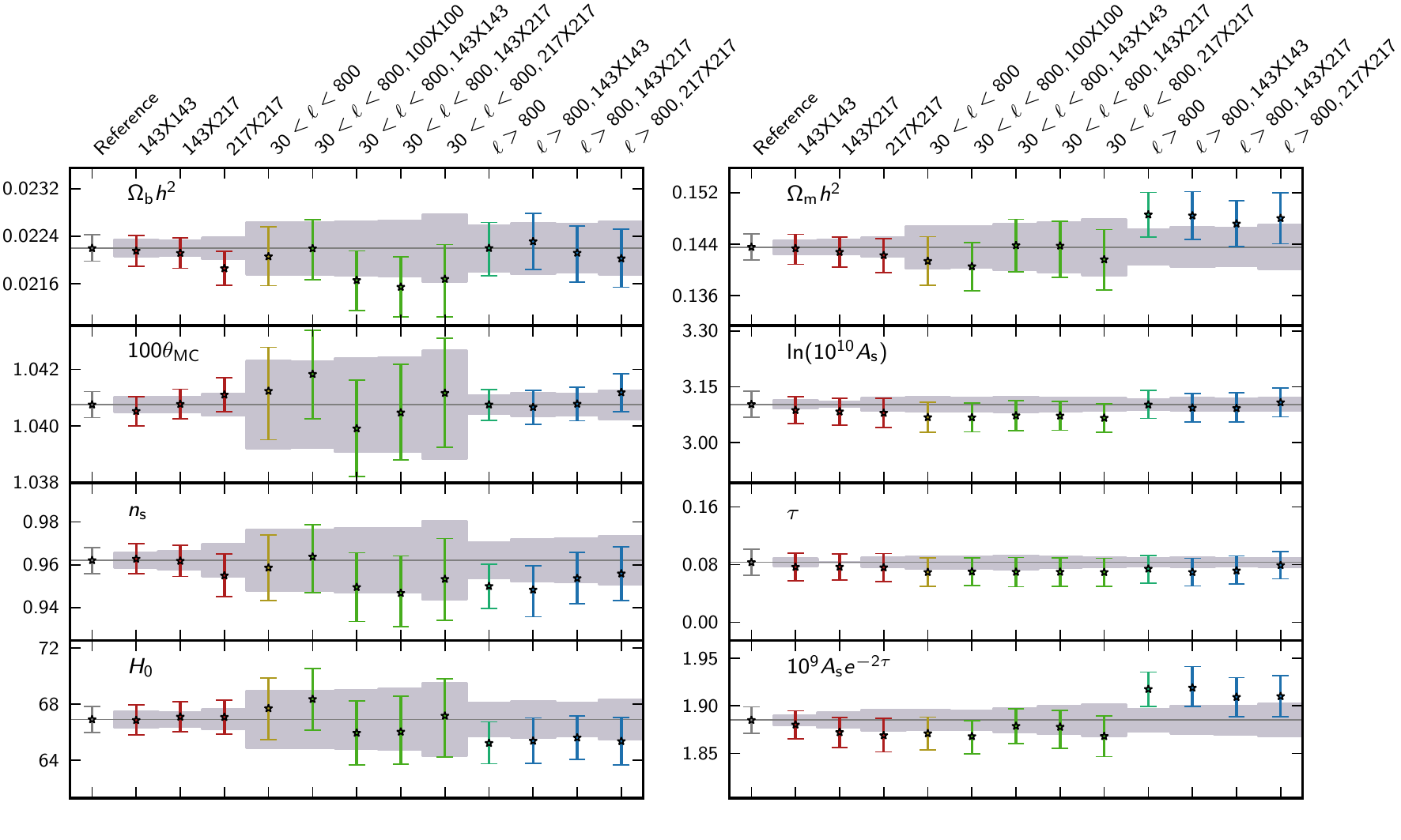}}
\end{center}
\vspace{-0.5cm}

\caption{Constraints on cosmological parameters from data derived from
individual frequencies. The data used is $30\,{<}\,\ell\,{<}\,2500$ unless
otherwise labeled, and in combination with a prior on $\tau$. The reference case
combines all frequencies. The constraints for $30\,{<}\,\ell\,{<}\,800$ and
$\ell\,{>}\,800$ are obtained with foreground parameters fixed to the best fit
of the reference case. The grey band shows the $\pm1\,\sigma$ expected shifts in
cosmological parameters with respect to the reference case \citep[calculated as
in equation~53 of][]{planck2014-a13}. For this test we use the \plikTT\
likelihood, as described in \cite{planck2014-a13}.  Results from individual
frequencies are in very good agreement.}

\label{fig:onefreq}
\end{figure*}

We also present here an additional test to verify that the shifts analysed in
the previous sections are consistently present in different frequency channels.
In order to do this, we estimated cosmological parameters from $\ell\,{<}\,800$
and $\ell\,{>}\,800$ using one frequency spectrum at a time, i.e.,
the $143\times143$, $143\times 217$, or $217\times217$ combinations.
Due to the low resolution of the
$100\times100$ data, for this case we only estimate parameters for
$\ell<800$ . We only use the {\tt Plik} likelihood at $\ell\,{>}\,30$
in combination with a prior on $\tau$.  As shown
in Fig.~\ref{fig:onefreq} we find very good agreement between
the different cases, suggesting that the shifts are not induced by one
particular frequency.
This confirms the findings of \cite{planck2014-a13}.

\begin{figure}[htbp!]
\begin{center}
\resizebox{\columnwidth}{!}{\includegraphics{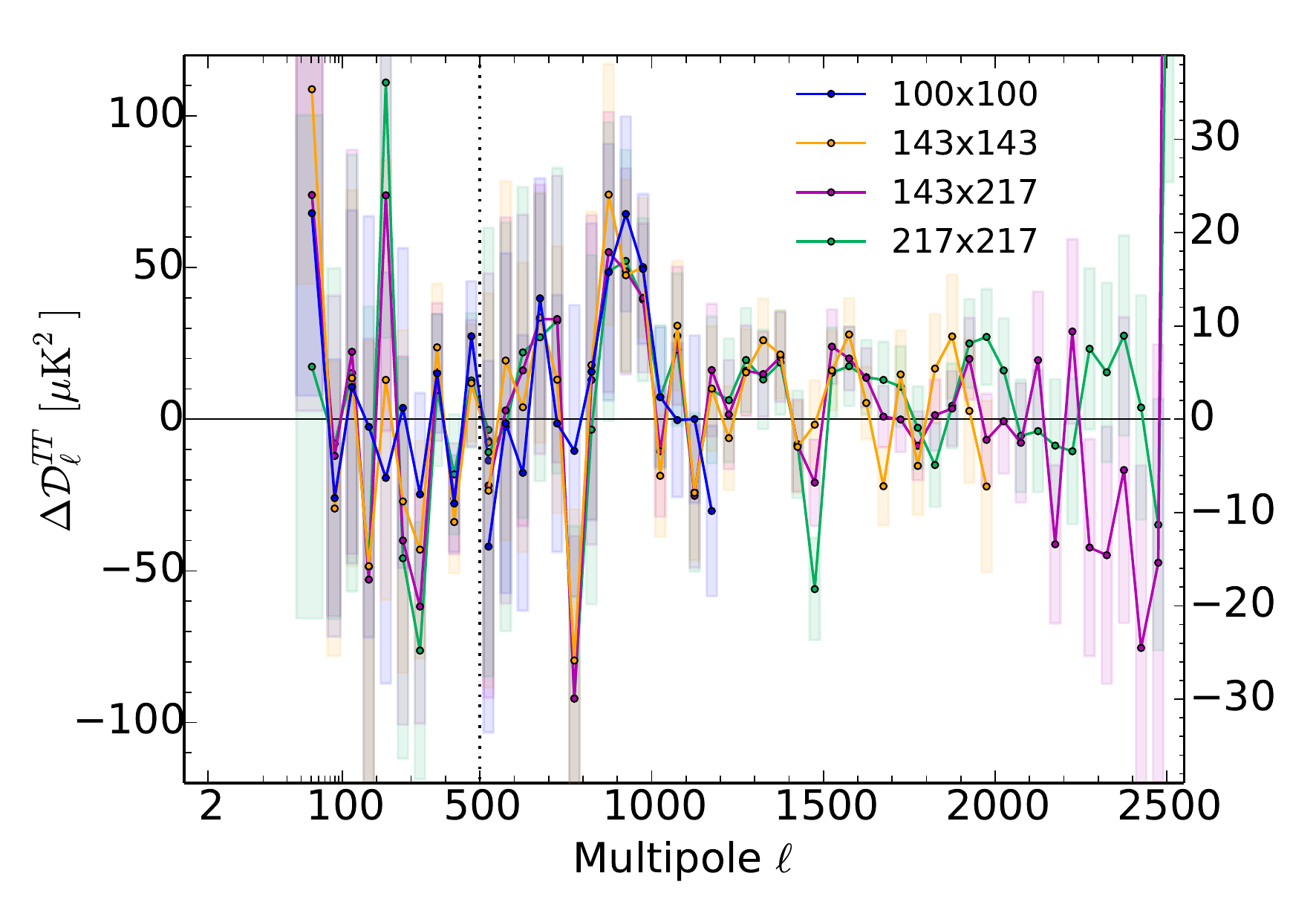}}
\end{center}
\vspace{-0.5cm}
\caption{Residuals for different frequency combinations with respect to the
$\ell\,{=}\,2$--800 best-fit
model. For each frequency we only show the $\ell$ range used in the \Planck\
likelihood. Although these data subsets are noisy, the oscillatory-like
feature seems consistent across frequencies.}
\label{fig:perfreq}
\end{figure}

In Fig.~\ref{fig:perfreq} we also show the frequency residuals with respect to
the best fit of the $\ell\,{<}\,800$ case. We find that the features
identified in Sect.~\ref{sec:shifts} to be driving the shifts are present in
all frequency channels. This also confirms the findings of section~5 of
\cite{planck2014-a13}, which showed good agreement in the comparison of the
inter-frequency residuals.

\subsection{Impact of the $\tau$ prior}
\label{sec:tauprior}

While this paper was being prepared, an updated analysis of \Planck HFI
large-scale polarization data was released
\citep{planck2014-a10}. These results give somewhat smaller
values of the optical depth to reionization, with smaller uncertainties than
from previous results.  The tightest constraint derived is
$\tau\,{=}\,0.055\pm0.009$, with slightly different values resulting from other
choices of data combination and treatment, e.g., $\tau\,{=}\,0.058\pm0.012$ in
\citet{planck2014-a25}.  By comparison, the prior we have been using is
$\tau\,{=}\,0.07\pm0.02$ (which was picked to correspond roughly to previous
\Planck LFI results). This tightening of the error bar and change in the central
value affects the significance of the parameter shifts we have been discussing.
Although this paper could have been written from the beginning with this updated
constraint on $\tau$, we chose not to and instead discuss its impact separately
here because: (1) it does not have a very big impact on the main results of this
paper; (2) the parameter shifts that have been discussed extensively to this
point in the community were the ones coming from the earlier $\tau$ constraint;
and (3) we can more clearly isolate and discuss the effect of the new prior in
this way.

As discussed in \cite{planck2014-a10}, the lower value of $\tau$ leads to some
shifts in \LCDM parameters from the full $\ell$-range. At fixed $A_{\rm
s}e^{-2\tau}$, the main effect of lowering $\tau$ is to reduce $A_{\rm s}$ and
hence reduce the gravitational lensing potential and associated smoothing of the
peaks. A secondary effect of changing $\tau$ at very low $\ell$'s (e.g., see
Fig.~\ref{fig:Derivs}) is too small with respect to the error bars at these
these multipoles to have an appreciable effect. The $\ell\,{<}\,800$ data are
largely insensitive to the peak smoothing, so no other parameters besides $\tau$
and $A_{\rm s}$ are affected (and we note that $A_{\rm s}$ alone is not one of
the six parameters with which we compute the significance of the shifts).
Conversely, the $\ell\,{>}\,800$ data {\it do\/} have sensitivity to
gravitational lensing, hence other parameters try and shift to compensate for
the decreased smoothing of the peaks.  The way that they do this is exactly
along the degeneracy direction discussed in Sect.~\ref{sec:lensing}, which gives
extra peak smoothing and involves increasing $\omega_{\rm m}$ and $A_{\rm
s}e^{-2\tau}$, while reducing $n_{\rm s}$ and $\omega_{\rm b}$. This leads to,
for example, a decrease in $H_0$ of about $0.5\,\hunits$. This is in the
direction of making the shifts slightly more significant.

The exact level of agreement when using the updated constraint on $\tau$ is
summarized in Table~\ref{tab:significances_lowtau}. These numbers come from
running simulations identical to those which led to
Table~\ref{tab:significances} except that we use a prior on $\tau$ of
$0.055\,{\pm}\,0.010$ instead. In practice this means that the prior applied to
each simulation is different, as well as the fiducial model from which the
simulations are drawn, since this model is obtained with $\tau$ fixed to the
mean of the prior (as discussed in Sect.~\ref{sec:simulations}). Generally, the
effective agreement changes by between $-0.1$ and $0.3\,\sigma$, thus slightly
worse. In any case, the differences due to the lower value of $\tau$ do not
qualitatively alter the main conclusions from this paper, and
Table~\ref{tab:significances_lowtau} should be considered our best estimate of
the level of agreement.

Given that we have seen a lower $\tau$ prior increase the significance of the
shifts, we might also ask if a higher $\tau$ prior can reduce them. Indeed, the
\planckTTonly\ data alone do prefer a higher value of $\tau$
\citep{planck2014-a03,Couchot15}, so one might be tempted to think that
perhaps the parameter shifts reflect a
tension between the values of $\tau$ from \planckTTonly\ and from large scale
polarization. To some extent this is true, and we have checked the significance
of the shifts between $\ell\,{<}\,800$ and $\ell\,{<}\,2500$ with a prior of
$\tau=0.10\pm0.02$, finding that they are reduced from $1.4\,\sigma$ to
$1.0\,\sigma$. This is consistent with the results of \cite{Addison15}, who also
showed that a higher value of $\tau$ can reduce the size of the shifts, although
given the results from \Planck HFI polarization it is very unlikely that $\tau$
actually being significantly higher than thought could be a realistic solution
to any tension that might be present.

\section{Comparison with other data sets}
\label{sec:compare}

\begin{figure*}[htbp!]
\begin{center}
\resizebox{\textwidth}{!}{\includegraphics{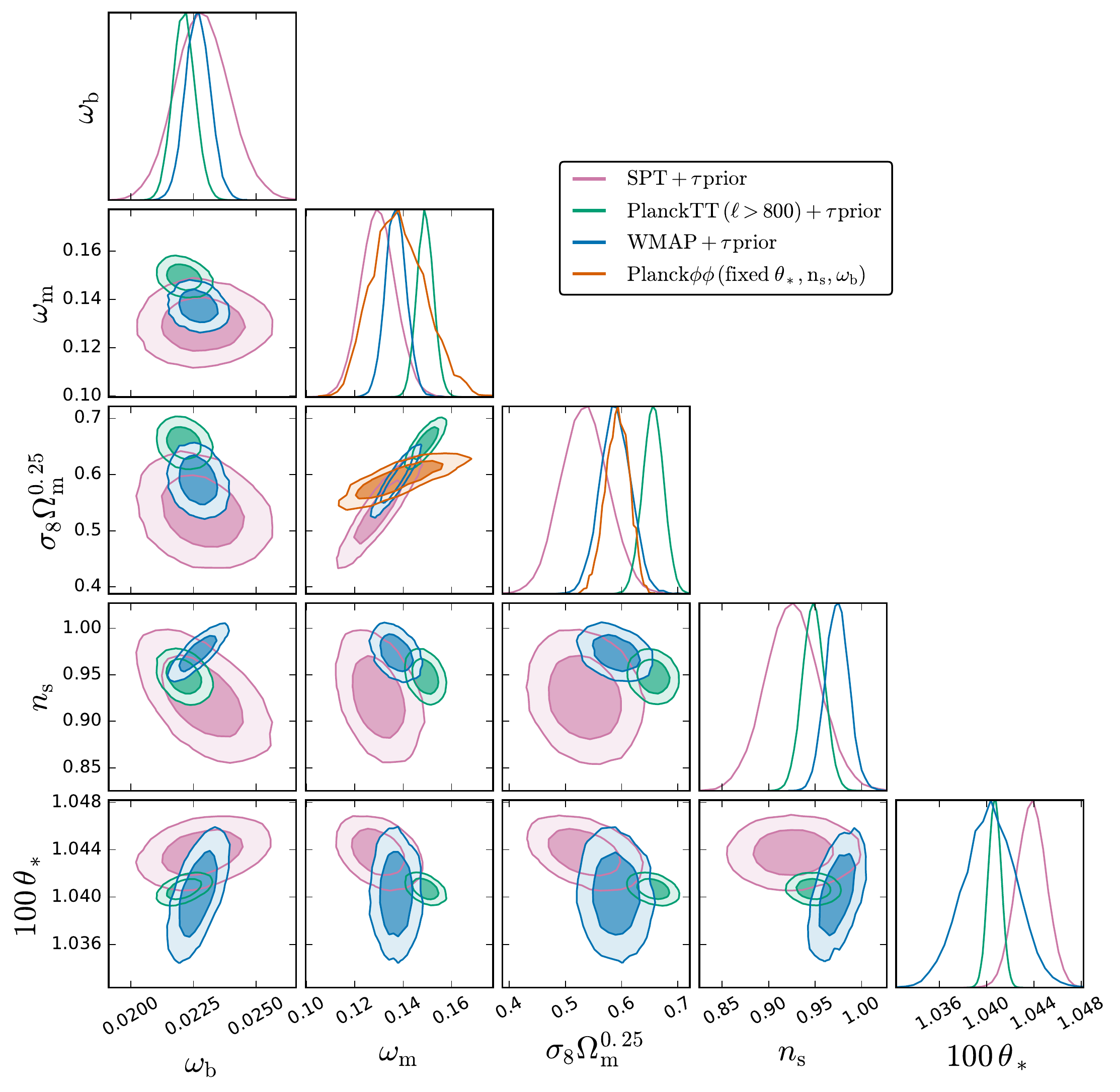}}
\end{center}
\vspace{-0.5cm}

\caption{Constraints on \LCDM parameters from: SPT data from \cite{Story13} in
pink; PlanckTT $\ell\,{>}\,800$ in green; and  WMAP in blue. Except for the
latter data set, which has no sensitivity to $\tau$, all others have been
combined with a prior $\tau=0.07\,{\pm}\,0.02$. The significance of parameter
shifts between these three approximately uncorrelated data sets can be roughly
calculated using Eq.~(\ref{eq:badchi2}). We find no strong evidence of
discrepancies, with SPT and WMAP agreeing at the $1.7\,\sigma$ level, \Planck
$\ell\,{>}\,800$ and WMAP agree even better at $1.1\,\sigma$, while \Planck
$\ell\,{>}\,800$ and SPT agree with each other at $2.1\,\sigma$. Also plotted in
orange is Planck$\phi\phi$ with $\theta_\ast$, $\omega_{\rm b}$, and $n_{\rm s}$
fixed to the \Planck\ best-fit values. This data set, across the two parameters
it constrains, is also not in significant tension with the others.
Sect.~\ref{sec:compare} discusses these comparisons in more detail.}

\label{fig:cmb_compare}
\end{figure*}

Having considered the internal consistency of the \planckTTonly\ data
themselves, as well as implicitly considering the comparison with WMAP,
we now extend our discussion to a number of other CMB data sets.
Although many measurements and analyses of the CMB have been made that have a
bearing on agreement with \Planck\
\citep[e.g.,][]{Calabrese13,Story13,Das14,Louis14,Naess14,George15}, it is
impossible here to discuss them all in detail.  We thus limit ourselves only to
those that are the most constraining on \LCDM parameters and therefore have the
power to test the level of consistency most stringently. We will specifically
consider the \Planck\ $TE$, $EE$, and $\phi\phi$ power spectra, as well as
measurements of the $TT$ damping tail from \citet{Story13}.

\subsection{Comparison with \textit{Planck} polarization}
\label{sec:polarization}
The first analysis of \Planck\ high-$\ell$ $TE$ and $EE$ spectra was presented
in \citet{planck2014-a13}. Consistency between parameters obtained from $TE$ and
$EE$ with those obtained from $TT$ was discussed in \citet{planck2014-a15},
which showed that error bars on \LCDM parameters obtained from $TE$ alone are of
similar magnitude to those from $TT$, and the best-fit values are generally
within $0.5\,\sigma$. For example, from \plikTE +\tauprior\ we find
$H_0=(67.9\,{\pm}\,0.93)\hunits$ as compared to $(66.9\,{\pm}\,0.95)\hunits$
from \plikTT+\tauprior.  The $EE$ constraints are considerably noisier, but
generally within $1\,\sigma$, with \plikEE+\tauprior\ giving
$H_0=(70.0\,{\pm}\,2.8)\hunits$, for example.  Because cosmic
variance partially correlates the $TE$ and $EE$ constraints with those from
$TT$, determining the exact level of consistency requires simulations. This
study was discussed in appendix C.3.6 of \citet{planck2014-a13},
where it was found that the cosmological parameters obtained from $EE$ and $TE$
are in agreement with those obtained with $TT$. Given that there are still
some residual systematic effects in
the polarization spectra, which prevented them from being used for the baseline
parameters for the 2015 \Planck\ release \citep{planck2014-a13}, we stop at
this point, rather than performing any more sophisticated tests.  Further
comparisons will be made following the next \Planck\ data release.

\subsection{Comparison with SPT}

The tightest constraints on \LCDM parameters obtained from the $TT$ damping tail
with a single experiment other than \Planck\ come from the South Pole Telescope
\citep[SPT, as presented in][]{Story13}. As such, assessment of the level of
consistency between the two is of great interest. Disagreement between the two
data sets has been claimed as an argument that the parameter shifts we
have been discussing are not of cosmological origin \citep{Addison15}. Although
a more detailed comparison is outside of the scope of this paper, we perform a
few basic tests of compatibility here, showing that any tension between
\Planck\ and SPT is not very statistically significant.

On their own, the SPT data are not very constraining on \LCDM parameters because
the sky coverage is about a factor of 10 times smaller than \Planck's. If we
limit \Planck to $\ell\,{>}\,800$, roughly the same multipoles measured by SPT,
the errors on all \LCDM parameters are twice as large or more, as can be seen by
comparing the green and pink contours in Fig.~\ref{fig:cmb_compare}. Combining
SPT with WMAP yields somewhat tighter \LCDM constraints, although still larger
than \Planck's full-$\ell$ range. It is not straightforward to compare \Planck
and WMAP+SPT because both \Planck and WMAP are cosmic variance limited at low
multipoles and hence very correlated. Instead, we will limit ourselves to data
sets that are uncorrelated and use Eq.~(\ref{eq:badchi2}), which we will apply
to the five parameters shown in Fig.~\ref{fig:cmb_compare}. This will suffer
from all of the problems mentioned in Sect.~\ref{sec:expectations}, but will
still give us a rough idea of the level of agreement. For WMAP+SPT versus
\Planck $\ell\,{>}\,800$ we find $\chi^2\,{=}\,12.0$, which is equivalent to a
2.1\,$\sigma$ fluctuation. SPT alone compared to \Planck $\ell\,{>}\,800$
yields $\chi^2\,{=}\,11.9$, also equivalent to 2.1\,$\sigma$. We can
additionally compare SPT to the \Planck full multipole range, which gives
$\chi^2\,{=}\,12.3$, equivalent to 2.2\,$\sigma$. Although we cannot
compare WMAP+SPT and \Planck directly, we already know from \cite{Kovacs13} that
WMAP and \Planck agree extremely well over the common multipole range.
Therefore, we would expect WMAP+SPT and \Planck parameters to be consistent to a
similar level as the numbers just quoted.

Additionally, we point out that despite the impression sometimes given, both
implicitly and explicitly, that the \Planck high-$\ell$'s are ``anomalous'' with
respect to parameters derived from WMAP, the same and more can be said of the
SPT parameters. Again using Eq.~(\ref{eq:badchi2}) and the five \LCDM parameters
shown in Fig.~\ref{fig:cmb_compare}, WMAP and SPT agree to within $1.7\,\sigma$,
while WMAP and \Planck $\ell\,{>}\,800$ are in {\it better} agreement,
$1.1\,\sigma$. Of course, given the significances we have seen in this section,
the point is that we find no strong evidence for disagreement between any of
these different CMB data sets.

\subsection{Comparison with \textit{Planck} lensing}
\label{sec:Plancklensing}

Finally, we consider the level of
agreement with the power spectrum of the gravitational
lensing reconstruction from \Planck\ data. It has previously been noted that
there is some tension between this data set and \planckTTonly\
\citep{planck2014-a15,planck2014-a17,Addison15}.

One way to quantify agreement is via constraints on the $A_{\rm L}$ parameter.
As described in Sect.~\ref{sec:lensing}, this scales the gravitational lensing
potential used in the calculation of the $TT$ spectrum. A similar parameter,
usually called $A_{\phi\phi}$, can be introduced when computing constraints from
PlanckTT+lensing, this time scaling the lensing potential used in the lensing
likelihood (but not the one used in the $TT$ spectrum calculation). We find
$A_{\rm L}=1.21\,{\pm}\,0.10$ from \TT, compared to
$A_{\phi\phi}=0.95\,{\pm}\,0.04$, a difference of $2.6\,\sigma$. This
comparison, however, is somewhat misleading because $A_{\phi\phi}$ and $A_{\rm
L}$ are a rescaling of the lensing potential with respect to two different
models. If we remove these intermediary models and compare directly the lensing
power preferred by the two data sets, for example $C_\ell^{\phi\phi}$ at
$\ell=100$, agreement is instead 2.3\,$\sigma$.

Another way to compare these data sets, which has the advantage that it assumes
\LCDM unlike the previous case, is to simply analyse each data set independently
given the \LCDM model and compare constraints on parameters. These constraints
are shown in Fig.~\ref{fig:cmb_compare}, in orange for \planckTTonly\
and in green for
lensing (the lensing data assume a fixed $\theta_\ast$, although are largely
insensitive to the exact value). The parameter most often compared is $\sigma_8
\Omega_{\rm m}^{0.25}$ because it is a good proxy for the amplitude of the
lensing potential and is most tightly constrained by the lensing data. Here, we
find $\sigma_8 \Omega_{\rm m}^{0.25}=0.600\,{\pm}\,0.011$ from Planck$\phi\phi$
and $0.623\,{\pm}\,0.013$ from \planckTTonly,
a difference of $1.3\,\sigma$. We note
that this agreement becomes even better with the addition of the lower prior on
$\tau$ discussed in Sect.~\ref{sec:tauprior}.

As pointed out by \cite{Addison15}, despite this good agreement over the full
$\ell$-range, the constraint on $\sigma_8 \Omega_{\rm m}^{0.25}$ from just the
$\ell\,{>}\,1000$ data is in tension with lensing at $2.4\,\sigma$. Unlike for
the full $\ell$-range, however, constraints from $\ell\,{>}\,1000$ on a second
parameter, $\omega_m$, are now comparable to those from lensing, hence it makes
sense to include this in the comparison. This slightly reduces the tension to
$2.2\,\sigma$.

\cite{Addison15} further pointed out that the quantity $\sigma_8 \Omega_{\rm
m}^{0.25}$ is internally inconsistent within the \Planck temperature data
themselves at a level of $2.9\,\sigma$ between $\ell\,{<}\,1000$ and
$\ell\,{>}\,1000$. We find instead $2.5\,\sigma$. The most likely source of
difference is that we use \texttt{plik\_lite}, which we believe gives the more
correct result, since it imposes more reasonable priors on the foreground
parameters and thus reflects more realistically our knowledge of foreground
contamination.

To conclude this section, although it is possible to single out specific
parameter differences, overall we find no significant evidence of any
strong discrepancies between the \planckTTonly\ and Planck$\phi\phi$ data.

\section{Conclusions}
\label{sec:conclusions}

The main goals of this paper have been threefold: (i) to isolate the features in
the \Planck\ $\ell\,{>}\,800$ temperature power spectrum that cause the shifts
in parameters away from the $\ell\,{<}\,800$ (or similarly WMAP) parameters;
(ii) to assess the consistency of these shifts with expectations; and (iii) to
provide an explanation of the physics behind why the parameters are shifting.
In our view, such a physical explanation and this ``opening of the likelihood
black box'' serves to assuage some of the concern that one might initially have
about the apparently unlikely nature of some of the shifts, and hence increases
the confidence one places in the \Planck\ data. While some discussions of points
(i) and (ii) have already appeared in the literature, we have greatly expanded
and clarified them here.

In particular, we have made extensive use of numerical simulations in order to
evaluate the consistency of the results obtained from a large number of
different multipole ranges. This allowed us to properly account for the
correlations between the different $\ell$ ranges and compute the exact
posterior distribution of the expected parameter shifts, avoiding the use of a
Gaussian approximation, contrary to what was
done in previous studies. In evaluating
the probability of a shift in the most deviant parameter out of the six
$\Lambda$CDM ones, we also pointed out the importance of taking into account
look-elsewhere effects (i.e., marginalizing over the set of parameters).

We have found that the cosmological parameters inferred from $\ell\,{<}\,800$
versus the full multipole range $\ell\,{<}\,2500$ in the context of the \LCDM
model are consistent with each other within approximately 10\,\% PTE. We find
similar significance levels when evaluating the probability of shifts in the
most deviant parameters, when comparing high-$\ell$ data with low-$\ell$,
or when splitting at multipoles other than $\ell\,{=}\,800$.
Table~\ref{tab:significances} and Fig.~\ref{fig:stats_lsplit_scan} summarize
these results. In light of the recent \Planck\ results on the reionization
optical depth \citep{planck2014-a10,planck2014-a25}, we find that using a lower
and tighter prior of $\tau=0.055\pm0.010$ has a mild impact on the significance
levels of the parameter shifts, increasing them by about $0.3\,\sigma$.

The discussion of point (iii), i.e., explaining the physics underlying the
shifts, has not previously existed at all. While we point out that
the interpretation
of the shifts is not unique, we provide one possible explanation by connecting
features in the spectra with shifts in parameters. We find that when reducing
the lever arm of the data by only using the larger angular scales
($\ell\,{<}\,800$), cosmological parameters are more strongly affected by the
low-$\ell$ deficit, i.e., the apparent lack of power at
$\ell\,{<}\,30$. To decrease
power at $\ell\,{<}\,30$, $n_{\rm s}$ increases, $A_{\rm s}e^{-2\tau}$ is then
lowered to reduce power at $\ell\,{\ga}\,500$, $\Ommh$ decreases to compensate
the induced change of power below $\ell\,{\simeq}\,500$, while $\Ombh$ increases
to reduce the amplitude of the second peak (which was raised by the decrease in
$\Ommh$). The Hubble constant is in turn pulled high to keep the angular size of
the horizon unchanged.

On the other hand, we find that the small-scale results are influenced by the
preference for a larger smoothing of the power spectrum peaks and troughs at
$\ell\,{\ga}\,1000$. While at face value it might seem like this smoothing is
the sign of an excess amplitude of gravitational lensing, we find that most of
the shifts in the \LCDM\ parameters serve not to increase the lensing potential,
but rather to fit these features through non-lensing related effects. While
neither the peak smoothing nor low-$\ell$ features are statistically very
significant, and could just be statistical fluctuations in the data, we show
that they can explain a large part of the observed parameter shifts.

In summary, we have identified the main features of the data leading to the
observed parameter shifts and explained the physics of why the parameters of
the $\Lambda$CDM model adjust in the way they do to fit these features.
Further, we find that these shifts are not in strong disagreement with
expectations for the size of such differences among a set of parameters;
thus there is no requirement to explain such shifts with either systematic
effects or new physics.

\begin{acknowledgements} The Planck Collaboration acknowledges the support of:
ESA; CNES, and CNRS/INSU-IN2P3-INP (France); ASI, CNR, and INAF (Italy); NASA
and DoE (USA); STFC and UKSA (UK); CSIC, MINECO, JA, and RES (Spain); Tekes,
AoF, and CSC (Finland); DLR and MPG (Germany); CSA (Canada); DTU Space
(Denmark); SER/SSO (Switzerland); RCN (Norway); SFI (Ireland); FCT/MCTES
(Portugal); ERC and PRACE (EU). A description of the Planck Collaboration and a
list of its members, indicating which technical or scientific activities they
have been involved in, can be found at
\href{http://www.cosmos.esa.int/web/planck/planck-collaboration}{\texttt{http://www.cosmos.esa.int/web/planck/planck-collaboration}}.
Part of the analysis for this paper was run on computers operated by
WestGrid (\url{www.westgrid.ca}) and Compute Canada
(\url{www.computecanada.ca}).
This work was also supported by the Labex ILP (reference ANR-10-LABX-63).  We
thank all of the users of Cosmology@Home for donating computing time in
support of this work, and in particular the top contributors,
MaDcCow (Thomas Wooton), 25000ghz (Roberto
Piantoni), and Rally1965, as well as the top team, BOINC.Italy.
\end{acknowledgements}

\bibliographystyle{aat}
\bibliography{Planck_bib,plancklcdm}

\appendix

\section{A more exhaustive set of tests}
\label{app:uberstats}

The main focus of this paper has been on shifts between parameters derived
from $\ell\,{<}\,800$ data and those from $\ell\,{<}\,2500$ data.
We considered this the most interesting choice because $\ell\,{=}\,800$ evenly
splits the Fisher information on \LCDM parameters in the \planckTTonly data;
additionally, we focused on low-$\ell$ parameters
versus full-$\ell$ parameters (as opposed to low-$\ell$ versus high-$\ell$),
since this is most directly relevant for the issue of WMAP versus
\Planck\ parameter shifts.

Despite this decision,
we would like to know if our particular choice of $\ell_{\rm split}\,{=}\,800$
greatly affected results, either making them seem more or less consistent than
otherwise. Additionally, in terms of a generic test of the \Planck\ data, there
are many other data splits that one might consider to test the consistency even
more stringently. We present results from a more exhaustive set of such tests in
this appendix. More specifically, we look at three different ways of
splitting the data:
\begin{enumerate}
\item $\ell\,{<}\,\ell_{\rm split}$ vs. $\ell\,{<}\,2500$;
\item $\ell\,{<}\,\ell_{\rm split}$ vs. $\ell\,{>}\,\ell_{\rm split}$;
\item $\ell\,{<}\,\ell_{\rm split}$ vs. $\ell\,{>}\,\ell_{\rm split}+50$.
\end{enumerate}
We do this at several different values of $\ell_{\rm split}$ across the range
allowed by our simulations. For each case, we compute the $\chi^2$ and {\texttt
max-param} statistics.

Of course, since we are now explicitly scanning over statistical tests,
we need to
account for a posteriori corrections to interpret the significance of any
outliers we find. This is the same effect already discussed in the context of
searching for a maximally discrepant parameter, but now for finding a maximally
discrepant partitioning of the data. It is straightforward to calculate these
corrections based on the suite of simulations. For each realization, we search
for the most discrepant result as a function of $\ell_{\rm split}$. We then
compare the result on the real data against this distribution and compute a PTE
as before.

\begin{figure*}
\begin{center}
\resizebox{\textwidth}{!}{\includegraphics{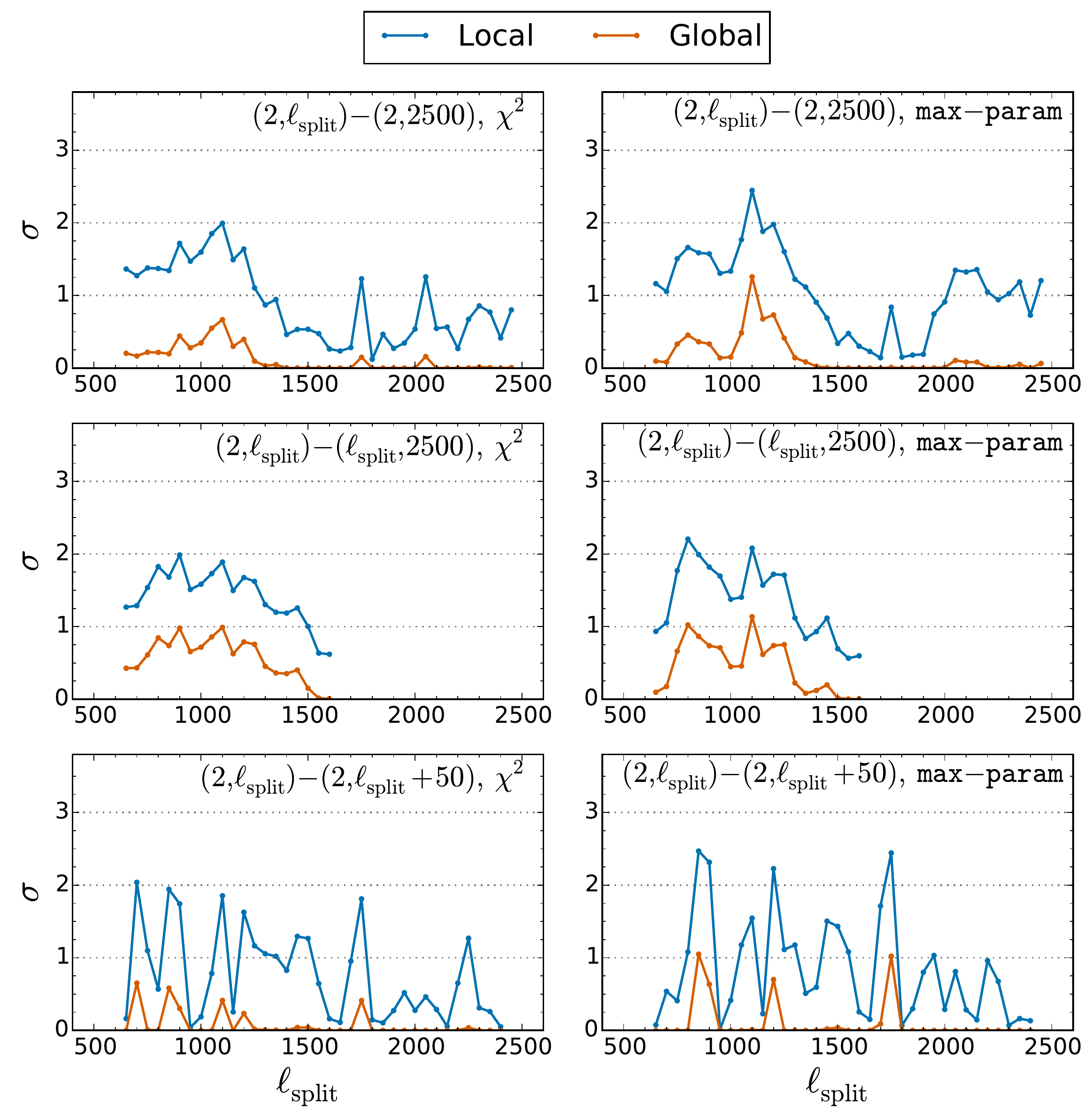}}
\end{center}
\vspace{-0.5cm}

\caption{Significance levels (in units of effective $\sigma$) of the parameter
shifts between two multipole ranges, according to a given statistic, as a
function of $\ell_{\rm split}$. The specific choice of the two multipole ranges
and the statistic used are labelled on each panel.  The blue line is the
``local'' significance, calculated as described in Sect.~\ref{sec:results}. The
orange line is the ``global'' significance which should be used to interpret the
significance of any outliers we find (see Appendix~\ref{app:uberstats} for
further description).}

\label{fig:stats_lsplit_scan}
\end{figure*}

We have computed results varying $\ell_{\rm split}$ between 650 and 2500 with a
step size $\Delta\ell=50$. The results are shown in
Fig.~\ref{fig:stats_lsplit_scan}. The blue line shows the raw (so-called
``local'') significance for each case, computed exactly as described in
Sect.~\ref{sec:results}. The significance shows considerable scatter, as one
might expect due to noise, with no outlier above roughly $2.5\,\sigma$. We see
that any other choice of $\ell_{\rm split}$ in the vicinity of 800 would have
given the same qualitative results that we have focused in the main body of this
paper.

If we search for the $\ell_{\rm split}$ which gives the largest local
significance, we need to account for the look-elsewhere effect to interpret the
true significance of this outlier. This is given by the orange line and labeled
``global.'' For example, if for some $\ell_{\rm split}$ we find a local
significance of $2\,\sigma$, then the global significance is the fraction of
simulations for which we find a shift at {\it any} $\ell_{\rm split}$ with a
local significance exceeding $2\,\sigma$. Generally speaking, this
marginalization lowers the significance of any outliers we find by around
1\,$\sigma$. To be clear, we are not claiming the actual significance of the
shifts presented in the main body of the paper are lower by $1\,\sigma$, since
we did not choose $\ell_{\rm split}\,{=}\,800$ based on finding a most
discrepant data split. Nevertheless, if we now look through
Fig.~\ref{fig:stats_lsplit_scan} for outliers (for example the roughly
$2.5\,\sigma$ outlier in the top right panel at $\ell_{\rm split}\,{=}\,1100$),
it is clear that the true significance is somewhat lower. The conclusion after
this wider set of tests is that we find no evidence for any inconsistency in the
data that was hidden by our specific choice of data partitioning.

\section{The low-\Bell approximation}
\label{app:lowl}
The simulations used in this paper make use of an approximate low-$\ell$
likelihood, as discussed in Sect.~\ref{sec:simulations}. Our main check of this
approximation, as described in that section, is to estimate parameters from
$\ell\,{<}\,800$ with the actual \texttt{Commander} likelihood swapped out for
our approximate likelihood applied to the \texttt{Commander} CMB map. The
$\ell\,{<}\,800$ case is important because it gives more weight to the low
multipoles than, for example, $\ell\,{<}\,2500$; hence it is a more stringent
test of the approximation. In either case, we find that all $\Lambda$CDM
parameters are within 0.05\,$\sigma$ and thus that the approximation is good
enough.

Of course, this test relies on one particular realization of the CMB (namely,
our actual CMB sky), and it is technically possible that
this realization randomly conspired to make our approximation seem better than
it actually is. In this appendix we therefore describe a further test that
looks at many different realizations.

If our low-$\ell$ approximation is correct, it should be the case that the mean
of the best-fit values from the simulations recovers the input fiducial
parameters, and the scatter in the simulations should be the same as the
posterior constraints from an MCMC chain run with the \texttt{Commander}
likelihood. An error in the approximation at low $\ell$, even just in the error
bars, could manifest itself as both a bias in the mean of the best-fit
parameters and a scatter that does not match the true posterior.

\begin{figure*}
\begin{center}
\resizebox{\textwidth}{!}{\includegraphics{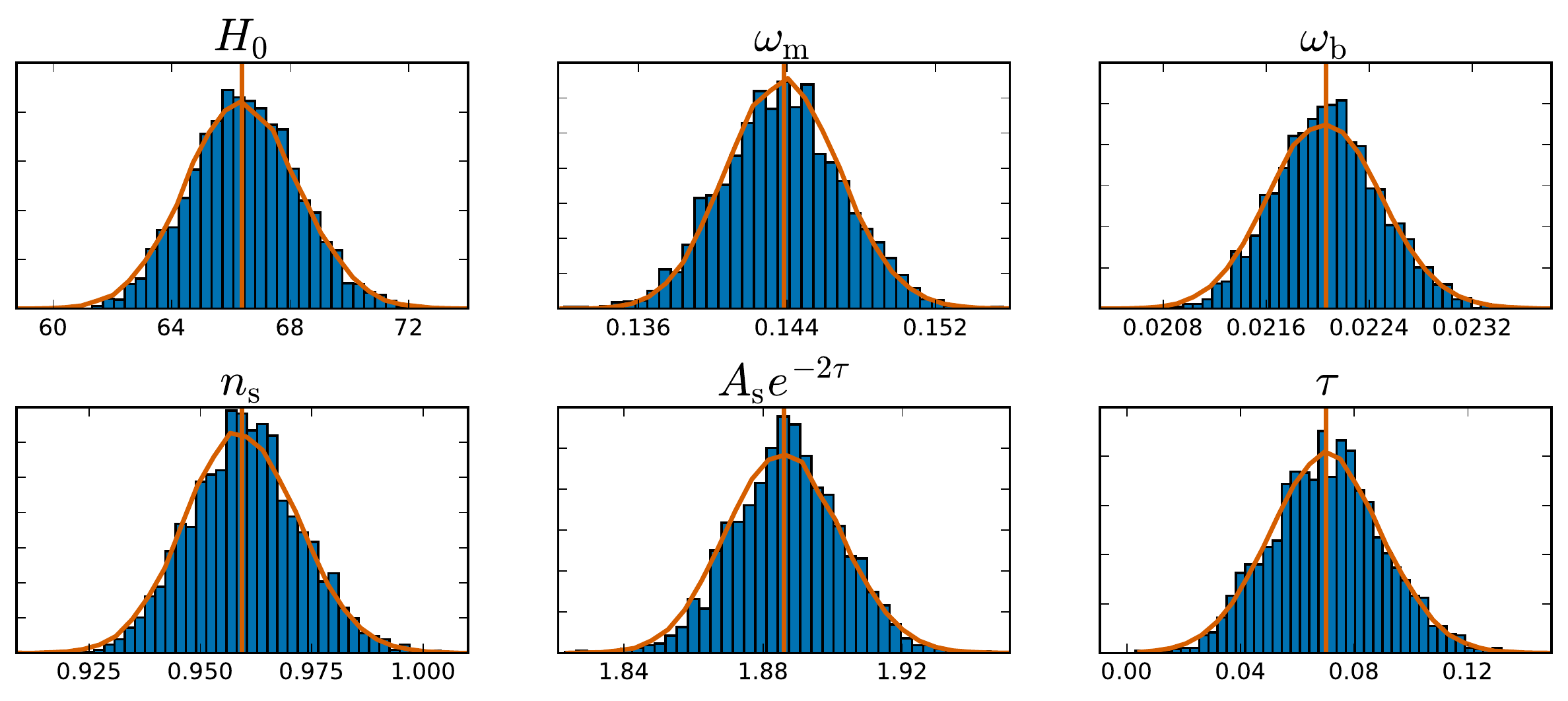}}
\end{center}
\vspace{-0.5cm}

\caption{Histograms showing the distribution of best-fit $\ell\,{<}\,800$
parameters from simulations performed using our low-$\ell$ approximation. The
vertical line is the input fiducial model and the contours show the posteriors
from an $\ell\,{<}\,800$ chain using the actual \texttt{Commander} likelihood at
low $\ell$. The unbiased recovery of the fiducial parameters and agreement with
the posteriors is a stringent test of the validity of our low-$\ell$
approximation. We note that these simulations, unlike the ones used in the main
body of the paper to determine significance levels,
have the prior on $\tau$ handled
slightly differently, so as to allow us to use them as a test of the low-$\ell$
approximation (see Appendix~\ref{app:lowl} for discussion).}

\label{fig:sims_sample_tau}
\end{figure*}

In Fig.~\ref{fig:sims_sample_tau} we show a distribution of the best-fit values
from simulations for the $\ell\,{<}\,800$ case, along with the input fiducial
values and the posteriors from a chain (which have been re-centred on the
fiducial values). However, there is one detail different about these simulations
than the ones used in the main body of the paper. Whereas those all have the
same prior on $\tau$ applied (so as to be consistent with what is done to the
real data), these simulations have a different prior for each realization;
the prior is still Gaussian with a width of 0.02, but its mean has been randomly
sampled from $0.07\pm0.02$ itself. This is akin to having drawn realization of
the low-$\ell$ polarization data, and although it has no bearing on the accuracy
of the low-$\ell$ approximation, it is necessary in order that the scatter
actually matches the posterior. We find then, as expected, that the simulations
are centred on the fiducial values to within the scatter expected from the
finite number of simulations, and the distribution does indeed track the
posterior constraint. We therefore conclude that our low-$\ell$ approximation is
sufficient and our previous determination of its accuracy on the real data was
not affected by our particular realization of the CMB. We stress that this is
not an easy test to pass; for example, we have checked that had we used the
traditional $f_{\rm sky}$ approximation this test would have failed noticeably.

\end{document}